\documentclass[useAMS,usenatbib]{mn2e}
\usepackage{graphicx}
\usepackage{amssymb,amsmath}
\usepackage{rotating}

\newcommand{\apj}{ApJ}

\newcommand{\mnras}{MNRAS}

\newcommand{\MC}{\multicolumn}
\newcommand{\kms}{km~s$^{-1}$}

\newcommand{\HI}{H{\sc i}}
\newcommand{\HII}{H{\sc ii}}
\newcommand{\sunn}{$_{\odot}$}

\newcounter{qub}
\newcommand{\qq}{\addtocounter{qub}{1}\arabic{qub}}

\title[Void XMP gas-rich dwarfs: spectroscopy at BTA]
{XMP gas-rich dwarfs in nearby voids: results of BTA spectroscopy}
\author[S.~Pustilnik, E.~Egorova, A.~Kniazev, Y.~Perepelitsyna, A.~Tepliakova,
 A.~Burenkov, D.~Oparin]
{S.A.~Pustilnik,$^{1}$\thanks{E-mail: sap@sao.ru (SAP)} E.S.~Egorova,$^{2}$
 A.Y.~Kniazev,$^{3,4,2}$  Y.A.~Perepelitsyna,$^{1}$
\newauthor A.L.~Tepliakova,$^{1}$ A.N.~Burenkov,$^{1}$ D.V.~Oparin$^{1}$ \\
$^1$ Special Astrophysical Observatory of RAS, Nizhnij Arkhyz,
      Karachai-Circassia 369167, Russia \\
$^2$ Sternberg Astronomical Institute of Moscow State University, Moscow,
Russia \\
$^3$ South African Astronomical Observatory, PO Box 9, 7935 Observatory, Cape Town,South Africa \\
$^4$ Southern African Large Telescope Foundation, PO Box 9, 7935 Observatory, Cape Town, South Africa
}

\begin{document}

\label{firstpage}

\date{Accepted: 2021 July 13; Revised: 2021 June 27; in original form 2020 October 19}

\pagerange{\pageref{firstpage}--\pageref{lastpage}} \pubyear{2021}

\maketitle

\begin{abstract} 

We present the second part of results of the ongoing project aimed at
searching for and studying eXtremely Metal-Poor (XMP) -- adopted as those
with Z$_{\rm gas}$ $\lesssim$ Z\sunn/30, or with 12+log(O/H) $\lesssim$
 7.21~dex -- very gas-rich blue dwarfs in voids.
They were first identified in the course of the 'unbiased' study of the galaxy
population in the nearby Lynx--Cancer void. These very rare and unusual
galaxies seem to be the best proxies of so-called Very Young Galaxies
(VYGs) defined recently in model simulations by Tweed et al. To date, for
16 pre-selected void XMP candidates, using the Big Telescope Alt-azimuth
(BTA), the SAO 6-m telescope, we have obtained spectra suitable for the
determination of O/H. For majority of the observed galaxies, the principal
line [O{\sc iii}]$\lambda$4363, used for the direct
classical T$_{\rm e}$ method of O/H determination, is undetected. Therefore,
to estimate O/H, we use a new 'strong-lines' method by Izotov et al.
This appears to be the most accurate empirical O/H estimator for the range of
12+log(O/H)$\lesssim$ 7.4--7.5. For objects with higher O/H, we use
the semi-empirical method by Izotov \& Thuan with our modification accounting
for variance of the excitation parameter O$_{\mathrm 32}$.
Six of those 16 candidates are found, with confidence, to be XMP dwarfs.
In addition, eight studied galaxies are less metal-poor, with
12+log(O/H)=7.24--7.33, and these can also fall into the category of VYG
candidates. Taking into account our recently published work and the previously known
(nine prototype galaxies) XMP gas-rich void objects, the new findings increase
the number of this type of galaxy known to date to a total of 19.
\end{abstract}


\section[]{Introduction}
\label{sec:intro}
\setcounter{figure}{0}

This paper is the third in the series devoted to searching for and studying
the most metal-poor dwarfs in the nearby voids. In the introductions to
the two preceding papers
\citep[hereafter referred to as PEPK20 and PKPE20][]{PEPK20,XMP.SALT}
we described in detail
the previous work and motivation for the ongoing project. Therefore, here
we only briefly outline the main points of the earlier searches for the
most metal-poor galaxies and the results obtained so far.

The prototype galaxy with the extremely low gas metallicity
(Z$_{\rm gas}$ $\sim$ Z\sunn/30)\footnote{The solar value of 12+log(O/H)=8.69
dex is adopted according to \citet{Asplund09}}
 was discovered about half a century ago \citep{Searle72}.
This is the famous blue compact dwarf galaxy IZw18 (= MRK~116).
Being remarkable for its intense star-formation burst, IZw18 remained for
a long time the focus of studies in many wavelengths, including the
{\it Hubble Space Telescope (HST)}
study of its resolved stellar population. Early ideas about IZw18 suggested
that it is a very young galaxy that formed its first stars several hundred
thousand years ago.
When its red giant branch (RGB) stars were detected with the help of the {\it HST},
its age was increased to at least $\sim$1--2~Gyr.
Meanwhile, all the accumulated observational data were
reconsidered carefully by \citet{PO12}, who found that a large contribution
of the overall nebular emission should be accounted for in various photometric
parameters. They concluded that however an older stellar population may be present,
the major part of the stellar mass in IZw18 formed during the last $\lesssim$1 Gyr.
For its companion, IZw18C, the respective age estimate is $\lesssim$0.2~Gyr.

The search for IZw18 analogues has continued since then, mostly among
star-forming emission-line galaxies. An important step in this
long-lasting process was the discovery of SBS0335--052E and its companion
SBS0335--052W, which have similar and lower gas metallicity
\citep{Izotov1990, Pustilnik97, Izotov09}.

Because of the results of massive spectroscopy surveys of
faint galaxies, mostly of the Sloan Digital Sky Survey \citep[SDSS; ][]{DR7, DR14}.
there has been a substantial advancement in the discovery of
eXtremely Metal-Poor (XMP) galaxies.
Several groups used the SDSS to search for very metal-poor candidates
and to check them with the high-quality follow-up spectroscopy
\citep[e.g.][]{ITG12, Izotov16, Guseva17, Sanchez16}.
In particular, about 300 galaxies were found with the
gas metallicity of Z $\lesssim$Z\sunn/10,
or 12+log(O/H) $\lesssim$7.7~dex \citep{Izotov16}.
However, the number of galaxies with
Z$_{\rm gas} \lesssim$ Z\sunn/30 found in this way
to date does not exceed $\sim$10 objects.

Several alternative methods, based on the colour and morphology selection of XMP
candidates from the SDSS image data base, resulted in only singular detections
\citep[e.g., ][]{LittleCub, Hsyu2018}.

One of the proposed methods to search for XMP dwarfs was suggested by
\citet{PaperI} and \citet{PaperVII}.
It is based on the discovery of the metallicity deficiency
(on average, by a factor of $\sim$1.4) of galaxies in voids in comparison
with similar objects in typical groups of the Local Volume (LV).
It was also found that $\sim$1/3 of the faintest void late-type dwarfs
show a strongly reduced gas metallicity. Namely, their values of O/H are
lower by a factor of 2--5 relative to those in galaxies of the same luminosity
residing in typical groups of the LV (hereafter, the reference
sample from \citet{Berg12}).

Several dwarfs with extremely low gas O/H
[12+log(O/H) $\lesssim$ 7.0--7.2],
and low or modest star formation rate (SFR), were found in the
nearby Lynx--Cancer void at distances of 10--25 Mpc
\citep{J0926, PaperII, PaperVII, Hirschauer16}.
The majority of these also show blue colours of the outer parts, consistent
with the time since the onset of the main SF episode of $\lesssim$1--3~Gyr,
and the extreme gas-mass fractions M$_{\rm gas}$/M$_{\rm bary} \sim$
0.98--0.99 \citep[e.g.][]{Triplet,U3672}. However, because they have low
luminosities, mostly low surface brightness (LSB) and faint emission lines,
the discovery of such galaxies in typical redshift surveys is limited as
a result of severe observational selection effects.

This discovery of unusual void dwarfs revealed that
they are almost exclusively the least luminous blue LSB galaxies in
the Lynx--Cancer void, and encouraged the search for similar objects
in other nearby voids, with the help of
the recently published 'Nearby Void Galaxies' (NVG) sample
by \citet[][hereafter PTM19]{PTM19}.
This sample was used to separate a group of 60 NVG low-luminosity
late-type dwarfs whose properties meant that they resembled
 prototype gas-rich XMP
galaxies \citep{PEPK20}. For these 60 selected void dwarfs, we performed
long-slit spectroscopy to determine their gas O/H. The first part of
the spectroscopic results obtained with the Southern African Large Telescope
(SALT) was presented in \citet{XMP.SALT}. Here we present the second
part of this study, based on spectroscopy with the Big Telescope Alt-azimuth
(BTA), the 6-m telescope of the Special Astrophysical Observatory of Russian
Academy of Sciences (SAO RAS).

The rest of this paper is arranged as follows. In Section~\ref{sec:observe},
we describe the observations and data processing. The results of spectral
observations of 20 galaxies are presented in Section~\ref{sec:OH_estimates}.
In Section~\ref{sec:discussion}, we discuss the obtained results along
with other available information. Finally, in Section~\ref{sec:conclusions},
we present our conclusions. In Appendices~A1 and A2, we present checks of two
indirect methods of O/H estimates used in this work. Tables with line
intensities,
derived physical parameters and oxygen abundances are presented in Appendix~B.
In Appendix~C, (online only), we present the finding charts and plots
of the obtained one-dimensional spectra.

\setcounter{qub}{0}
\begin{table*}
\begin{center}
\caption{Journal of BTA spectral observations}
\label{tab:journal}
\hoffset=-2cm
\begin{tabular}{r|l|l|l|l|r|l|c|c} \hline  \hline \\ [-0.2cm]
\MC{1}{r|}{} &
\MC{1}{c|}{Name} &
\MC{1}{c|}{Date} &
\MC{1}{c|}{Grating} &
\MC{1}{c|}{Exp. time}&
\MC{1}{c|}{PA} &
\MC{1}{c|}{Galaxy coordinates}&
\MC{1}{c|}{$\theta$ (arcsec)}&
\MC{1}{c}{Airmass}  \\


\MC{1}{r|}{} &
\MC{1}{c|}{(1)} &
\MC{1}{c|}{(2)} &
\MC{1}{c|}{(3)} &
\MC{1}{c|}{(4)} &
\MC{1}{c|}{(5)} &
\MC{1}{c|}{(6)} &
\MC{1}{c|}{(7)} &
\MC{1}{c}{(8)} \\
\hline 
\qq&AGC102728          & 2019.10.22 & 1200B   &4$\times$1200&  99.0& J000021.4+310119  & 1.5 & 1.05 \\
   & "                 & 2019.10.22 & 1200R   &2$\times$900 &  99.0& "                 & 1.1 & 1.06 \\
\qq&PGC000083          & 2019.10.22 & 1200B   &4$\times$900 & 131.0& J000106.5+322241  & 1.3 & 1.04 \\
   & "                 & 2019.10.28 & 1200B   &2$\times$600 &  41.5& "                 & 1.5 & 1.03 \\
   & "                 & 2020.01.20 & 1200B   &2$\times$900 &  59.0& "                 & 1.4 & 1.37 \\
   & "                 & 2020.01.20 & 1200R   &2$\times$900 &  59.0& "                 & 1.4 & 1.84 \\
\qq&PISCESA            & 2017.11.16 & 1200B   &2$\times$1200&  35.5& J001446.0+104847  & 1.3 & 1.52 \\
\qq&AGC411446          & 2017.09.13 & 1800R   &4$\times$900 &  98.0& J011003.7--000036 & 1.7 & 1.39 \\
   & "                 & 2017.11.16 & 1200B   &3$\times$1200& 177.0& "                 & 1.2 & 1.39 \\
\qq&PISCESB            & 2017.11.16 & 1200B   &2$\times$1200&  35.9& J011911.7+110718  & 1.0 & 1.44 \\
\qq&AGC122400          & 2019.10.22 & 1200R   &2$\times$600 & 20.0 & J023122.1+254245  & 1.2 & 1.19 \\
   & "                 & 2020.01.20 & 1200B   &2$\times$900 & 53.0 & "                 & 1.5 & 1.31 \\
   & "                 & 2020.08.19 & 1200B   &3$\times$900 &140.0 & "                 & 1.2 & 1.13 \\
\qq&AGC124609          & 2019.10.26 & 1200B   &2$\times$900 &  60.0& J024928.4+344429  & 1.1 & 1.10 \\
\qq&KKH18              & 2018.01.14 & 1800R   &3$\times$900 & 235.0& J030305.9+334139  & 1.9 & 1.19 \\
\qq&AGC189201          & 2019.01.01 & 1800R   &$2\times$900 &  42.0& J082325.6+175457  & 3.0 & 1.33 \\
   &  "                & 2019.10.26 & 1200B   &$2\times$900 & 137.0&  "                & 1.3 & 1.27 \\
   &  "                & 2019.10.26 & 1200R   &$1\times$900 & 137.0&  "                & 1.3 & 1.25 \\
   &  "                & 2020.01.19 & 1200B   &$3\times$900 & 173.6 &  "               & 1.2 & 1.33 \\
   &  "                & 2020.01.19 & 1200R   &$1\times$600 & 173.6 &  "               & 1.0 & 1.33 \\
\qq&J0823+1758         & 2020.01.19 & 1200R   &$1\times$600 & 163.0& J082335.0+175813  & 1.0 & 1.11 \\
   &  "                & 2020.01.20 & 1200B   &$3\times$900 & 163.0&  "                & 1.5 & 1.12 \\
   &  "                & 2020.01.20 & 1200R   &$2\times$900 & 163.0&  "                & 1.5 & 1.33 \\
\qq&J1012+3946         & 2018.04.15 & 1800R   &1$\times$900 &  70.0& J101259.9+394617  & 1.8 & 1.11 \\
\qq&AGC208397          & 2018.04.15 & 1800R   &3$\times$900 &  10.0& J103858.1+035227  & 3.0 & 1.31 \\
\qq&AGC239144          & 2018.04.15 & 1200B   &$4\times$900 & 109.0& J134908.2+354434  & 1.2 & 1.19 \\
   &  "                & 2019.01.01 & 1800R   &2$\times$900 & 120.5&  "                & 3.3 & 1.07 \\
   &  "                & 2020.01.19 & 1200B   &4$\times$900 & 118.2&  "                & 1.7 & 1.14 \\
   &  "                & 2020.01.19 & 1200R   &2$\times$900 & 118.2&  "                & 1.7 & 1.08 \\
\qq&J1440+3416         & 2020.01.20 & 1200B   &1$\times$900 & 119.0& J144431.6+341601  & 1.1 & 1.25 \\
   &  "                & 2020.05.14 & 1200B   &2$\times$1200& 125.6&  "                & 2.6 & 1.02 \\
   &  "                & 2020.05.14 & 1200R   &1$\times$600 & 125.6&  "                & 2.6 & 1.02 \\
\qq&J1444+4242         & 2017.04.17 & 1200B   &3$\times$1200&  41.0& J144449.8+424254  & 2.4 & 1.12 \\
\qq&PGC2081790         & 2018.04.15 & 1200B   &4$\times$900 &  11.3& J144744.6+363017  & 1.3 & 1.01 \\
\qq&J1522+4201         & 2017.09.13 & 1200B   &2$\times$900 &  62.0& J152255.5+420158  & 1.4 & 1.57 \\
\qq&J2103-0049         & 2017.09.13 & 1200B   &5$\times$900 &  05.0& J210347.2--004950 & 1.7 & 1.44 \\
\qq&AGC321307          & 2017.09.13 & 1800R   &2$\times$900 &  48.5& J221404.7+254052  & 1.6 & 1.14 \\
   &  "                & 2017.11.16 & 1200B   &1$\times$1200&  23.6&  "                & 1.7 & 1.07 \\
\qq&AGC334513          & 2019.10.28 & 1200B   &2$\times$900 &  96.0& J234850.4+233527  & 1.5 & 1.07 \\
   &  "                & 2019.10.28 & 1200R   &2$\times$600 &  96.0&  "                & 1.5 & 1.07 \\
   &  "                & 2020.01.19 & 1200B   &4$\times$900 &  49.2&  "                & 1.2 & 1.47 \\
   &  "                & 2020.01.19 & 1200R   &2$\times$900 &  49.2&  "                & 1.3 & 1.74 \\
\hline
\qq&J0823+1748         & 2020.01.19 & 1200R   &$1\times$600 & 163.0& J082310.6+174826  & 1.0 & 1.11 \\
\qq&AGC322279          & 2017.09.13 & 1800R   &2$\times$900 & 167.0& J220316.8+174744  & 1.4 & 1.14 \\
\qq&AGC332939          & 2017.09.13 & 1800R   &2$\times$900 &  56.0& J230816.7+315406  & 1.7 & 1.11 \\
   &  "                & 2017.11.16 & 1200B   &3$\times$1200&  25.0&  "                & 1.7 & 1.03 \\
\hline \hline \\[-0.2cm]
\end{tabular}
\end{center}
\end{table*}

\section[]{Observations and data reduction}
\label{sec:observe}

The spectral observations of mainly the Northern part of the XMP candidate
sample were conducted with the BTA, which is the SAO 6-m telescope, and
the multimode focal
reducer SCORPIO  \citep{SCORPIO} in the telescope's prime focus.
In the period from 2017 September to 2019 January (five nights), we used SCORPIO
in the original version in the long-slit mode (width of 1.0~arcsec, length
of $\sim$6~arcmin). In the period from 2019 October to 2020 August,
we observed during seven more nights with the upgraded version of SCORPIO.
This upgraded instrument allows us, in particular, to change
the pre-installed grisms during observations of the same target, without any time loss.
In this period, we used the long-slit mode with a slit width of 1.2~arcsec.

The details of observations for individual galaxies are presented
in Table~\ref{tab:journal}, in which we give the galaxy names, the dates of
observations, the grisms used, exposure times (in s), position
angles of the long slit (PA, in degrees), J2000 coordinates of the target
galaxies, seeings $\theta$ (in arcsec) and airmass.

For the main programme, during the dark time, when conditions allowed, we used
grism VPHG1200B with the full range of 3650--5450~\AA\ and FWHM = 5.5~\AA.
The CCD 2K$\times$2K detector was EEV-42-40 with the pixel size of 13.5$\mu$,
corresponding to 0.18~arcsec on the detector. We used the binning factor
of 2 along the slit. Because our main goal was to obtain the gas metallicity
in the observed galaxies,
the long slit was oriented close to the current parallactic angle, in order
to minimize the effect of atmospheric dispersion \citep{Filip82}.
The emission lines of these spectra were used
to derive the electron temperature $T_{\rm e}$ and the oxygen abundance O/H.
When it was suitable, during the period since 2019 October, along with
grism VPHG1200B we also used grism VPHG1200R. This was done to acquire the spectrum
of a studied galaxy in the red (full range 5680--7430~\AA, FWHM = 5.5~\AA) at
exactly the same slit position as for grism VPHG1200B.

For the back-up programme (for cases of 'poor' seeing and/or Moon time), we used
grism VPHG1800R with the range of $\sim$6100--7100~\AA\ and the spectral
resolution of FWHM $\sim$ 3.5~\AA. Since 2019 October, grism VPHG1800R was
unavailable, so for such cases we used grism VPHG1200R.
These spectra allowed us to check the coincidence of optical radial velocity
with that derived from \HI\ data for those \HI\ sources that were identified
only because of the close radio and optical positions for
 the Arecibo Legacy Fast ALFA (ALFALFA) blind \HI\
survey \citep{ALFALFA}.

Besides, these spectra were used for estimates of the strength of
lines [N{\sc ii}]$\lambda$6584 and doublet [S{\sc ii}]$\lambda$$\lambda$6716,6731
relative to that of H$\alpha$. This information was subsequently used to
rank the observed galaxies, more or less confidently, as XMP candidates for
the follow-up observations with grism VPH1200B. A couple of galaxies
(numbers 22 and 23 in Tables~\ref{tab:journal} and \ref{tab:BTA_OH}) were observed at
a time when the selection of the void XMP candidates was not finalized. They
were identified later as residing outside the nearby voids from PTM19.

The processing of the spectra and the measurement of emission-line fluxes with
the use of both {\sc IRAF}\footnote{{\sc IRAF}: the Image Reduction and Analysis
Facility is distributed by the National Optical Astronomy Observatory,
which is operated by the Association of Universities
for Research in Astronomy, Inc. (AURA) under cooperative agreement with the
National Science Foundation (NSF).}
and {\sc MIDAS}\footnote{{\sc MIDAS} is an acronym for the European Southern
Observatory package -- Munich Image Data Analysis System}
were described in detail in \citet{PaperVII}. Below, we
describe the main steps of all the procedures used. The standard pipeline included
removal of cosmic ray hits, bias subtraction, flat-field correction,
wavelength calibration and night-sky background subtraction. By using
data of spectrophotometric standards  observed on the same nights, all
spectra were transformed to absolute fluxes. The individual  one-dimensional
(1D) spectra of the studied \HII\ regions were then extracted by summing up without
weighting of several rows along the slit. Typically, we extracted 6--12
binned pixels ($\sim$2--4~arcsec).

We note that after the continuum in the 1D spectra is drawn and the
line fluxes are measured as described in \citet{PaperVII}, we use
the iteration procedure from \citet{ITL94}, which simultaneously
determines the extinction coefficient, C(H$\beta$), and the equivalent
width of Balmer absorption, EW(abs), in the underlying stellar continuum.
See also Sect.~\ref{sec:OH_estimates} for the use of the model continuum
for several spectra of E+A type.

The plots of the 1D spectra obtained are presented in
Figures S4--S5.  
in the supplementary data section
(online Appendix~C).
The measured relative fluxes for the presented spectra are available in
Appendix~B (Tables B1--B7). 

As in our previous paper (PKPE20), the majority of the observed candidate XMP
dwarfs are very faint and/or of low surface brightness. Thus, it can be a challenge
to obtain their independent spectra and to point the slit to the right \HII\ region.
To make it easier to provide independent checks of our data,
we provide in Figs S1 and S2  
the slit
positions for each of the galaxies in our programme overlaid on the galaxy images taken
from the data bases of the Dark Energy Camera Legacy Survey \citep[DECaLS;][]{DECaLS},
the Panoramic Survey Telescope \& Rapid Response System  \citep[PanSTARRS PS1;][]{PS1-database}
or the SDSS \citep{DR7, DR14}.
In Fig.~S3, 
we also show the slit positions for three
additional observed galaxies that are not in the XMP void candidate list.

\setcounter{qub}{0}
\begin{table*}
\caption{Candidate XMP dwarfs observed at the BTA and new O/H data. Parameter M(\HI)/L$_{\rm B}$ is in solar units.}
\begin{tabular}{r|l|l|r|r|r|r|l|l|l}
   &Name           & Galaxy coordinates&V$_h$ & D &B$_{\rm tot}$&$M_{\rm B}$&M(\HI)/& 12+log(O/H)   & Notes   \\
   &               &                 & \kms  &(Mpc)&(mag)      &(mag)      &$L_{\rm B}$&      &         \\ %
   &  (1)          & (2)              & (3) &(4)  & (5) & (6)   & (7)   &   (8)         &  (9)     \\ %
\hline
\qq&AGC102728      & J000021.4+310119 & 566 & 8.8 &19.40&--10.53& 2.44  & ...                   &Very faint H$\alpha$  \\
\qq&PGC000083      & J000106.5+322241 & 542 & 9.1 &17.14&--12.89& 3.13  & 7.15$\pm$0.03 (s,c)   &Average of 2 knots \\
\qq&PiscesA        & J001446.0+104847 &  235& 5.6 &17.98&--11.31& 2.22  & 7.26$\pm$0.05 (s,c)   &         \\
\qq&AGC411446      & J011003.7--000036& 1137&15.9 &19.58&--11.54& 4.80  & 7.00$\pm$0.05 (s,c)   &         \\ %
\qq&PiscesB        & J011911.7+110718 &  616& 8.9 &17.63&--12.23& 2.80  & 7.31$\pm$0.05 (s,c)   &         \\ %
\qq&AGC122400      & J023122.1+254245 & 938 & 15.5&18.48&--12.97& 2.27  & 7.19$\pm$0.12 (s,c)   &         \\
\qq&AGC124609      & J024928.4+344429 &1588 & 25.0&18.20&--14.10& 1.51  & 7.89$\pm$0.02 (dir)   &         \\
\qq&KKH18          & J030305.9+334139 & 210 &  4.8&16.70&--12.56& 0.98  & ...                 &I(NII)/I(H$\alpha$)=0.03  \\ %
\qq&AGC189201      & J082325.6+175457 &1475 & 23.4&19.24&--12.76& 3.94  & 7.31$\pm$0.04 (dir)   &         \\ %
\qq&J0823+1758     & J082335.0+175813 &1509 & 23.8&19.49&--12.54& ...   & 7.33$\pm$0.05 (s,c)   &         \\ %
\qq&J1012+3946     & J101259.9+394617 &1340 & 21.9&18.32&--12.43& 3.80  & ...                 &I(NII)/I(H$\alpha$) uncertain   \\ %
\qq&AGC208397      & J103858.1+035227 & 763 & 11.9&19.95&--10.59& 5.60  & ...                 &I(NII)/I(H$\alpha$)=0.03         \\ %
\qq&AGC239144      & J134908.2+354434 &1366 & 20.4&19.06&--12.54& 3.17  & 7.25$\pm$0.06 (s,c)   &Average of 2 knots \\ %
\qq&AGC249590      & J144031.6+341601 &1489 & 21.7&18.45&--13.28& 2.67  & 7.23$\pm$0.05 (s,c)   &         \\ %
\qq&J1444+4242     & J144449.8+424254 & 634 & 10.9&19.69&--10.54& 1.30  & 7.16$\pm$0.04 (s,c)   &Average of 2 knots \\ %
\qq&PGC2081790     & J144744.6+363017 &1226 & 18.1&18.04&--13.29& 2.11  & 7.32$\pm$0.04 (s,c)   &         \\ %
\qq&J1522+4201     & J152255.5+420158 & 608 &  9.8&17.74&--12.31& 0.43  & 7.30$\pm$0.05 (s,c)   &         \\ %
\qq&J2103--0049    & J210347.2--004950&1411 & 17.4&17.44&--14.07& 1.30  & 7.21$\pm$0.04 (s,c)   &         \\ %
\qq&AGC321307      & J221404.7+254052 & 1152& 16.2&18.15&--13.27& 2.37  & 7.89$\pm$0.10(mse,c)&         \\ %
\qq&AGC334513      & J234850.4+233527 & 1662& 23.8&18.89&--13.25& 3.49  & 7.24$\pm$0.05 (s,c)   &         \\ %
\hline
\qq&J0823+1748     & J082310.6+174826 &27600&376.4&20.15&--17.86& ...   & ...                 &Distant I(NII)/I(H$\alpha$)$\sim$0.01 \\ %
\qq&AGC322279      & J220316.8+174744 &1272 & 17.2&18.39&--12.98& 2.37  & ...                 &Non-void I(NII)/I(H$\alpha$)=0.19   \\ %
\qq&AGC332939      & J230816.7+315406 & 692 & 10.7&17.18&--13.22& 1.61  & 7.69$\pm$0.06 (dir) &Non-void \\ %
\hline
\end{tabular}
\label{tab:BTA_OH}
\end{table*}

\section[]{Results of spectral observations and O/H estimates}
\label{sec:OH_estimates}

Tables with measured line intensities, derived electron temperatures and
oxygen abundances are presented in Appendix~B
(Tables~B1--B7). 

Because of very low fluxes of emission lines in the observed \HII\ regions
and their low metallicities, the principal faint auroral  line
[O{\sc iii}]$\lambda$4363, used for the determination of the electron
temperature, was not detected in the majority of our targets.

The line [O{\sc iii}]$\lambda$4363 was reliably detected only
in the spectra of three galaxies. For these, O/H was estimated via the
direct method and was marked as 12+log(O/H)(T$_{\rm e}$).
The respective formulae with the most updated atomic data
are given in \citet{Izotov06a}. All calculations for the O/H estimate
using the direct method are made in a similar way to those in \citet{Sgr}
and \citet{PaperVII}.

For the 14 remaining objects with detected oxygen lines, we used the two
methods described below, which are based on fluxes of the strong oxygen
lines:  $R_{\mathrm 23}$, the flux ratio of the sum of
oxygen lines
[O{\sc ii}]$\lambda$3727 and [O{\sc iii}]$\lambda\lambda$4959,5007
to the flux of H$\beta$; and the parameter O$_{\mathrm 32}$, defined here as the
flux ratio of [O{\sc iii}]$\lambda$5007 to [O{\sc ii}]$\lambda$3727.

Recently, \citet{Izotov19DR14} and \citet{Izotov21} suggested the variant of
an empirical 'strong-lines' method, which shows small internal
scatter for the range of the lowest values of O/H
[12+log(O/H) $\lesssim$ 7.5]. This improvement is made by taking into
account the dependence of log(O/H) on both parameters, $R_{\mathrm 23}$
and O$_{\mathrm 32}$. The respective relation is:
$$12+log(O/H) = 0.950 \times log(R_{\mathrm 23} - a_{\mathrm 1}\times O_{\mathrm 32}) + 6.805, $$
where $a_{\mathrm 1}$ =  0.080 -- 0.00078 $\times O_{\mathrm 32}$.
While the strong-lines method is suitable only for the lowest metallicities,
as the authors show, for this range, it provides a low internal scatter
of $\sim$0.05~dex.
We mark O/H estimated with this strong-lines method
as 12+log(O/H)(s).
The above relation empirically accounts for the large scatter in the
ionization parameter $U$ in various \HII\ regions. This leads to a reduction in the
relatively large internal scatter inherent to the other methods based
on strong oxygen lines.

For 13 of 14 void objects with detected strong oxygen
lines, but without the [O{\sc iii}]$\lambda$4363 line,
the line ratios are indicative of that low level of O/H.
Hence,  for those objects, we used  the strong-lines
method of \citet{Izotov19DR14} and its recent slightly modified
version \citep{Izotov21} as the most reliable.
We also perform  our own analysis in Appendix~A1,
illustrated in Figure~\ref{fig:SL-dir},
addressing both the issues of its small offset relative to O/H(dir)
and  its internal scatter.

For one of the void galaxies, J2214+2540, its strong oxygen lines
indicate 12+log(O/H) $>$ 7.5 and hence other methods of estimating O/H
are needed. A so-called semi-empirical (hereafter, se) method by \citet{IT07}
is based on a correlation between T$_{\rm e}$ and parameter $R_{\mathrm 23}$.
The functional relation between the two variables was fitted based on
calculations for the grid of \HII\ region models in a paper by \citet{SI03}.
The se method first calculates the empirical estimate of T$_{\rm e}$ from
the measured $R_{\mathrm 23}$ and then uses this T$_{\rm e}$ similar to
the classical direct method.
This method was checked and calibrated in particular by \citet{IT07}
and \citet{PaperVII}.
Its internal scatter was estimated as rms $\sim$0.07--0.08~dex. The method is
applicable for 12+log(O/H) $\lesssim$ 7.9.

In the course of the review process for this paper, it was drawn to our attention
that for some galaxies
with a well-measured flux of [O{\sc iii}]$\lambda$4363 and the related
good accuracy estimate of the electron temperature T$_{\rm e}$ in the O$^{++}$
zone,  the difference between T$_{\rm e}$(dir) and
T$_{\rm e}$(R$_{\mathrm 23}$) may reach several thousand K. This in turn
might lead to a bias in the estimate of O/H via the se method.

We address this issue in more detail in Appendix \ref{ssec:SE-method} and then
apply the results of the modified semi-empirical method (hereafter mse)
to the current observations. This mse method will also be applied to
our earlier data for void galaxies with 12+log(O/H) $ > $ 7.5,
when we will summarize all their metallicity data.

For a few of the observed galaxies, Balmer absorption lines were clearly
visible in the blue-UV range. We modeled their underlying continuum with the
{\sc ULySS} package\footnote{http://ulyss.univ-lyon1.fr} \citep{Koleva2009}.
This model continuum fitted Balmer absorption lines, and thus
corrected  to a first approximation the flux of H$\beta$ emission.
For these objects, the EW(abs) derived in the next step via iterations,
as described above with the procedure from \citet{ITL94},
relates in fact to the residual EW(abs), which is already mainly
accounted for by the {\sc ULySS} fitting (e.g., in galaxies J2103--0049
and J2214+2540).

In Table~\ref{tab:BTA_OH} we present the following parameters. Columns 1
and 2 give the name and J2000 RA and declination of the galaxy. In column 3, we give
the heliocentric velocities as adopted from HyperLEDA data base. For J0823+1758
and J0823+1748, we present radial velocities which were first
measured by us (see Section~\ref{ssec:individual} for details).
The distances in Mpc (column 4) derived from the HST data via the
tip of the RGB (TRGB) for four galaxies are taken from HyperLEDA;
for the remaining 16 galaxies,
they are taken from our NVG catalogue (PTM19) based on the
velocity field model by \citet{Tully08}. In columns 5 and 6, we give the total
apparent and respective absolute blue magnitudes. The latter are calculated
from the apparent magnitudes and the adopted distances, taking into account
the Galaxy extinction from \citet{Schlafly11}.

 The apparent blue magnitudes B$_{\rm tot}$ were estimated by us via
measurements of their $g, r$ magnitudes if their images were available in the
above-mentioned public data bases. Then/ these $g, r$ magnitudes were
transformed to the $B$ band via the relation of \citet{Lupton05}.
In the remaining cases, we took B$_{\rm tot}$ from HyperLEDA.
In column 7, we also give the ratio of hydrogen mass to blue luminosity
M(\HI)/L$_{\rm B}$, in solar units.  In column 8, the adopted value
of 12+log(O/H) is presented as derived in this paper.
In column 9, we give some notes.

Tables with measured line intensities, derived electron temperatures and
oxygen abundances are presented in Appendix~B.
 Besides 12+log(O/H)(dir), we show 12+log(O/H)(s,c),
12+log(O/H)(mse,c) and 12+log(O/H)(se,c).
In Table~\ref{tab:BTA_OH},
for the range of 12+log(O/H) $\lesssim$ 7.4~dex, when the strong-lines method
is applicable, we adopted the value of 12+log(O/H) (s,c).
As shown in Appendices~\ref{ssec:SL-method} and \ref{ssec:SE-method}, this
estimate has a significantly
smaller scatter
in comparison to those of 12+log(O/H) (mse,c) and 12+log(O/H) (se,c).

\section[]{Discussion}
\label{sec:discussion}

\subsection{Notes on individual galaxies}
\label{ssec:individual}

For some of our observed XMP dwarf candidates, it is worth giving
additional information or comments.

\subsubsection{\bf AGC102728 = J0000+3101}
 Only a faint H$\alpha$ line was seen in
one of the positions along the long slit. This galaxy is resolved into
individual stars at the HST images. \citet{Tikhonov19} derived its
 distance of 8.84$\pm$0.68~Mpc based on the TRGB method.
An alternative TRGB distance for this galaxy of D=12.4~Mpc was recently
derived in \citet{McQuinn21}.
In the colour HST image,  several faint red galaxies are seen in the
surroundings of this galaxy; these could belong to a distant group
or a cluster. Our slit position crossed one of these galaxies projected
on to the body of AGC102728. A clear emission line was
detected at $\lambda$6863~\AA\ corresponding to a redshifted
 H$\alpha$ line with $z$ = 0.04571.

\subsubsection{\bf PiscesA = J0014+1048}
This galaxy was included, among others, in the list
of about 80 candidate low-metallicity dwarfs selected on their morphological
and colour properties by \citet{James17}. However, for two different positions
of the long slit, these authors did not detect emission lines in their
spectra. As our results show, the reason is that PiscesA seems to have only one
substantially bright \HII\ region. It was identified in our BTA acquisition
images with the medium-width SED665 filter (see Fig.~S1, top row). 

\subsubsection{\bf AGC124609 = J0249+3444}
This void galaxy has a bright
 high excitation \HII\ region with a normal metallicity
for its luminosity. Its spectrum
displays a strong line of He{\sc ii}~4686. Besides, its high ratio
O$_{\mathrm 32}$ = 8.3, along with R$_{\mathrm 23} =$ 8.7, is indicative
of an \HII\ region with possibly leaking Lyman continuum
\citep[e.g.][]{Izotov18b}. As such, this galaxy appears to be
one of the nearest and least massive known Ly-c leaker candidates and
deserves further observation using  high signal-to-noise (S/N) spectroscopy.
This would help us to study the diversity of Ly-c leaker properties and
 give us the
opportunity to study this phenomenon with a high spatial resolution.

\subsubsection{\bf J0823+1758 and J0823+1748}
 These galaxies were not in our list of
the pre-selected XMP candidates (PEPK20), or in any data base
as objects with known radial velocity. They were noticed as possible counterparts of
a void galaxy AGC189201 in the course of a visual inspection of
its surroundings, because of to the relative angular proximity and their
blue colours and morphology.
Both galaxies were first observed in red during the Moon time. This
allowed us to classify the first galaxy as a real counterpart and a new void
object (with $\delta$V = +34~\kms\ relative to that of AGC189201), while
the second galaxy appeared to be a distant object at a redshift of z $\sim$ 0.092.

\subsubsection{\bf J1522+4201}
For this object, we have no red part of the spectrum.
To estimate its C(H$\beta$) and O/H, we adopted the flux of the H$\alpha$ line,
based on the Balmer ratio I(H$\alpha$)/I(H$\beta$) = 3.0 from its SDSS spectrum
(SpecObjID = 1889302432475801600), acquired on the same region within a 3-arcsec
round aperture. This galaxy appeared also among 66 very metal-poor candidates
separated in SDSS DR14 by \citet{Izotov19DR14}. To estimate its O/H,
they adopted the flux ratio of the line [O{\sc ii}]$\lambda$3727
and H$\beta$ to be $\sim$2.6. Our independent BTA spectrum reveals
a lower value of this ratio, $\sim$2.2. Both
values are consistent within uncertainties, however.

\subsection{The new lowest-metallicity void dwarfs}
\label{ssec:newXMP}

Among the 20 very metal-poor candidate galaxies, observed at BTA, we found
six galaxies with
Z$_{\rm gas} \lesssim$ Z\sunn/30, or 12+log(O/H) $\lesssim$ 7.21~dex.
They deserve further discussion as the search for such objects and their
follow-up study is the main goal of our observational programme.

\subsubsection{\bf PGC00083 = J0001+3222}
For this galaxy, we obtained good quality spectra
for two knots 'a' and 'b', with 12+log(O/H) = 7.17$\pm$0.05 and
7.13$\pm$0.05~dex, both derived via the strong-lines method
\citep{Izotov19DR14}. Their separation along the slit is $\sim$6.5~arcsec,
or $\sim$240~pc.
This galaxy is one of the nearest void XMP objects known. According to the velocity
field model of \citet{Tully08} adopted for galaxy distances in the
NVG catalogue (PTM19), its distance is 9.1~Mpc (i.e. this galaxy
resides in the LV). Indeed, it enters the Updated Nearby
Galaxy Catalog (UNGC) by \citet{UNGC} -- and the most updated version at
\mbox{http://www.sao.ru/lv/lvgdb/} -- with the adopted distance
of 9.4~Mpc (derived via the baryonic Tully--Fisher relation).
According to PTM19, the galaxy resides in void No.~25.
Its distance to the nearest luminous neighbour is $D_{\rm NN} \sim$4.1~Mpc.

The apparent tadpole morphology of this galaxy (see Fig.~S1), 
with a 'head' on the SW edge, is due to the chance projection of a background
reddish galaxy on to the edge of this disc-like dwarf. The slit position with
PA=41.5\degr\ with grism VPH1200B, crosses the centre of the 'head'.
Two clear emission lines are visible in its spectrum
([O{\sc ii}]$\lambda$3727 and H$\beta$) that allow us to determine its
redshift of $z$ = 0.0872. The line [O{\sc iii}]$\lambda$5007 appears just
outside the range, while [O{\sc iii}]$\lambda$4959 is not detected.

It is worth mentioning that there is another candidate XMP galaxy in our
programme, AGC102728 = J0000+3101 (described in Sec.~\ref{ssec:individual}),
situated in the close surroundings of J0001+3222.
Their mutual angular distance of $\sim$45~arcmin corresponds to
the projected distance of $\sim$120~kpc. Its radial velocity differs from
that of PGC00083 only by $\delta$ V = +24~\kms\ ($\sim$300~kpc). Its distance,
estimated in the frame of the same velocity field model, $D$ = 9.4~Mpc,
is consistent with the velocity-independent distance estimate of
$D$ (TRGB) $\sim$ 8.84~Mpc.
The distance estimates for both void galaxies are consistent with each other
within their cited uncertainties. Their minimal mutual distance
is too large to treat these low-mass objects as gravitationally bound.
If they
are not bound, they may belong to the same structure element, such as a void
filament.

\subsubsection{\bf AGC411446 = J0110--0000}
This dwarf, with 12+log(O/H) = 7.00$\pm$0.05~dex
derived via the strong-lines method of \citet{Izotov19DR14}, is the second
most metal-poor galaxy found so far in our search programme. Its extremely low
O/H was discovered for first time  with the BTA spectrum presented here. To
improve the spectra quality and the accuracy of O/H determination, we performed
follow-up SALT observations (PKPE20) with the resulting estimate of 12+log(O/H) (s,c)
 = 7.07$\pm$0.05~dex.
This value is fully consistent with our BTA determination. In fact, because
both estimates are obtained with the strong-lines method and have
similar uncertainties, the most robust estimate would be their weighted mean
($\sim$7.04~dex).

According to PTM19, the galaxy is situated in a large void No.~3
(Cet-Scu-Psc),
probably at the periphery of a small void group with the central spiral
galaxy NGC428 (M$_{\rm B}$ = --19.2). Its $D_{\rm NN} \sim$6.0~Mpc. We do not
extend its description much in this paper as we are preparing a separate
publication devoted to the complex study of this remarkable XMP dwarf
(Pustilnik et al., in preparation).

\subsubsection{\bf AGC122400 = J0231+2542}
This new void XMP object at $D$ = 15.5~Mpc, with
12+log(O/H) = 7.19$\pm$0.12~dex, has a larger O/H uncertainty because of
noisy signal for the [O{\sc ii}]$\lambda$3727 doublet.
The galaxy also falls in void No.~3 of PTM19. Its distance to the nearest
luminous neighbour is $D_{\rm NN} \sim$ 3.6~Mpc.

\subsubsection{\bf AGC249590 = J1440+3416}
This new void XMP dwarf at $D = 21.7$~Mpc, with
12+log(O/H) = 7.23$\pm$0.06~dex, resides in void No.~16 from PTM19, with
$D_{\rm NN} \sim$ 2.5~Mpc. This XMP dwarf is probably a member of the
void triplet with the SAB galaxy NGC5727 (M$_{\rm B}$ = --17.1) at
$\sim$17~arcmin (or $\sim$107~kpc in projection) and with $\delta V$=9~\kms.
The third triplet member is PGC2043836 with M$_{\rm B}$ = --15.3.

\subsubsection{\bf J1444+4242}
This new void XMP dwarf, with 12+log(O/H) = 7.16$\pm$0.04~dex,
resides in the LV at D=10.9 Mpc. It is a faint companion (4.5
mag fainter) of a larger host, a disc metal-poor dwarf UGC9497, at the
angular separation of $\sim$8.5~arcmin ($\sim$27~kpc).
It was discovered in the course of \HI\ mapping of the host UGC9497
(Chengalur et al., in preparation). For the BTA slit position close
to the major axis of the blue elongated LSB body, we detected two
emission-line knots separated by $\sim$3~arcsec, or 150~pc, with the similar
values of 12+log(O/H) (s,c) = 7.20$\pm$0.05 and 7.11$\pm$0.05~dex,
for knots 'a' and 'b', respectively.

It is of interest that besides the more massive host UGC9497 with the wide
range of estimates of 12+log(O/H) = 7.27--7.63~dex
\citep{Guseva03,Izotov19DR14},
there exists in the same space cell (at a distance of $\sim$1~Mpc) one more
very metal-poor dwarf J1522+4201, with 12+log(O/H) = 7.30~dex
(see Table~\ref{tab:BTA_OH}).
Not one of the three void galaxies discussed is yet included in the UNGC.

\subsubsection{\bf J2103--0049}
 This new void XMP dwarf resides in void No.~25, at a
distance of 17.4 Mpc. Like the two other new void XMP dwarfs, J0001+3222
and J1444+4242, this object is situated close to the known XMP dwarf
J2104--0035 \citep{Izotov06b} at D $\sim$ 17.2~Mpc.
Their mutual projected distance of $\sim$22~arcmin corresponds to the linear
distance of $\sim$110~kpc, while the radial distances may differ by $\sim$200~kpc.
Like several other cases, they may belong to the same elements of a void
substructure, likely a void filament. As GMRT mapping reveals
\citep{Ekta2008}, the unusual \HI\ morphology of galaxy J2104--0035 itself bears
the traces of recent interaction or probable merging.

\subsection{Intermediate summary of new void XMP galaxies}
\label{ssec:summary-XMP}

During our ongoing project described in the Introduction,
as a result of both SALT and BTA spectroscopy programmes, we found 10
new XMP dwarfs.
Because the follow-up spectroscopy of the pre-selected $\sim$60 candidates
is only $\sim$75\% complete, we expect to present
several similar objects in forthcoming papers. In parallel, for
a majority of the new void XMP dwarfs, we are preparing the results of
the GMRT \HI\ mapping and of their multiband photometric study performed
on the available images in public data bases. Therefore, we postpone a more
advanced analysis of void XMP dwarfs until the completion
of these studies.

However, the current sample of 10 void XMP dwarfs is already
sufficiently large, so that it is worth making a preliminary
summary of  some of their properties.
In Table~\ref{tab:new_XMP-sum}, we collect their main observational
parameters, partly from the literature and partly from our data, published
or prepared for submission. We briefly discuss some of
the related issues.
The content of the columns is explained in the
table caption.

In the selection process of XMP candidates in PEPK20,
we initially imposed several limitations on their observational parameters,
based on the observed properties of the prototype XMP dwarfs. Furthermore,
by intention, we have widened the boundaries to be outside the parameter ranges
found in the prototype group. This is considered to be an attempt to probe the
occurrence of XMP dwarfs
in the regions of the parametric space not covered by the prototype objects.
We can summarize the current status of the new void XMP dwarfs as follows.

The absolute blue magnitudes, M$_{\rm B}$,  of the 10 new void XMP galaxies vary
between --10.5 and --14.07~mag, with a median value of --12.6~mag.
That is, the blue luminosity of this group varies by a factor of $\sim$25.
The total range of M$_{\rm B}$ values matches
well that of the nine XMP prototype objects, namely the eight XMP dwarfs with
known O/H from Table~1 of PEPK20 together with IZw18.

The mass of \HI, M(\HI) also varies in a large range, from 0.34 to
$\sim 7.9 \times$ 10$^{7}$~M\sunn\ (or logM(\HI) in the range 6.53--7.90),
that is by a factor of $\sim$23. The median value of
M(\HI) is 3.8 $\times$ 10$^{7}$~M\sunn.
As for the comparison of this parameter with the prototype group, for the new
XMP dwarfs, M(\HI) appears substantially lower. Both its upper and
lower boundaries have logM(\HI) that is about 0.4--0.6~dex
lower than those for the prototype objects. The
median M(\HI) of the prototype
group is $\sim$4 $\times$ 10$^{8}$~M\sunn, which is an order of magnitude
larger than for the newly found void XMP dwarfs.

A similar difference is visible for parameter M(\HI)/L$_{\rm B}$.
Even if we exclude the extreme value of 17.1 for the prototype dwarf
J0706+3020, three new XMP dwarfs have M(\HI)/L$_{\rm B}$ = 1.3--1.7, well
below the lower boundary of 2.4 for the prototype group.

The integrated colours $g-r$ of the new XMP dwarfs vary from --0.03
to $\sim$0.4, with the median value of $\sim$0.15.
This compares with the $g-r$ range of --0.08 to +0.19 for the prototype
group. Again, we see that the three new XMP dwarfs with $g-r$ = 0.25 -- 0.4 are
redder than the prototype XMP dwarfs.

Thus, we do see a large scatter and a shift of parameters of
the new XMP dwarfs (colours $g-r$, M(\HI)/L$_{\rm B}$) relative to those of the
prototype
group. This might indicate the diversity of the origin and/or evolutionary
status of some void XMP dwarfs. Further, more sensitive and more accurate
measurements of the most outlying XMP dwarf representatives should help
to elucidate this issue.

\begin{table*}
\caption{Parameters of new Nearby Void XMP dwarfs found at BTA and SALT}
\begin{tabular}{llrllrrrcllp{2.5cm}}
\hline
IAU name     & 12+log(O/H)           & $\frac{\rm MHI}{L_{\rm B}}$ &$\mu_{\rm B,0,c}$ & $M_{\rm B}$ &$(g-r)_0$  &$V_{\rm h}$& D    &logMHI    &logM$_{*}$ &$B_{\rm tot}$& \MC{1}{c}{Notes~~~~~~~~~~~~~~~~~~~} \\
	     & dex                   &        &          & mag         & mag   &      & Mpc  &         &          & mag        &       \\
 ~~~~1       & 2                     & 3~~    & ~~4      & 5           &  6    & 7    & 8    & 9~~~    & ~~10     & 11~~       & 12~~  \\
\hline
J0001+3222   & 7.15$\pm$.03$\dagger$ &   3.1  &  23.5:   & -12.86      &  0.13 & 542  &  9.1 &  7.48 &6.19 & 17.14     & AGC103567,PGC00083  \\
J0110--0000  & 7.00$\pm$.05$\dagger$ &   4.8  &  23.4    & -11.54      &  0.07 &1137  & 15.9 &  7.53 &5.66 & 19.58     & AGC411446 $*$       \\
J0112+0152   & 7.17$\pm$.05$\dagger$ &   1.9  &  ...     & -13.09      &  0.04 &1089  & 15.4 &  7.62 &6.02 & 18.23     & AGC114584 $*$       \\
J0231+2542   & 7.19$\pm$.12$\dagger$ &   2.3  &  25:     & -12.45      &  0.17 & 938  & 15.5 &  7.73 &6.16 & 18.92     & AGC122400           \\
J0256+0248   & 6.96$\pm$.06$\dagger$ &   1.7  &  24:     & -11.58      &  0.10 & 794  & 12.4 &  7.22 &5.88 & 19.46     & AGC124629           \\
J1038+0352   & 7.15$\pm$.05$\dagger$ &   5.7  &  24:     & -10.59      &--0.03 & 763  & 11.9 &  7.18 &5.60 & 19.95     & AGC208397        \\
J1259--1924  & 7.20$\pm$.08$\ddagger$&   3.1  &  25:     & -12.16      &  0.4: & 827  &  7.3 &  7.77 &6.64 & 17.50     & PGC044681        \\
J1440+3416   & 7.23$\pm$.05$\dagger$ &   2.7  &  24:     & -13.28      &  0.10 &1489  & 21.7 &  7.90 &6.62 & 18.45     & AGC249590 $**$ \\
J1444+4242   & 7.16$\pm$.04$\dagger$ &   1.3  &  24:     & -10.54      &  0.25 & 634  & 10.9 &  6.53 &5.71 & 19.11     & Pair with UGC9497    \\
J2103--0049  & 7.21$\pm$.05$\dagger$ &   1.3  &  ...     & -14.07      &  0.38 &1411  & 17.4 &  7.83 &7.33 & 17.44     & PGC1133627   \\
\hline
\multicolumn{12}{p{16.2cm}}{The meaning of Table columns. Col.~1: Short
IAU-type name; Col.~2: gas O/H in units 12+log(O/H) and its 1$\sigma$
error; } \\
\multicolumn{12}{p{16.2cm}}{Col.~3: M(HI)/$L_{\rm B}$  in solar units;
  Col.~4: corrected for the MW extinction and inclination, the central
  surface brightness  $\mu_{\rm B,0,c}$ } \\
\multicolumn{12}{p{16.2cm}}{in mag/$\Box{\arcsec}$; Col.~5: Absolute $B$-band
  magnitude;  Col.~6: corrected for MW extinction total $g-r$ colour; Col.~7:
  } \\
\multicolumn{12}{p{16.2cm}}{heliocentric radial velocity in \kms; Col.~8:
 adopted distance in Mpc; Col.~9: log of galaxy HI-mass in M\sunn;
Col.~10:} \\
\multicolumn{12}{p{16.2cm}}{log of stellar mass in M\sunn; Col.~11: total
$B$-band magnitude; Col. 12: Notes and common names (see sect.~
    \ref{ssec:summary-XMP} for detail).  } \\
\multicolumn{12}{p{16.2cm}}{Low accuracy data are marked with (:). $*$ - near group NGC428. $**$ - in triplet.} \\
\multicolumn{12}{p{16.2cm}}{$\dagger$ O/H derived via the 'Strong-lines'
   method by \citet{Izotov19DR14, Izotov21}, with correction of --0.01~dex.
   See text.} \\
\multicolumn{12}{p{16.2cm}}{$\ddagger$ this object has too low value of
  O$_{\mathrm 32}$ in comparison to the 'calibrator' sample.} \\
\end{tabular}
\label{tab:new_XMP-sum}
\end{table*}

Of the 10 new XMP void dwarfs, eight are typical late-type LSB galaxies with
the range of central surface brightness $\mu_{\rm B,0,c}$ between 23.4
and $\sim$25 (in mag~arcsec$^{-2}$). The two remaining XMP dwarfs with
the absent $\mu_{\rm B,0,c}$, appear brighter, seemingly because of
the enhanced overall star formation during the recent epoch.
On this parameter, they match the prototype group well, in which at least
a half of objects have $\mu_{\rm B,0,c}$ in the range of 24.1 --
25.4~mag~arcsec$^{-2}$, while the remaining XMP dwarfs, with clear
traces of recent disturbance, are brighter.

The majority of new O/H values from this work and PKPE20 match well
in the plot 12+log(O/H) versus $M_{\rm B}$ (see Fig.~\ref{fig:ZvsMBTe})
with the distribution of the low-luminosity galaxies from the Lynx--Cancer
void. The lowest O/H dwarfs also fill the region of parameters occupied
by the earlier known XMP gas-rich objects.

The issue of 'clustering' or the 'local' environment of void XMP dwarfs is
difficult to describe in terms of a simple numerical parameter.
We illustrate this property qualitatively, going from the strong and
evident interaction, which indicates an advanced merger,
to the opposite case of the confident isolation and absence of visible
disturbing agent.

In the prototype XMP group we have four members of the certain or probable
mergers:  one object is a member of an interacting pair and three
galaxies are  fairly well isolated dwarfs.
Among the newly found XMP dwarfs, one (J1444+4242) belongs to a pair
and another (J1440+3416) is a probable member of a void triplet.
Two more objects are probable members
of the distant periphery of a void group. The remaining six dwarfs are
well isolated, despite the fact that three of them have distant neighbours
marking
probable elongated void structures. That is, in total, about a half or
more of XMP void dwarfs do not show the presence of evident disturbing neighbours.

\subsection{Issue of the reduced gas metallicity in void galaxies}
\label{ssec:void.ZL}

Recently \citet{McQuinn20} discussed the relation, 12+log(O/H) versus
$M_{\rm B}$, for void galaxies -- based on data from
\citet{PaperVII} and \citet{Eridanus} -- in comparison with that for the
reference sample
of the late-type LV galaxies in a more typical environment
\citep{Berg12}. They used for comparison 49 void galaxies with
O/H derived via the direct method. Their conclusion, based on a linear
regression on the void sample data (their fig.~7, right), was that
the reference and void sample relations are consistent with each other.
This result implies that there is no significant difference in the average
metallicity for a fixed luminosity of the void and reference sample
galaxies.

\begin{figure*}
\includegraphics[width=11.0cm,angle=-90,clip=]{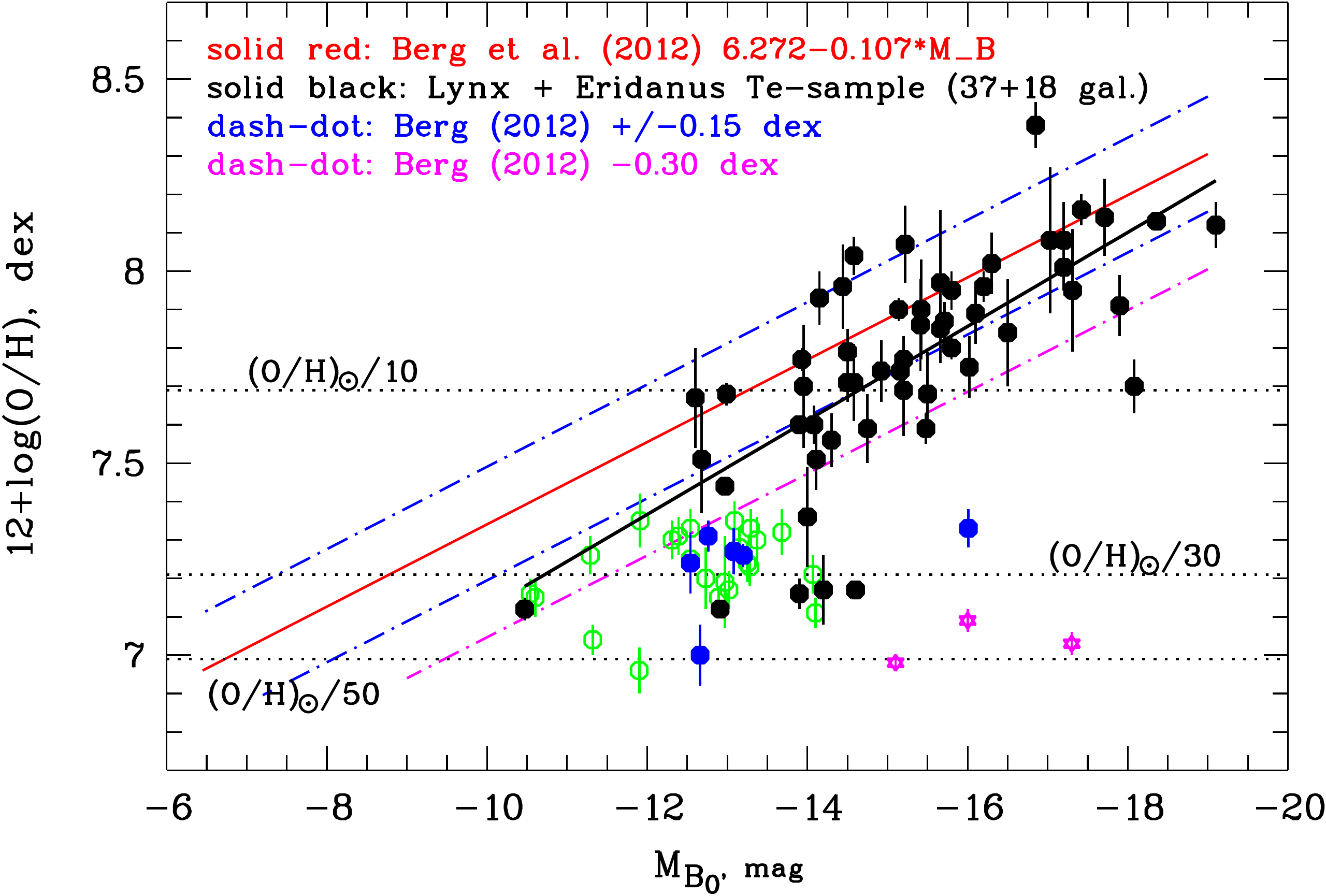}
\caption{
Positions of all galaxies with known direct O/H (black
filled octagons), in Lynx--Cancer (37 objects, \citet{PaperVII}),
and Eridanus (18 objects, \citet{Eridanus}) voids.
The solid black line shows
the linear regression on all 55 void galaxies. It runs somewhat steeper
than the reference relation (solid red) of \citet{Berg12}. As a result,
for more luminous
galaxies, the difference with the reference sample is  subtle. However,
for the range of M$_{\rm B} \gtrsim -14$, the great majority of void galaxies
sit well below the reference relation. Green empty octagons show void
galaxies from PKPE20 and this work with O/H estimated via the 'strong-line'
method of \citet{Izotov19DR14}. Blue octagons are void dwarfs with O/H(dir)
from the literature and 'strong-line' O/H for several dwarfs in the Lynx--Cancer void.
Three distant record-low
XMP BCGs J0811+4730, J1234+3901 and J2229+2725 from papers by Izotov et al.
(2018a, 2019, 2021) are shown as purple stars for comparison.
See discussion in Sec.~\ref{ssec:void.ZL}.
}
\label{fig:ZvsMBTe}
\end{figure*}

We believe, however, that this issue needs a more advanced statistical analysis,
which we perform below.
To the naked eye,  the void data in Figure~\ref{fig:ZvsMBTe}
(black solid octagons)
reveal a clear asymmetry (i.e. a downward shift) in their distribution
relative to the reference sample linear regression line (solid red line).
Hence, a more advanced approach should
check the null hypothesis of whether the distribution of void galaxy
data is consistent with the regression line describing the trend
of the reference sample galaxies. This can be readily checked with the
Student criterion \citep[$t$-statistics; see, e.g.,][]{Korn1968};
 for more detail, see \citet[][pp.23--25, 178]{Bolshev1983}.
If the null hypothesis is rejected at the adopted confidence level,
then the visible shift of O/H for the void galaxy sample can be
treated as an estimate of the difference of the two sample 'means'.

Our sample of galaxies in the Lynx--Cancer and Eridanus voids, with O/H(dir) from
\citet{PaperVII} and \citet{Eridanus}, is  different from that used by
\citet{McQuinn20}. Namely, to the 30 galaxies from table~3 in \citet{PaperVII},
we add six galaxies with O/H(dir) from their table~2 and IZw18, which is also
included in the Lynx--Cancer void sample. The total number
of galaxies with O/H(dir) in the two voids is 55 (37+18).

We apply the Student criterion to compare two samples via their residuals
relative to the linear regression line for the reference sample (Sample~1).
If we subtract the linear regression value for each data point of Sample~1,
we obtain a related sample Sample~1$^{\prime}$ of N~1 = 38 points with  mean
M~1$^{\prime}$  = 0 and standard
deviation S~1$^{\prime}$ = 0.15~dex in log(O/H), the latter value as given
in \citet{Berg12}. For the void data (Sample~2, N~2 = 55 points), for each galaxy
with its absolute magnitude M$_{\rm B}$,
we compute the residual $\Delta$ log(O/H) of their real 12+log(O/H)
relative to the Sample~1 regression line. These residuals comprise the
Sample~2$^{\prime}$ with its calculated mean
M~2$^{\prime}$ =  --0.139$\pm$0.027~dex and its standard deviation
S~2$^{\prime}$ = 0.1988~dex. If both Sample~1$^{\prime}$
and Sample~2$^{\prime}$ have asymptotically normal distribution and the same
value of the mean, then the absolute value of the difference of two sample
means $|$M~1$^{\prime}$ -- M~2$^{\prime}$$|$  divided by the combined standard
deviation estimate s$_{\mathrm 1,2}$,
$|$M~1$^{\prime}$ -- \mbox{M~2$^{\prime}$$|$/s$_{\mathrm 1,2}$} is distributed as $t$
statistics $t(Q,N1+N2-2)$, where

\begin{equation}
s_{\mathrm 1,2} = \sqrt\frac{(N1+N2) \times (N1 \times (S1)^2 + N2 \times (S2)^2)}{N1 \times N2 \times (N1+N2-2)}
\end{equation}

Here, $Q$ is the probability that $t$ exceeds the selected value, related
to the selected probability. With the numbers in hand, we calculate
that s$_{\mathrm 1,2}$ = 0.03848 and
\mbox{$|$M1$^{\prime}$ -- M2$^{\prime}$ $|$/s$_{\mathrm 1,2}$} = 0.139/0.03848 = 3.62.
Since this is larger than t(Q=0.05\%,N=91) = 3.402 (Table 3.2 from
\citet{Bolshev1983}), this means that the $t$-criterion rejects the null
hypothesis on the same 'mean' of Samples~1$^{\prime}$ and 2$^{\prime}$  at
the confidence level of $ C = 1 - P $ = 0.999. The computed difference of the two
means --0.139~dex implies that void galaxies have a mean value of gas O/H
$\sim$30\% lower than the reference sample.
While the above confidence level of 0.999 is high, there is a caveat in
the test, as the assumption of normal distribution of log(O/H) residuals
for the void galaxy sample is not justified.

Therefore, we  additionally apply a non-parametrical statistical criterion
``2$\times$2~contingency table'' \citep{Bolshev1983}. This is often used
in biology and applied studies. In \citet{Pustilnik95} and \citet{PaperIV}
we described it in detail and used it in astronomy.
The goal of the method is to check whether the two considered  properties in
a sample of objects are independent, or are related. In this test, we consider the
reference sample galaxies and the void sample with O/H(dir) data as
representatives of the common population.

The first property comprises $Y$  (i.e. a void galaxy) and 'without' $Y$
(i.e. a LV reference sample object).
The second property comprises $Z$, which has the a log(O/H) residual
$>$ --0.15~dex (conditionally, 'normal' O/H),
and \mbox{'without' $Z$}, which has this parameter $<$ --0.15~dex (conditionally 'low' O/H;
i.e. a stronger outlier towards a smaller O/H).
Then, to test the null hypothesis that the two properties in the sample
are independent of each other, we must compose a 2$\times$2 table with
the general format as in Table~\ref{tab:2x2} (see the first panel).

\begin{table}
\caption{General format of a 2$\times$2 contingency table (first panel)
 and its version for checking void galaxy O/H deficiency (second panel)}
\vspace{1mm} \hspace{-9mm} \centerline{
\begin{tabular}{lccc} \hline
Property     & $Y$     & without $Y$  & Sum   \\
	     & void    & non-void     &       \\ \hline\hline
$Z$          & $m$     & $n-m$        & $n$   \\
'normal' O/H & 34      & 34           &  68   \\ \hline
without $Z$  & $M-m$   & $N-n-(M-m)$  & $N-n$ \\
'low'  O/H   & 21      &  4           &  25   \\ \hline \hline
Sum          & $M$     & $N-M$        & $N$   \\
             & 55      & 38           &  93   \\ \hline
\end{tabular}
} \vspace{1mm}
\label{tab:2x2}
\end{table}

Here, $m$, $n-m$,  $M-m$,  and $N-n-(M-m)$ are the
numbers of galaxies in the sample having the property combinations
($Y$,\,$Z$), (without $Y$,\,$Z$), ($Y$,\,\mbox{without $Z$})  and
(without $Y$,\,without $Z$), respectively. As shown in \citet[][pp.~77--78]{Bolshev1983},
if the properties  $Y$ and $Z$ are independent of each other,
the probability of obtaining the contingency table with such numbers
is described by a hypergeometric distribution. At $N > $ 25, this
can be well approximated by the so-called incomplete beta function
$I_x (a,b)$, where the parameters $x$,~$a$,~and~$b$ can be
expressed in terms of the numbers in the 2$\times$2 contingency
table by the formulas (27)--(30) in \citet[][p.~74]{Bolshev1983}.
See also these formulas in the appendix of \citet{Pustilnik95}.

If there is no correlation between the properties $Y$ and $Z$,
the occupation numbers in the contingency table should
correspond to a low probability of rejecting the Null hypothesis.
For our choice, we have the following numbers (see Table
\ref{tab:2x2}).

The probability of the 2$\times$2 table,
 with the occupation numbers 34, 34, 21 and 4,
calculated with the appropriate formulas for the incomplete
beta function, corresponds to the confidence level of rejecting
the null hypothesis of \mbox{$C = 0.9979$}. This implies
that there is a statistical relation for our sample of galaxies
between the property to reside in voids and the property to have a
reduced metallicity relative to the reference sample objects
with the same blue luminosity. Selecting the border between
the 'normal' and 'low' O/H as a residual log(O/H) $>$ --0.22~dex,
we obtain a 2$\times$2 table with the occupation numbers 42, 38, 13 and 0.
Its respective confidence level is \mbox{$C = 0.99967$}.

As PEPK20 and \citet{McQuinn20} recently discussed, there can be various reasons
for such a reduced metallicity. To elucidate the nature of this phenomenon, one
needs detailed studies of a substantially large sample of such objects.
However, independent of the results of
such a study, there is clear evidence that this phenomenon is related to
the residence in voids.

We illustrate in Fig.~\ref{fig:ZvsMBTe} the significance of the difference discussed
above between the two samples. The reference relation for LV galaxies
from \citet{Berg12} is shown by a red solid line. The similar linear
regression on 55 galaxies from the two mentioned voids with O/H(dir) is
shown by a black solid line. We also show as green empty octagons the NVG galaxies
from PKPE20 and from this work, which have reliable O/H estimates O/H(s,c) via the
strong-lines method of \citet{Izotov19DR14} (see discussion in
Appendix~\ref{ssec:SL-method}).

We recall that these NVG objects
represent the pre-selected 60 dwarf galaxies, which are candidate XMP galaxies from the
'complete' subsample of 380 dwarfs with M$_{\rm B} >$ --14.3
in the NVG sample (PEPK20). Therefore, we do not compare this group
with the reference sample. Their discovery just supports our earlier
finding that among the lowest luminosity void dwarfs there exists a
substantial fraction of gas-rich objects with highly reduced metallicity.
While the amount of data on void galaxy metallicities has increased
substantially, a discussion  of the detailed distribution and possible connection
with other parameters still awaits a larger data set.

\subsection{Relation of void XMP dwarfs to very young galaxies}
\label{ssec:VYGs}

One of the goals of this project is the search for new unusual void
dwarfs resembling very young galaxies (VYGs) as predicted by \citet{Tweed18}.
They are defined as objects that formed more than half their stellar
mass during the last 1 Gyr.

Two record-low XMP galaxies, 12+log(O/H) = 6.98 and 7.02~dex, found by
\citet{Izotov18a} and \citet{Izotov19} show no tracers of the old stellar
population. This fact is consistent with their
extremely low metallicity if one suggests that both properties
are related to the short time elapsed since the
beginning of the main SF episode.

We arrived at a similar conclusion on the probable existence of
young dwarfs in voids in our work devoted to the study the galaxy
population of the Lynx--Cancer void \citep{PaperIV,PaperVII}.
Probably it is not by chance that one of the most studied void XMP dwarfs
assigned to the VYG type and the first prototype XMP dwarf is a blue compact
galaxy IZw18=MRK~116 \citep{PO12}. Such actively star-forming
blue and UV-excess objects have attracted attention in the course of pioneering
surveys of the sky.
However, such outstanding star-forming (candidate) VYGs are very rare.
This follows from only a few findings of XMP objects from over the
whole emission-line galaxy sample in the SDSS DR14 \citep{Izotov19DR14}.

From the results of our search programme, we find that in the volume limited
by the nearby voids (described in PTM19), the majority of the XMP dwarfs
resembling the predicted VYGs are mostly blue LSB galaxies
with a much lower SF efficiency than that in actively star-forming
galaxies. Therefore, in the context of the search for VYGs, obtaining
their census in the local Universe and comparison with model simulations,
our project looks perspective and deserves further development.
While the least-massive dwarfs in voids still
need more detailed predictions of their properties in model simulations,
some general conclusions on the later formation of void galaxies are
presented in \citet{Peper21}.

\section[]{Conclusions}
\label{sec:conclusions}

The NVG sample provides us with a new opportunity to search for the
unusual XMP gas-rich dwarfs in voids. We performed the selection with
the use of publicly available galaxy properties, based on their
similarity to the properties of the known small group of XMP dwarfs.
As a result, we formed a list of 60 void XMP candidates with the
$M_{\rm B}$ range of [--10,--14.3] for their follow-up  careful study.
For 26 of them, we conducted spectroscopy with SALT, which resulted in the
discovery of five new XMP dwarfs with 12+log(O/H) $\lesssim$ 7.21~dex.
Here we present five more new XMP void dwarfs observed at the BTA
and one XMP (J0110--0000) discovered at the BTA but first published in the
paper on the SALT results.
That is, to date we have discovered 10 new void XMP dwarfs.
In addition, at both telescopes, we found 13 very metal-poor
void dwarfs with  7.24~dex $\lesssim$ 12+log(O/H) $\lesssim$ 7.33~dex.

Summarizing the presented results and the related discussion, we draw the
following conclusions.

\begin{enumerate}
\item
In the framework of the ongoing project to search for new unusual void dwarfs
among the pre-selected 60 XMP candidates residing in nearby voids, using the BTA
we perform long-slit spectroscopy of 20 candidates in addition to 26
galaxies already observed at SALT. Two of these 20 objects are common with
those observed at SALT.
\item
For 16 of 20 galaxies observed at the BTA, we derive estimates of their gas O/H.
Only two of them have a 'normal' 12+log(O/H) for their luminosity, both of
7.89~dex.
The remaining 14 of our void galaxies, with 12+log(O/H) $\leq$ 7.33~dex, appear
very metal-poor. Six of them, with 12+log(O/H) = 7.00 -- 7.23~dex, fall into
the category of XMP galaxies, defined here as objects with Z$_{\rm gas} \lesssim$
Z\sunn/30.
Of them, J0110--0000, already presented in our previous paper PKPE20, was
initially found using the BTA.
\item
The O/H values of void XMP dwarfs are reduced by a factor of 2.5--4
relative to the expected values for similar galaxies in the LV
reference sample of \citet{Berg12}.
Their colours show a significant scatter. However, for half of them,
we find blue colours of outer parts and extremely large gas-mass
fraction (0.97--0.99). These properties are similar to those
of the prototype XMP group, including those from the Lynx--Cancer void.
This finding extends the group of the {\bf nearby} candidates for
VYGs to a dozen and allows us to better study their
statistical properties, including their similarity and diversity.
\item
The remaining eight new void dwarfs, with measured values of O/H,
fall in the adjacent range of 12+log(O/H) = 7.24--7.33~dex. Such low
metallicity dwarfs are still very rare, especially in the LV
and the adjacent space. The division to XMP objects as those with
Z $\lesssim$ Z\sunn/30 is conditional.
It will not be surprising if similar objects
will be identified in the group of a slightly less metal-poor galaxies.
More detailed studies of this group will give insights into their evolutionary
state and the possible relation to the most extreme XMP representatives.
\item
In Table~3, we summarize some of the known parameters of all 10 nearby void XMP dwarfs
 found to date at SALT and BTA.
The comprehensive multiwavelength study of the XMP void dwarfs already found
will advance our understanding of galaxy formation and evolution
and the specifics of star formation in such atypical conditions. This
will also provide chances to confirm the discovery of the predicted rare
VYGs.
\end{enumerate}

\section*{Acknowledgements}
The authors thank D.~I.~Makarov, R.~I.~Uklein, A.~F.~Valeev and
A.~S.~Vinokurov for the help with observations at the BTA. We
 also acknowledge the feedback and constructive
criticism of an anonymous referee, which helped us to improve the paper.
The reported study was funded by the Russian Foundation for Basic Research
(RFBR)  according to the research project
No.~18-52-45008-IND$\_{\rm a}$.
AYK acknowledges support from the National Research Foundation (NRF) of
South Africa.
ESE acknowledges support from the RFBR grant No.~18--32--20120.
The initial phase of this work was performed as a part of the government
contract of SAO RAS approved by the Ministry of Science and Higher Education
of the Russian Federation.
Observations with the BTA are also supported by this Ministry
(including agreement No.05.619.21.0016, project ID RFMEFI61919X0016).

We appreciate the work of V.~L.~Afanasiev, before his untimely passing,
and A.~V.~Moiseev and their colleagues
on the substantial upgrade of SCORPIO that allowed us to conduct
the long-slit spectroscopy at the BTA more efficiently.
We also appreciate the allocation of
DDT at the BTA, which helped us to confirm some of the newly found XMP objects.

We acknowledge the important role of the ALFALFA blind \HI\ survey.
A large part of the XMP dwarf candidates in our project were selected because
they became known as nearby gas-rich galaxies after ALFALFA.
The use of the HyperLEDA\footnote{http://leda.univ-lyon1.fr}
data base during several stages of this study is greatfully acknowledged.
This research has made use of the NASA/IPAC Extragalactic Database (NED),
which is operated by the Jet Propulsion Laboratory, California Institute
of Technology, under contract with the National Aeronautics and Space
Administration.

We acknowledge the use of the SDSS data base for this work.
Funding for the Sloan Digital Sky Survey (SDSS) has been provided by the
Alfred P. Sloan Foundation, the Participating Institutions, the National
Aeronautics and Space Administration, the National Science Foundation,
the US Department of Energy, the Japanese Monbukagakusho, and the Max
Planck Society. The SDSS web site is http://www.sdss.org/.
The SDSS is managed by the Astrophysical Research Consortium (ARC) for the
Participating Institutions.
We also acknowledge the use for this work of public archival data from the
Dark Energy Survey (DES), the Pan-STARRS1 Surveys (PS1) and the PS1
public science archive.

\section*{Data Availability}

The data underlying this article are available in Appendices~B and C.
Appendix~C is available only in the online version of the paper.

\appendix

\section{Check of the strong-line and
the semi-empirical methods of Izotov et al. (2019, 2007)}
\label{sec:IT07}

\subsection{The strong-lines method}
\label{ssec:SL-method}

To compare O/H derived via the strong-lines method of \citet{Izotov19DR14}, we
selected only data where the good S/N lines [O{\sc iii}]$\lambda$4363
and [O{\sc ii}]$\lambda$3727 were both available in the spectra.

In particular, we selected 15 of 66 regions in table~A.2 of
\citet{Izotov19DR14}, based purely on the SDSS DR14 spectra,
as satisfying our criteria.
We added 71 data points for 43 regions from the literature and this work,
which include the great majority of published
direct O/H estimates with 12+log(O/H) $\lesssim$ 7.5~dex and the accuracy
of $\sigma$ log(O/H) $\lesssim$0.08--0.09~dex.
They include, among others, three record-low metallicity XMP objects from
\citet{Izotov18a,Izotov19,Izotov21}, and four regions in DDO68 from
\citet{DDO68,LBV,IT07,ITG12,Berg12,DDO68_OH}; Little Cub \citep{LittleCub},
Leo~P \citep{Skillman13}, J0926+3343 \citep{J0926} and AGC198691
\citep{Hirschauer16}.
Data for two regions of IZw18 are adopted from \citet{IT98}, for several regions
in both SBS0335--052W and SBS0335--052E from
\citet{Thuan05,Papaderos06} and \citet{Izotov09}, for  UGC772 from
\citet{Izotov06b,ITG12} and \citet{IT07}, and for two regions in UGCA292
 from \citet{vZee00}.
For other objects with 12+log(O/H) $\lesssim$ 7.5~dex, the data are
retrieved
from \citet{Izotov06a, ITG12, Izotov16, Izotov20}, \citet{SAO0822, HS2134}
and \citet{Guseva09}.

\begin{figure}
\includegraphics[width=5.0cm,angle=-90,clip=]{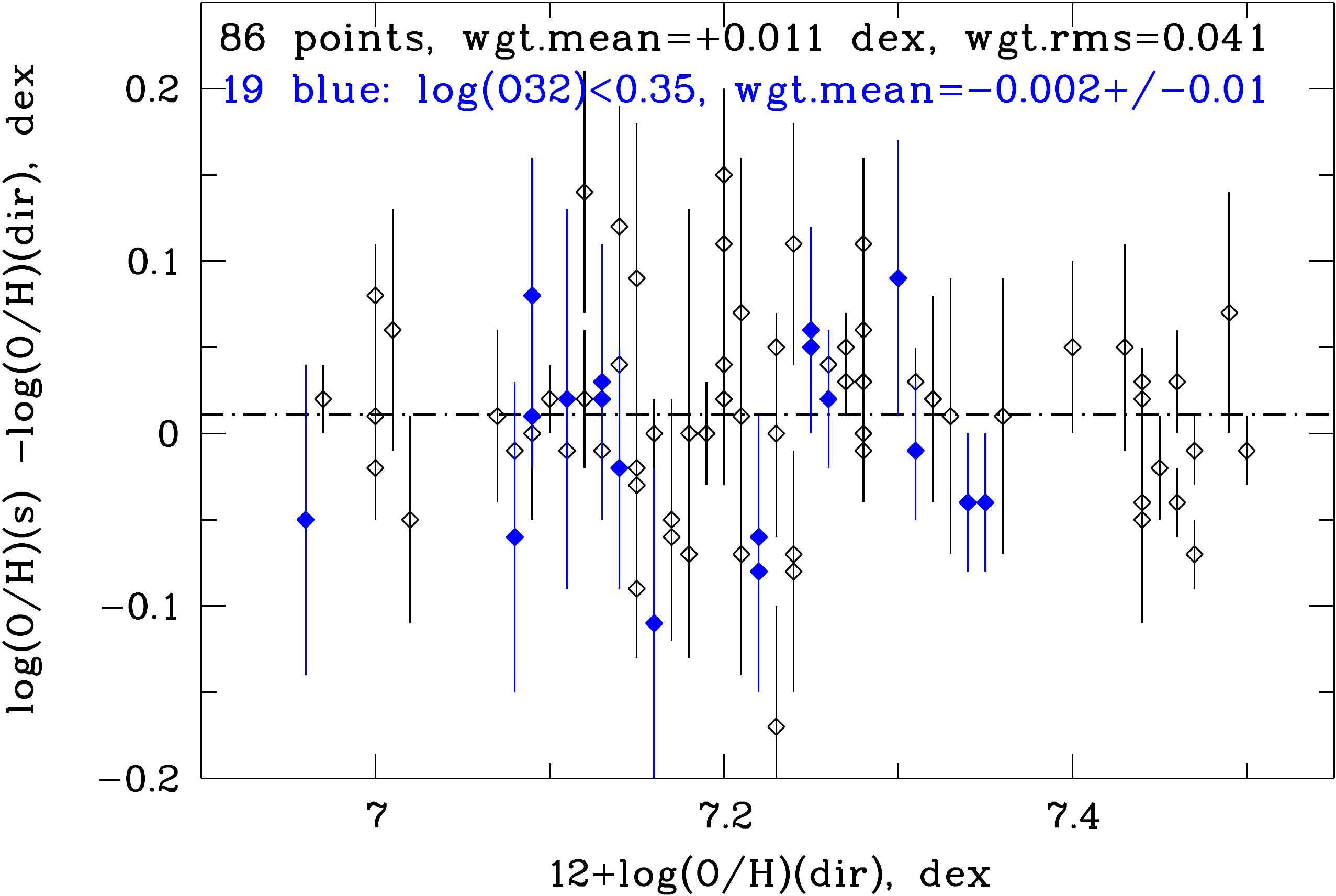}
\caption{
Relation between the difference of log(O/H)(s) (strong-lines) and
log(O/H)(dir) versus
12+log(O/H)(dir) for 86 points in the range of 6.98--7.5~dex
(open lozenges) with their error bars.
The horizontal dash-dotted line shows the weighted mean value of 0.011~dex
on all 86 points with log(O$_{\mathrm 32}$) in the range [--0.3,+1.7].
19 of them (blue), with log(O$_{\mathrm 32}$) of --0.3 to +0.35 (with the
weighted mean of --0.002$\pm$0.01).
See text for references on O/H data and further discussion.
}
\label{fig:SL-dir}
\end{figure}

In Figure~\ref{fig:SL-dir} we plot the differences of
log(O/H)(s) -- log(O/H)(dir) versus 12+log(O/H)(dir) for 86
points (different observations) of 55
different \HII\ regions in 37 galaxies with 12+log(O/H) $<$ 7.5~dex. In
several popular XMP galaxies such as IZw18, SBS0335--052E, SBS0335--052W
and DDO68,
O/H values in their \HII\ regions were obtained on several independent
observations.

The slope of the linear regression does not differ from zero within
uncertainties.
Because there is no trend of log(O/H)(s) -- log(O/H)(dir) with
12+log(O/H)(dir) in the
considered range of O/H, we estimate the mean value of
 log(O/H)(s) -- log(O/H)(dir) on
the whole sample. The horizontal dash-dotted line shows the weighted mean
(+0.011$\pm$0.0044~dex) with the weighted rms of 0.041~dex on all 86 points.
This difference is a little smaller than the similar value of +0.04~dex
from paper of \citet{Izotov19DR14}. Due to larger statistics and the use
of the weighted mean, our value of the offset of 12+log(O/H)(s) relative
to 12+log(O/H)(dir) should be more reliable. Therefore, in statistical
studies, where both types of O/H are used, O/H (s) and O/H (dir),
in order to minimize possible bias, we suggest the use of the parameter
12+log(O/H)(s,c) = 12+log(O/H)(s) -- 0.011~dex.

It is worth of mentioning that the great majority of our studied void dwarf
galaxies have spectra with low excitation. Therefore, to apply the
suggested correction to our new XMP void dwarfs, it is useful to check the
absence of possible bias among the control sample of XMP galaxies with known
O/H(dir) and  those with 'low' log(O$_{\mathrm 32}$).
Accounting for the whole range of log(O$_{\mathrm 32}$) of [--0.3,+1.7]
for the used 86 points, we choose the lower 1/3 of the range, that is
\mbox{--0.3 $<$ log(O$_{\mathrm 32}$) $<$ 0.35} (or O$_{\mathrm 32}$
$<$ 2.24). All our  new XMP void galaxies have that low O$_{\mathrm 32}$.
The respective 19 points with log(O$_{\mathrm 32}$) $<$ 0.35
of the whole 86 points
are shown by blue symbols in Fig.~\ref{fig:SL-dir}. Their weighted mean
of \mbox{--0.002$\pm$0.010~dex} does not differ within its uncertainty from
the general weighted mean of +0.011~dex. Therefore, for our
low-O$_{\mathrm 32}$ new void XMP galaxies, we use the parameter
12+log(O/H)(s,c) defined above.

\subsection{The semi-empirical method}
\label{ssec:SE-method}

As mentioned in Section~\ref{sec:OH_estimates}, the values
of T$_{\rm e}$ calculated via the direct method (i.e. with the
use of auroral line [O{\sc iii}]$\lambda$4363) and those
calculated with R$_{\mathrm 23}$ within the se
method \citep{IT07} can differ by thousands of K and this might lead
to a bias in the values of O/H derived via the se method relative
to those of the direct method.

The difference in estimates of T$_{\rm e}$ can be related to the large
range of the ionization parameter $U$.
For a given value of R$_{\mathrm 23}$, $U$ can vary by two to three orders
of magnitude for \HII\ regions in various galaxies but this is not accounted
for in the se method.

Another factor affecting the scatter of the estimate of T$_{\rm e}$ via
R$_{\mathrm 23}$ can be the range of the effective temperatures T$_{\rm eff}$
of the central ionizing star (or cluster)  or the related parameter
(i.e. the hardness of ionizing flux).

However, as demonstrated by \citet[][see his fig.~9]{Skillman89}
with models of \HII\ regions (the range of $U$ = 0.0001--0.1 and the
central star T$_{\rm eff}$ of 38000, 45000 and 55000~K), the most important
factor determining the ionized gas temperature T$_{\rm e}$ is $U$.
In his grid, the variation of $U$ results in the largest variations of
T$_{\rm e}$ for
T$_{\rm eff}$=55000 (from $\sim$8000~K for 12+log(O/H) = 7.22~dex
to $\sim$4000~K at 12+log(O/H) = 7.92~dex).

For the central star with T$_{\rm eff}$=38000~K, the range of T$_{\rm e}$
variations is smaller: from $\sim$4500~K at 12+log(O/H) = 7.22~dex to
$\sim$2500~K at 12+log(O/H) = 7.92~dex.
At the same time, at the fixed value of $U$, the dependence of T$_{\rm e}$
on T$_{\rm eff}$ (between 38000 and 55000~K) is small for
the lowest $U$ = 0.0001 (with the range of the variance of 1000 --
1500~K) and substantially larger for
the highest values of $U$ = 0.1 (range of the variance of $\sim$3000~K).

The check of the effect of hardness to the empirical estimate of T$_{\rm e}$
from observational data is outside the scope of this work.
However, we can try to follow the possible effect of $U$ on T$_{\rm e}$ based on
the large amount of data available in the literature.
To do this at a first approximation, we draw the relation between
the difference $\delta$t$_{\rm e}$(dir,R$_{\mathrm 23}$) of the two
$t_{\rm e}$ estimates (hereafter $t_{\rm e}$=T$_{\rm e}$/10000):
$t_{\rm e}$(dir)
and of $t_{\rm e}$(R$_{\mathrm 23}$) (derived with the se method
of \citet{IT07}) and log(O$_{\mathrm 32}$).
Parameter O$_{\mathrm 32}$ is defined as the flux ratio of the lines
[O{\sc iii}]$\lambda$5007 and [O{\sc ii}]$\lambda$3727.
O$_{\mathrm 32}$ is an observational proxy of $U$ (e.g.,
 \citet[][fig.~7]{Skillman89}; \citet{McGaugh91}.

\begin{figure}
\includegraphics[width=5.2cm,angle=-90,clip=]{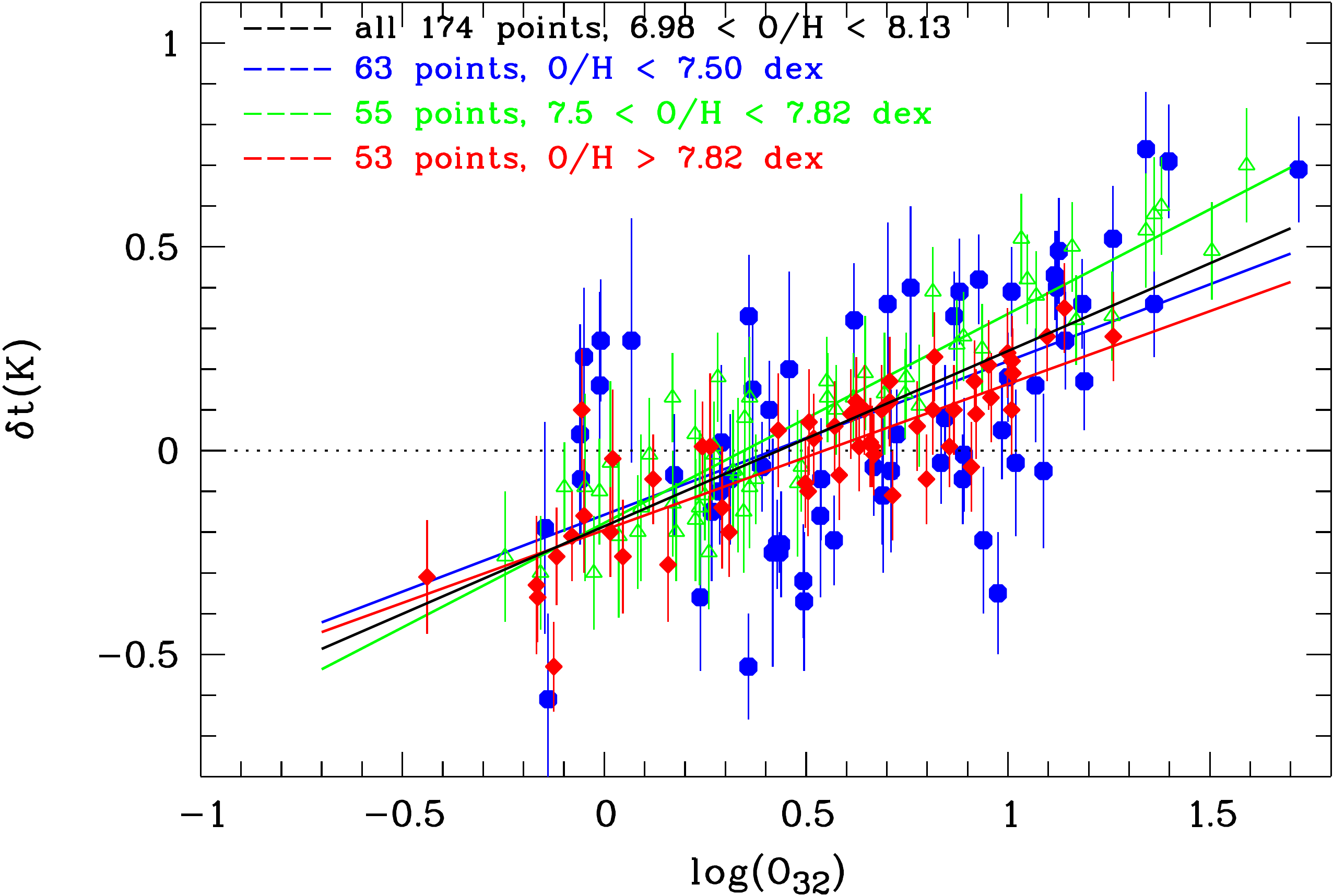}
\includegraphics[width=5.2cm,angle=-90,clip=]{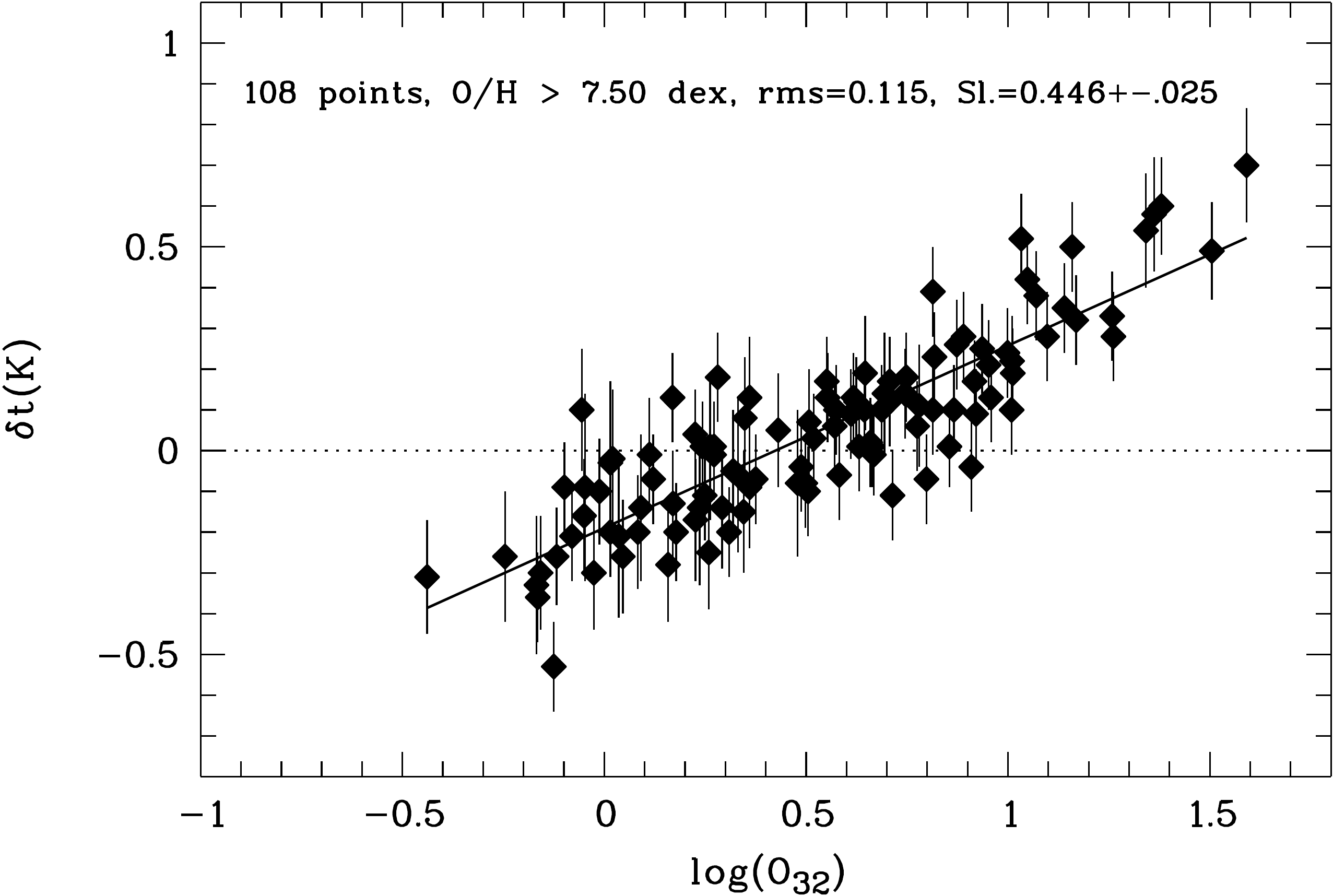}
\caption{Relation between the difference
$\delta$~t$_{\rm e}$(dir, R$_{\mathrm 23}$) (where t=T$_{\rm e}$/10000) and
log(O$_{\mathrm 32}$) for 174 measurements of \HII\ regions in a range of
12+log(O/H) = 6.98--8.13~dex.
Error bars show the $\pm$1$\sigma$ uncertainties of t$_{\rm e}$(dir).
Top panel:
the straight lines show the linear regression fits for the whole sample
(black solid) and three subsamples.
The whole sample is broken into three approximately equal parts
 with 63, 55 and 53 points,
with 12+log(O/H) in the ranges of [6.98--7.5] (blue),
[7.5--7.82] (green) and [7.82--8.13]~dex (red). Their rms scatter
about regression lines are, respectively, 0.25, 0.099 and 0.103~K.
Bottom panel: linear regression line and the scatter about it
for 108 points with 12+log(O/H) $>$ 7.5~dex.
Its scatter about the regression is rms=0.116~K in comparison
to this parameter of 0.25~K for the remaining
63 points with 12+log(O/H) $<$ 7.5~dex.
}
\label{fig:dTe.fit}
\end{figure}

In Figure~\ref{fig:dTe.fit} we show a set of
$\delta$ t$_{\rm e}$(dir,R$_{\mathrm 23}$) for 174 different measurements
(135 \HII\ regions in 116 galaxies) versus their parameter log(O$_{\mathrm 32}$).
Their gas 12+log(O/H) and parameter log(O$_{\mathrm 32}$) are distributed in wide ranges:
 6.98--8.13~dex,  --0.4 to +1.7, respectively.
The data are collected from the papers already cited in Appendix~\ref{ssec:SL-method} for
the lowest O/H objects -- some of them include data for objects with 12+log(O/H)(dir)
$>$ 7.5~dex -- and from several additional papers for galaxies with
12+log(O/H)(dir) $>$ 7.5~dex. They include papers by \citet{vZee97, vZee06a, vZee06b,
Bresolin09}, \citet{HS0837, PaperII, PaperVII, XMP.SALT}, \citet{Eridanus}
and this work.

In the top panel, we split the whole sample into three approximately equal parts
on metallicity (conditionally, 'bottom', 'middle' and 'top' with the
borders at 12+log(O/H) = 6.98--7.5~dex, 7.5--7.82 and 7.82--8.13~dex)
to probe how the relation $\delta$~t$_{\rm e}$(dir,R$_{\mathrm 23}$) vs
log(O$_{\mathrm 32}$) depends on metallicity.
While the slopes of the linear regressions for the three subgroups show some
variance (from $\sim$0.36 to $\sim$0.51), the rms scatter of $\delta$~t for
the two higher O/H groups is reasonably small: 0.10--0.103~K. For the subgroup
with the lowest metallicities, the rms scatter $\delta$~t remainss large, $\sim$0.25~K.

The above 'small' rms$\sim$0.1~K implies that for \HII\ regions with
12+log(O/H) in
the range of 7.5--8.1~dex, taking into account a term proportional to
log(O$_{\mathrm 32}$) will mainly compensate for the difference between
t$_{\rm e}$(dir) and t$_{\rm e}$(R$_{\mathrm 23}$) for both
sufficiently high and low values of O$_{\mathrm 32}$, meaning that the
uncertainty of the derived t$_{\rm e}$ is  similar to that of the original
estimate of t$_{\rm e}$ from \citet{IT07}.

As for the
\HII\ regions with 12+log(O/H) $\lesssim$ 7.5~dex, the currently available
data on $\delta$~t$_{\rm e}(dir,R_{\mathrm 23}$) show too large scatter
about the respective linear regression. Therefore, the
application  of such a correction to t$_{\rm e}$(R$_{\mathrm 23}$)
bears a substantial additional error in t$_{\rm e}$, and, in turn,
into an estimate of O/H.  Fortunately, a purely empirical
formula involving both R$_{\mathrm 23}$ and O$_{\mathrm 32}$,
suggested by \citet{Izotov19DR14} (the so-called strong-lines method),
approximates O/H of
\HII\ regionswell  in this lowest metallicity domain, adding only
$\sigma \sim$ 0.04~dex into the related uncertainty of 12+log(O/H)
(see Appendix~\ref{ssec:SL-method}).

\begin{figure}
\includegraphics[width=5.2cm,angle=-90,clip=]{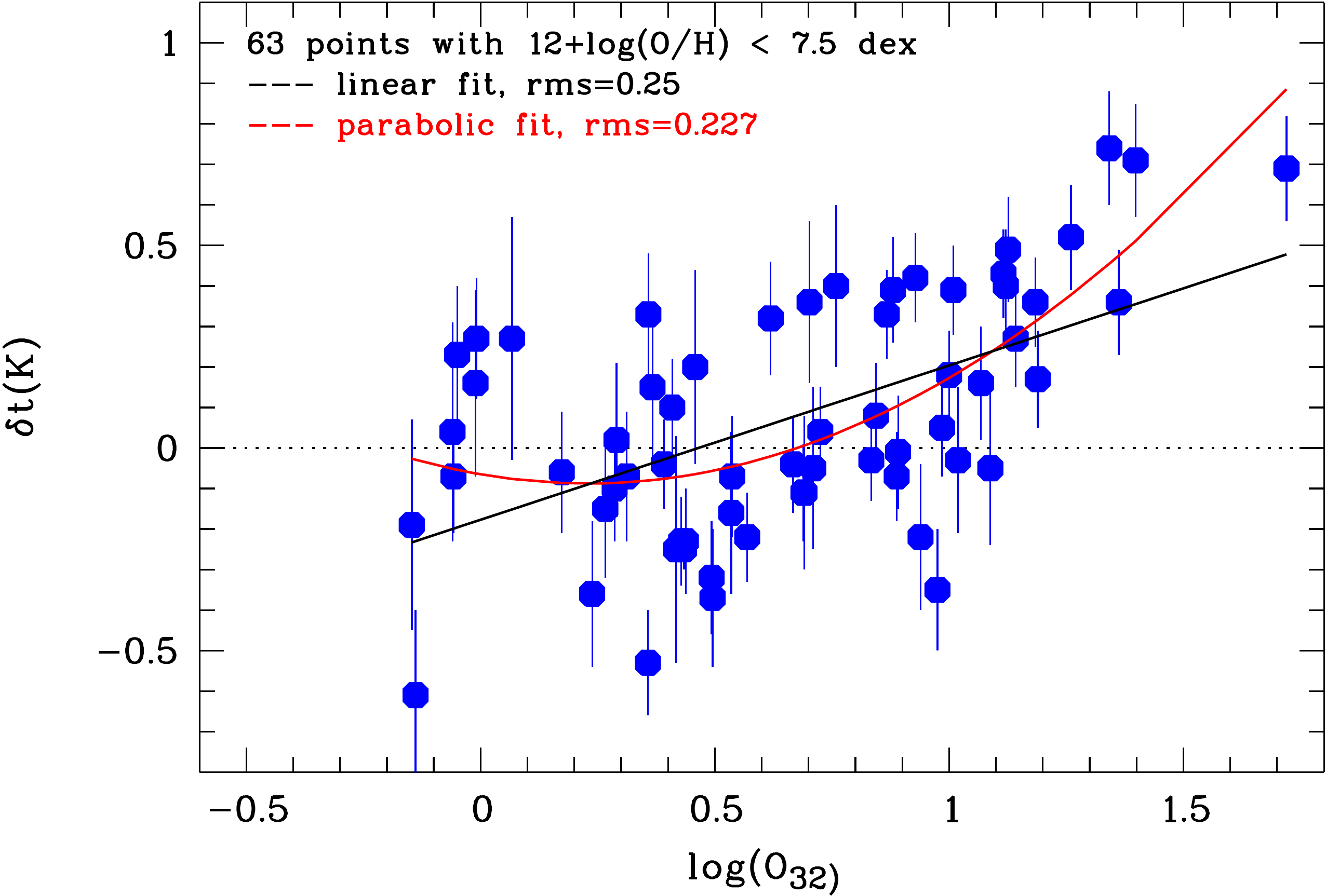}
\caption{A similar plot as in Figure~\ref{fig:dTe.fit} but for 63 points
with 12+log(O/H) $<$ 7.5~dex.
Both linear and parabolic fits are shown.
See text for further explanations and discussion.
}
\label{fig:dTe_lowO32.fit}
\end{figure}

For the range of 12+log(O/H) = 7.5--8.1~dex, we adopt the
linear regression in the bottom panel as:
$$ \delta~t_{\rm e}(dir,R_{\mathrm 23})= +0.4463\times log (O_{\mathrm 32}) - 0.1886, (A1) $$
with the rms = 0.116~K about this line,  $\sigma$(slope) = 0.026
and $\sigma$(constant) = 0.0176~K.
As one can see, $\delta$t$_{\rm e}(dir,R_{\mathrm 23})$ is close to zero at
log(O$_{\mathrm 32}$) $\sim$ 0.46 (or O$_{\mathrm 32} \sim$ 2.9). For the
range of log(O$_{\mathrm 32}$) $\sim$ 0.26--0.66,
$\delta$t$_{\rm e}(dir,R_{\mathrm 23})$ falls within $\pm$0.1~K, which
corresponds to the formal accuracy of the original se method \citep{IT07}.
That is, the original se method can give an acceptable estimate of
t$_{\rm e}$(dir) for O$_{\mathrm 32} \sim$ 1.8--4.6.
For O$_{\mathrm 32}$ outside this range, the systematic error in the estimate
of t$_{\rm e}$(dir) becomes larger, reaching --0.4~K and +0.5~K at the
extreme low and high values of O$_{\mathrm 32}$.

Thus, in order to use the se method of \citet{IT07} in the whole range of
the observed excitation parameter O$_{\mathrm 32}$, one should modify
the original formula for  $t_{\rm e}$ from \citet{IT07}:
$$ t_{\rm e} = -1.3685 \times \log(R_{\mathrm 23}) + 2.6258,   (A2) $$
adding the term log(O$_{\mathrm 32}$) from equation (A1) as follows:
$$ t_{\rm e} = -1.3685 \times log(R_{\mathrm 23}) + 2.4372 + 0.4463\times log (O_{\mathrm 32}) (A3) $$

As for the range of 12+log(O/H) $<$ 7.5, there is still a need for a better
estimate of $t_{\rm e}$, as the strong-lines method assumes the use
of corrected intensities of strong lines, which in turn depend on the adopted
value of $t_{\rm e}$. In Fig.~\ref{fig:dTe_lowO32.fit}, we show linear and
quadratic fits of $\delta$~t versus logO$_{\mathrm 32}$ for 63 points with
12+log(O/H) $<$ 7.5. As can be seen, there is a flattening in this relation
for values of logO$_{\mathrm 32}$ $<$ 0.5, which the parabolic fit catches
better. Indeed, although its rms scatter remains large, it certainly
reduces relative to that for the linear regression (0.227 versus 0.25~K).
Therefore, for the range of 12+log(O/H) $<$ 7.5, we adopt an alternative
formula for $t_{\rm e}$, including the quadratic fit in Fig.~\ref{fig:dTe_lowO32.fit}:

$$ t_{\rm e} = -1.3685 \times log R_{\mathrm 23} + 2.561 -0.199\times log O_{\mathrm 32} +0.437\times (log O_{\mathrm 32})^2 (A4) $$

The small values of $\delta$t for the lowest  O/H range and
logO$_{\mathrm 32}$ $<$ 0.5, imply that the se method in this range of
parameters should work without the substantial systematics.

\begin{figure}
\includegraphics[width=5.2cm,angle=-90,clip=]{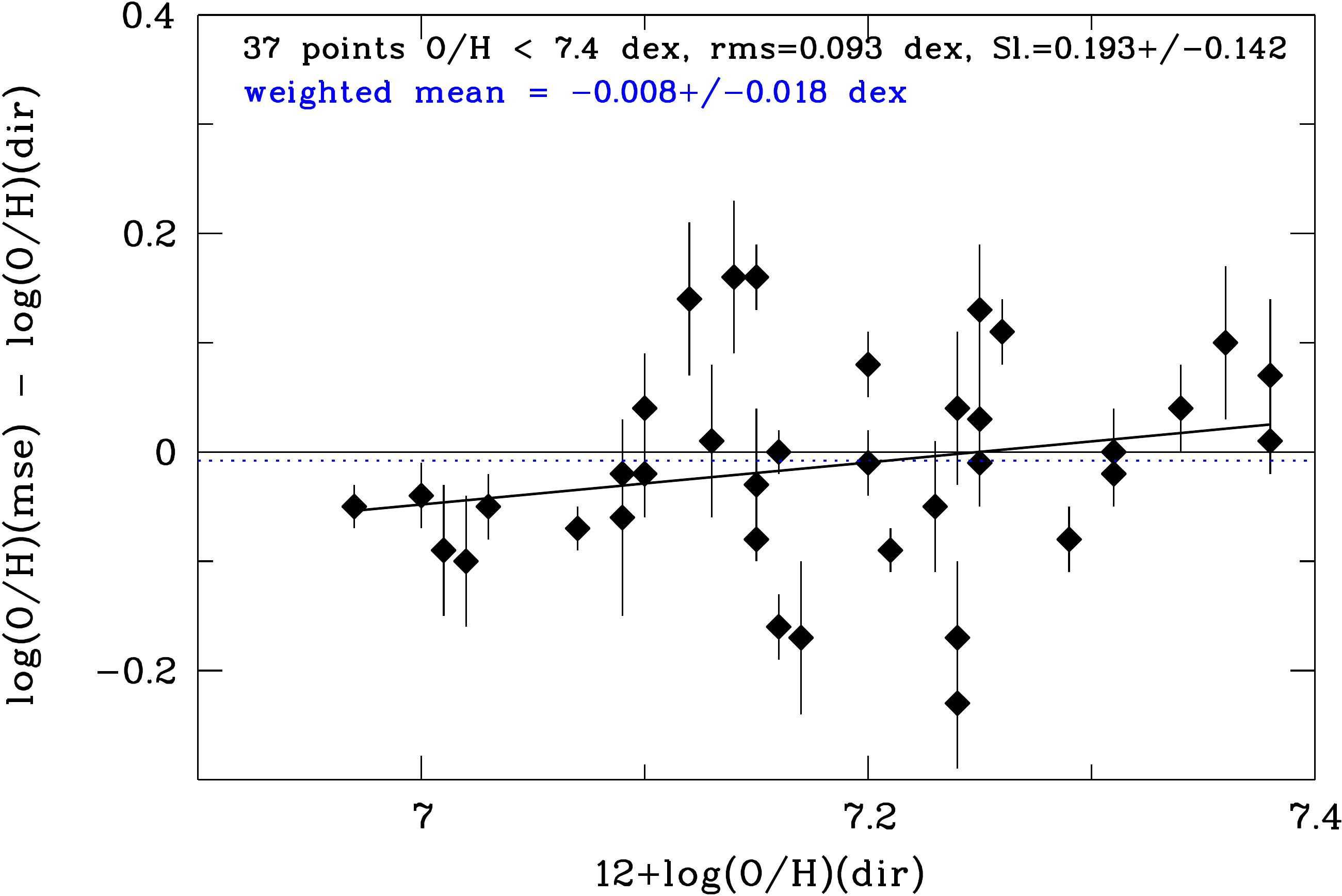}
\includegraphics[width=5.2cm,angle=-90,clip=]{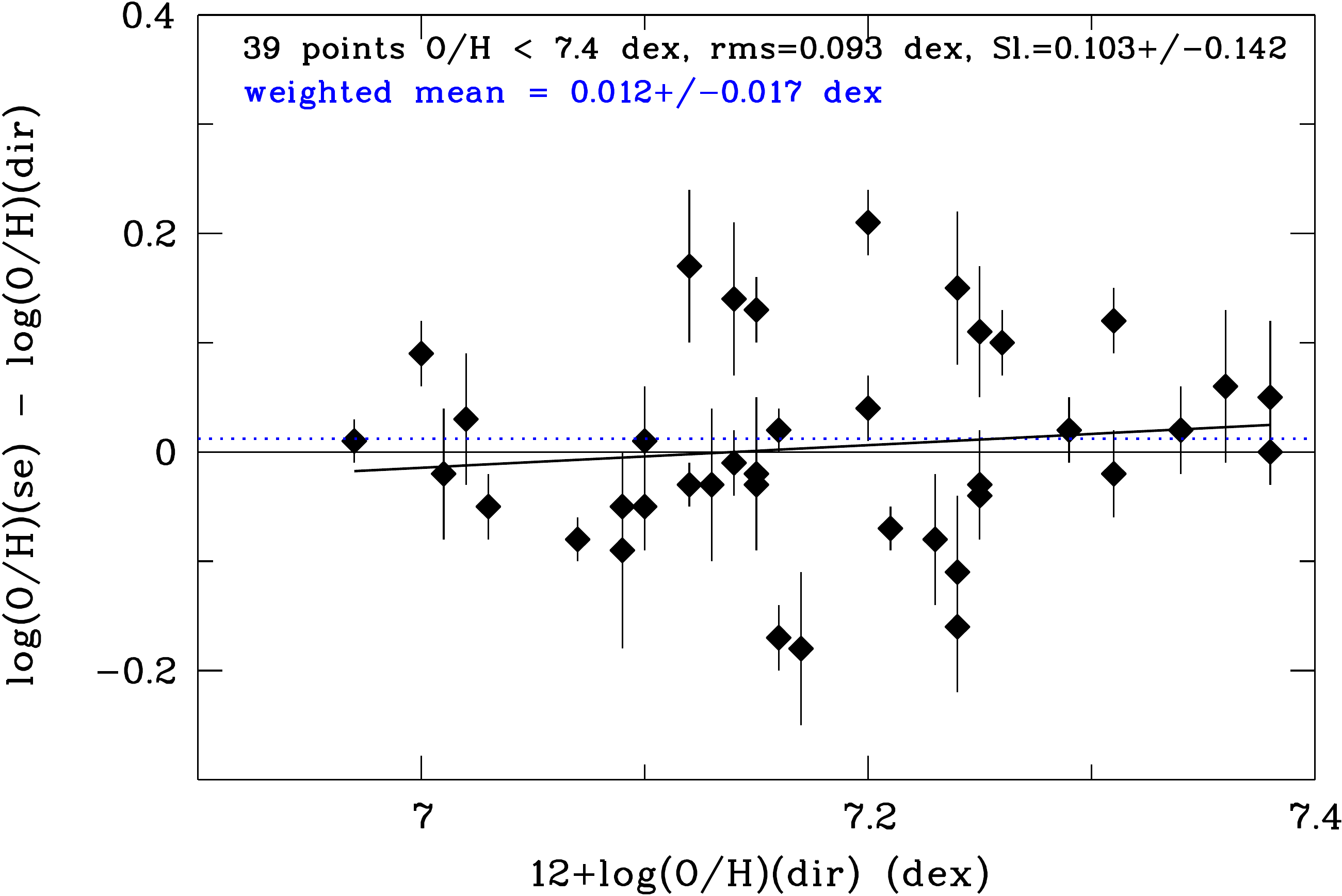}
\caption{Top panel:
relation of the difference of logO/H(mse) and logO/H(dir) versus
12+logO/H(dir) for 37 points with 12+log(O/H) $<$ 7.4~dex. The weighted
mean of the difference is $\sim$ --0.008~dex with the rms scatter of
$\sim$.093~dex.
Bottom panel: similar relation for the difference of logO/H(se) and logO/H(dir) vs
12+logO/H(dir) for 39 points with 12+log(O/H) $<$ 7.4~dex.  The weighted
mean difference is $\sim$ +0.012~dex with the rms scatter of $\sim$.093~dex.
}
\label{fig:OHmse_OHd_XMP}
\end{figure}

Similarly to the analysis in \citet{Izotov19DR14}, we limit the range of
12+log(O/H) $<$ 7.4, in order to have an opportunity to directly compare
our results with theirs. In Fig.~\ref{fig:OHmse_OHd_XMP},
we plot the difference of log(O/H)(mse) -- log(O/H)(dir) and
log(O/H)(se) -- log(O/H)(dir), respectively, versus 12+log(O/H)
on all available
data. The inclined solid lines in both plots show the fitted linear
regressions. In both cases, the slope does not differ significantly from zero.
Therefore, we calculate the weighted means (blue dotted lines) as a measure
of the mean difference of the mse and se methods relative to
12+log(O/H)(dir) in this O/H range. They are \mbox{--0.008$\pm$0.010~dex} for
the mse method, and \mbox{+0.012$\pm$0.017~dex} for the se method. The rms scatter for
both methods is very close, of $\sim$0.093~dex. We notice that for the se method,
\citet{Izotov19DR14} give the mean difference of log(O/H) $\sim$ +0.08~dex.

\begin{figure}
\includegraphics[width=5.2cm,angle=-90,clip=]{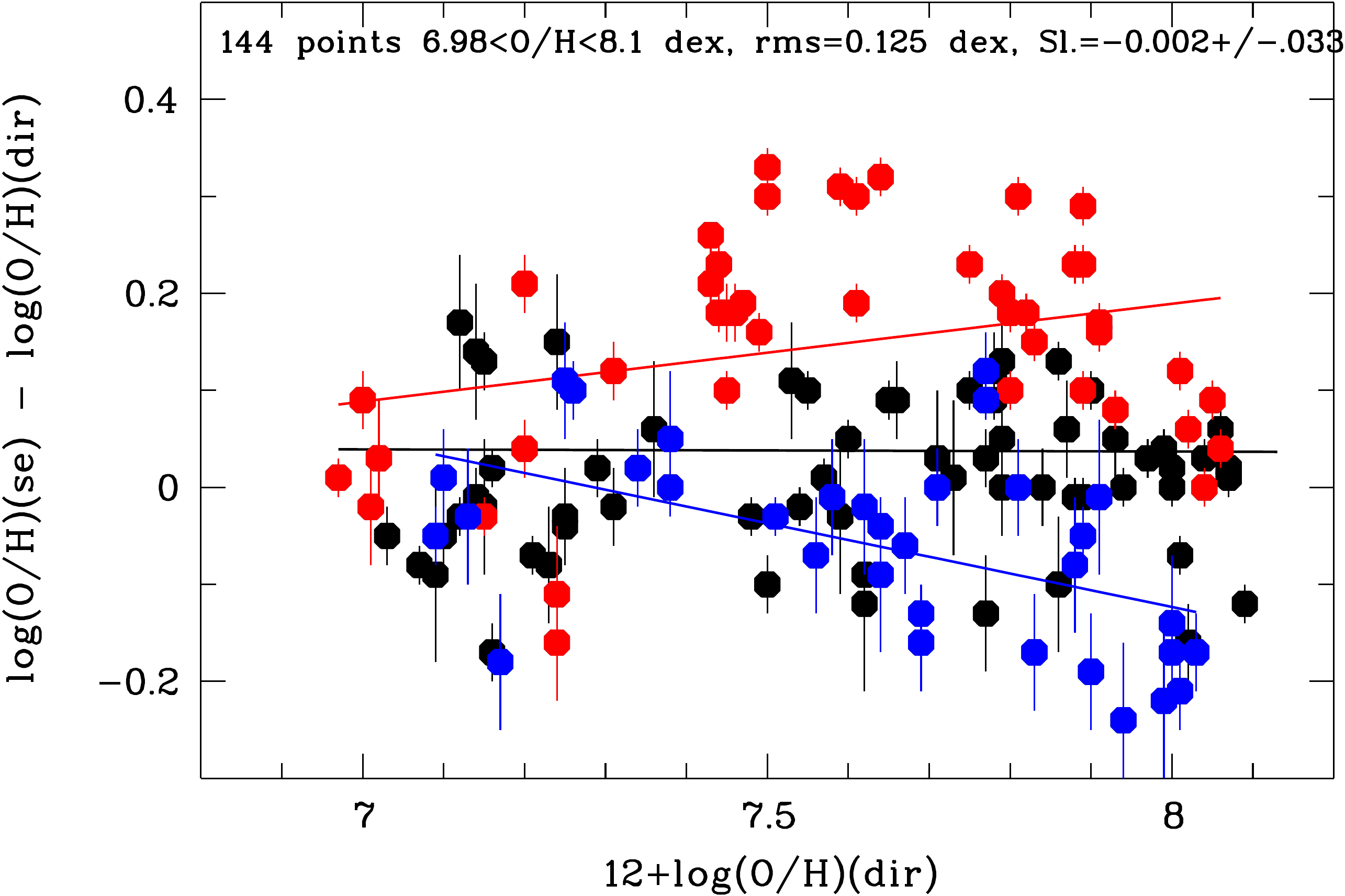}
\includegraphics[width=5.2cm,angle=-90,clip=]{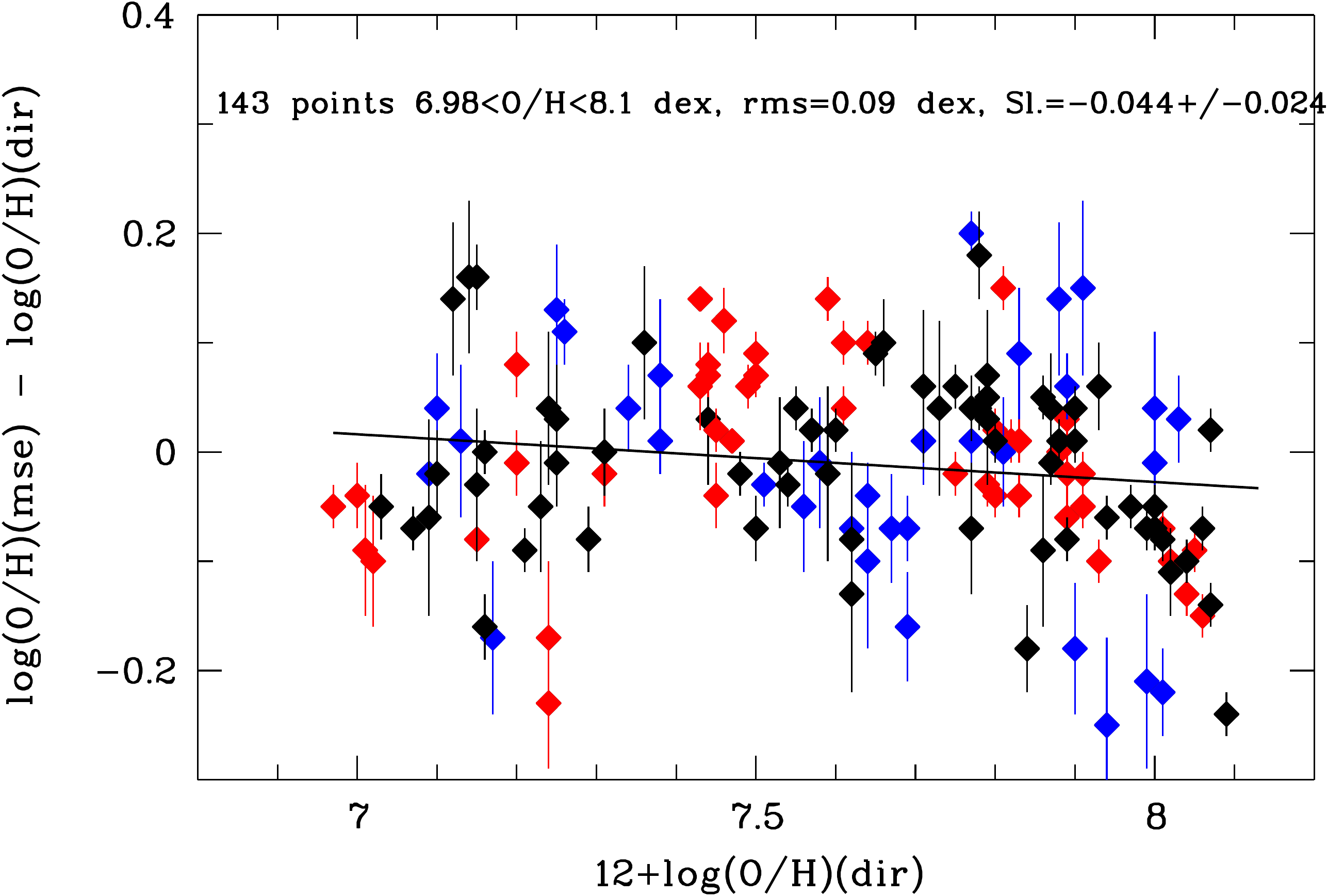}
\caption{ Top panel:
relation of the difference log(O/H)(se) -- log(O/H)(dir) versus
12+log(O/H)(dir) on a sample of 144 points from the literature, with
6.98 $<$ 12+log(O/H) $<$ 8.1~dex.
The black line shows the linear regression fit for the whole data
with the rms = 0.125~dex. Blue, black and red points
show separately the data with low, middle and high O$_{\mathrm 32}$, corresponding
to logO$_{\mathrm 32}$ in the ranges (--0.4, +0.2; +0.2, +0.8; and +0.8, +1.7)
respectively. Blue and red solid lines show respective linear regressions
for low and high O$_{\mathrm 32}$ subsamples.
Bottom panel: same as in top panel, but for the mse method for 143 points.
Its scatter on the linear regression is significantly smaller, rms = 0.09~dex.
}
\label{fig:OHse_OHd}
\end{figure}

Despite the above conclusion of a small difference of (O/H)(se) and
(O/H)(dir) in the lowest O/H range, in a wider range of O/H the difference
is well traced.
We illustrate this situation with the original se method in
Fig.~\ref{fig:OHse_OHd}.
Here the differences of log(O/H)(se) -- log(O/H)(dir) are
plotted versus 12+log(O/H)(dir). The total range of parameter
logO$_{\mathrm 32}$ from $\sim$ --0.4 to $\sim$ 1.7 is divided
into three ranges (--0.4,+0.2; +0.2,+0.8; and +0.8,+1.7). The points
for the lower, higher and intermediate logO$_{\mathrm 32}$ ranges are
shown in blue,   red, and  black.
Now it is clearly seen that the lowest logO$_{\mathrm 32}$ points show
the systematically reduced O/H relative to log(O/H)(dir), which appear
smallest for the lowest 12+log(O/H) $\lesssim$ 7.3, but $\sim$ 0.2~dex
lower than O/H(dir) for 12+log(O/H) $\gtrsim$ 8.0.
For points with high logO$_{\mathrm 32}$, the estimates of
log(O/H)(se) appear systematically elevated relative to log(O/H)(dir).
For the lowest O/H, 12+log(O/H) $\lesssim$ 7.3, they show large scatter
but the mean difference of $\lesssim$ 0.05~dex. For 12+log(O/H) $\sim$ 8.0
dex, the mean difference is $\sim$ +0.2~dex.

The se method, based on the modified equations (A3) and (A4) for the
estimate of t$_{\rm e}$, which includes the logO$_{\mathrm 32}$ term,
allows us to remove the systematics inherent to the original se
method by \citet{IT07}, as illustrated in Fig.~\ref{fig:OHse_OHd} (bottom panel).
As can be seen, blue and red points show the even scatter on the linear
regression line drawn on the whole sample. Moreover, as a result of
the removal of systematics, the general scatter on the linear regression
is reduced substantially: from rms = 0.125~dex in Fig.~\ref{fig:OHse_OHd} (top;
original se method) to rms = 0.090~dex for the mse method.
This indicates that the addition of a term with logO$_{\mathrm 32}$ to the
original formula for T$_{\rm e}$ from \citet{IT07} indeed improves the
accuracy of O/H for the indicated O/H range for a wide range of O$_{\mathrm 32}$.

For the practical use of the mse method, it is necessary to fit a relation of
12+log(O/H)(dir) versus 12+log(O/H)(mse) (Y versus X in the equation below) as follows:
$$ Y = 0.960(\pm 0.024)~X + 0.318(\pm 0.184)~~~~ (A5)  $$
The rms scatter of 143 analysed points on the linear regression is 0.090~dex.
Over the range of 12+log(O/H)(mse) = 7.0--8.1~dex,  the related correction,
that is log(O/H)(dir) -- log(O/H)(mse),
varies between +0.038~dex and --0.006~dex, respectively.
We call the parameter $Y$, calculated in equation (A5), a
corrected log(O/H)(mse,c) and apply it in further O/H estimates
as a well approximating log(O/H)(dir) without an additional offset.
The intrinsic accuracy of log(O/H)(mse,c) is 0.09~dex.

The residual scatter between 12+log(O/H)(mse) and 12+log(O/H)(dir) can
be attributed partly to the accuracy of the original O/H(dir), and partly
to the unaccounted factor of the ionizing radiation hardness. It is clear
from the models
\citep[e.g.,][]{Skillman89} that its effect on T$_{\rm e}$ can be comparable,
in principle, with the effect of the ionization parameter $U$ (or of its proxy
O$_{\mathrm 32}$).
However, there are very few data on the radiation hardness in various
\HII\ regions. Because of various selection effects inherent to samples with the
sufficiently strong line [O{\sc iii}]$\lambda$4363, used for O/H(dir),
their hardness parameter falls into a range that is not wide.
The above scatter could also arise from differences in the method used to
estimate the extinction coefficient CH$\beta$ (affecting
O$_{\mathrm 32}$) and the value of EW(abs) (affecting R$_{\mathrm 23}$)
that were used in our analysis of spectra from the literature.

For the majority of our void galaxy spectra, the value of logO$_{\mathrm 32}$
falls below 0.5. Therefore, one expects some corrections of
t$_{\rm e}$ calculated in papers by
\citet{PaperVII, XMP.SALT} and \cite{Eridanus}  with the
original se method, with a related small correction of $\Delta$ log(O/H).
We employ this mse method and present in
Tables~B1--B7 
the O/H estimates derived with this method, O/H(mse,c), for comparison derived with other methods.

\section{Tables with line fluxes and derived parameters}
\label{sec:tables_fluxes}

The tables in this appendix include the measured line fluxes F($\lambda$)
(relative to the flux of H$\beta$)  and the line intensities I($\lambda$),
corrected for extinction and the underlying stellar Balmer absorption,
with their errors.

The tables also include the measured EWs of the H$\beta$ emission  line
and the derived parameters: the extinction coefficient C (H$\beta$),
the equivalent width of the underlying stellar Balmer absorptions EW(abs),
the electron temperatures T$_{\rm e}$(O{\sc iii}) -- estimated with the mse
method (see Appendix~A2) -- if it was not possible with the direct method,
and T$_{\rm e}$(O{\sc ii}), the two temeperatures, corresponding to two qzones
of oxygen ionization. We also present the derived oxygen abundances
in two stages of ionization and the total value of O/H, including its
value in units of 12+log(O/H) as derived either via the direct method,
or the mse method, when [O{\sc iii}]$\lambda$4363 was not detected.

For the range of 12+log(O/H) $<$ 7.5~dex, we also present O/H(s,c) and
O/H(se,c) as described in Appendices~A1 and A2. The flux in the emission line
H$\beta$ is in units of 10$^{-16}$ erg~s$^{-1}$cm$^{-2}$. Its error
reflects the measurement uncertainty only and does not include probable
loss on the slit.
Electron densities n$_{\rm e}$ in \HII\ regions could not be estimated
when the doublet [S{\sc ii}]$\lambda\lambda$6716,6731
was outside the observed wavelength range. In this case, we adopted
n$_{\rm e}$ to be 10 per cm$^3$, typical of \HII\ regions in dIrr galaxies.

\begin{table*} 
\caption{Measured and corrected line intensities, and derived oxygen abundances}
\label{t:Intens1}
\begin{tabular}{lcccccc} \hline
\rule{0pt}{10pt}
& \MC{2}{c}{PGC000083a=J0001+3222a} & \MC{2}{c}{PGC000083b=J0001+3222b}  & \MC{2}{c}{PiscesA=J0014+1048}  \\ \hline
\rule{0pt}{10pt}
$\lambda_{0}$(\AA) Ion                  & F($\lambda$)/F(H$\beta$)&I($\lambda$)/I(H$\beta$) & F($\lambda$)/F(H$\beta$)&I($\lambda$)/I(H$\beta$) & F($\lambda$)/F(H$\beta$)&I($\lambda$)/I(H$\beta$) \\ \hline
3727\ [O\ {\sc ii}]\                           & 1.768$\pm$0.138 & 1.530$\pm$0.144     & 0.612$\pm$0.105 & 0.664$\pm$0.120     & 1.196$\pm$0.098 & 1.182$\pm$0.108     \\
4101\ H$\delta$\                               & ---               & ---               & 0.190$\pm$0.028 & 0.287$\pm$0.056     & 0.161$\pm$0.019 & 0.297$\pm$0.044     \\
4340\ H$\gamma$\                               & 0.352$\pm$0.023 & 0.475$\pm$0.042     & 0.402$\pm$0.028 & 0.475$\pm$0.042     & 0.359$\pm$0.025 & 0.457$\pm$0.039     \\
4861\ H$\beta$\                                & 1.000$\pm$0.049 & 1.000$\pm$0.061     & 1.000$\pm$0.049 & 1.000$\pm$0.053     & 1.000$\pm$0.040 & 1.000$\pm$0.045     \\
4959\ [O\ {\sc iii}]\                          & 0.264$\pm$0.020 & 0.222$\pm$0.020     & 0.498$\pm$0.027 & 0.474$\pm$0.027     & 0.551$\pm$0.021 & 0.497$\pm$0.021     \\
5007\ [O\ {\sc iii}]\                          & 0.880$\pm$0.035 & 0.741$\pm$0.035     & 1.360$\pm$0.056 & 1.287$\pm$0.056     & 1.689$\pm$0.059 & 1.519$\pm$0.058     \\
6563\ H$\alpha$\                               & 3.196$\pm$0.118 & 2.725$\pm$0.130     & ---               & ---               & 3.264$\pm$0.103 & 2.752$\pm$0.104     \\
6717\ [S\ {\sc ii}]\                           & 0.281$\pm$0.022 & 0.230$\pm$0.022     & ---               & ---               & ---               & ---               \\
6731\ [S\ {\sc ii}]\                           & 0.179$\pm$0.020 & 0.147$\pm$0.020     & ---               & ---               & ---               & ---               \\
  & & \\
C(H$\beta$)\ dex          & \MC {2}{c}{0.04$\pm$0.05} & \MC {2}{c}{0.17$\pm$0.06} & \MC {2}{c}{0.12$\pm$0.04} \\
EW(abs)\ \AA\             & \MC {2}{c}{3.10$\pm$0.27} & \MC {2}{c}{2.30$\pm$0.76} & \MC {2}{c}{3.85$\pm$0.37} \\
F(H$\beta$)               & \MC {2}{c}{1.68$\pm$0.06} & \MC {2}{c}{1.40$\pm$0.04} & \MC {2}{c}{7.76$\pm$0.21} \\
EW(H$\beta$)\ \AA\        & \MC {2}{c}{  17$\pm$ 1}   & \MC {2}{c}{  54$\pm$ 2}   & \MC {2}{c}{  38$\pm$ 1}   \\
& \\
$T_{\rm e}$(OIII)(K)\                & \MC {2}{c}{21033$\pm$1107~~}     & \MC {2}{c}{20128$\pm$1116~~}     & \MC {2}{c}{18478$\pm$1059~~}     \\
$T_{\rm e}$(OII)(K)\                 & \MC {2}{c}{15971$\pm$284 ~~}     & \MC {2}{c}{15619$\pm$67  ~~}     & \MC {2}{c}{14934$\pm$89  ~~}     \\
$N_{\rm e}$(SII)(cm$^{-3}$)\         & \MC {2}{c}{ 10$\pm$10 ~~}         & \MC {2}{c}{ 10$\pm$10 ~~}         & \MC {2}{c}{ 10$\pm$10 ~~}         \\
O$^{+}$/H$^{+}$($\times$10$^5$)\     & \MC {2}{c}{1.148$\pm$0.124~~}     & \MC {2}{c}{0.532$\pm$0.096~~}     & \MC {2}{c}{1.086$\pm$0.101~~}     \\
O$^{++}$/H$^{+}$($\times$10$^5$)\    & \MC {2}{c}{0.373$\pm$0.042~~}     & \MC {2}{c}{0.745$\pm$0.088~~}     & \MC {2}{c}{1.021$\pm$0.131~~}     \\
O/H($\times$10$^5$)\                 & \MC {2}{c}{1.521$\pm$0.131~~}     & \MC {2}{c}{1.277$\pm$0.131~~}     & \MC {2}{c}{2.107$\pm$0.165~~}     \\
12+log(O/H)(mse,c)\                  & \MC {2}{c}{~7.19$\pm$0.10~~}      & \MC {2}{c}{~7.12$\pm$0.10~~}      & \MC {2}{c}{~7.33$\pm$0.10~~}      \\
12+log(O/H)(se,c)\                  & \MC {2}{c}{~7.25$\pm$0.10~~}      & \MC {2}{c}{~7.05$\pm$0.10~~}      & \MC {2}{c}{~7.29$\pm$0.10~~}      \\
12+log(O/H)(s,c)\                   & \MC {2}{c}{~7.17$\pm$0.05~~}      & \MC {2}{c}{~7.13$\pm$0.05~~}      & \MC {2}{c}{~7.26$\pm$0.05~~}      \\
\hline
\end{tabular}
\end{table*}

\begin{table*} 
\caption{Measured and corrected line intensities, and derived oxygen abundances}
\label{t:Intens2}
\begin{tabular}{lcccccc} \hline
\rule{0pt}{10pt}
& \MC{2}{c}{AGC411446=J0110--00000}   & \MC{2}{c}{PiscesB=J0119+1107}    & \MC{2}{c}{AGC122400=J0231+3542}   \\ \hline
\rule{0pt}{10pt}
$\lambda_{0}$(\AA) Ion                  & F($\lambda$)/F(H$\beta$)&I($\lambda$)/I(H$\beta$) & F($\lambda$)/F(H$\beta$)&I($\lambda$)/I(H$\beta$) & F($\lambda$)/F(H$\beta$)&I($\lambda$)/I(H$\beta$) \\ \hline
3727\ [O\ {\sc ii}]\                           & 0.918$\pm$0.108 & 0.776$\pm$0.109     & 2.551$\pm$0.149 & 2.327$\pm$0.169     & 1.600$\pm$0.750 & 1.192$\pm$0.664     \\
4340\ H$\gamma$\                               & 0.227$\pm$0.024 & 0.386$\pm$0.058     & 0.300$\pm$0.062 & 0.446$\pm$0.143     & 0.114$\pm$0.059 & 0.448$\pm$0.464     \\
4861\ H$\beta$\                                & 1.000$\pm$0.044 & 1.000$\pm$0.055     & 1.000$\pm$0.055 & 1.000$\pm$0.091     & 1.000$\pm$0.202 & 1.000$\pm$0.340     \\
4959\ [O\ {\sc iii}]\                          & 0.282$\pm$0.033 & 0.239$\pm$0.033     & 0.300$\pm$0.025 & 0.258$\pm$0.025     & 0.543$\pm$0.138 & 0.361$\pm$0.137     \\
5007\ [O\ {\sc iii}]\                          & 0.848$\pm$0.044 & 0.717$\pm$0.044     & 1.006$\pm$0.051 & 0.862$\pm$0.051     & 1.657$\pm$0.263 & 1.097$\pm$0.261     \\
6548\ [N\ {\sc ii}]\                           & 0.033$\pm$0.065 & 0.028$\pm$0.065     & ---               & ---               & ---               & ---               \\
6563\ H$\alpha$\                               & 3.111$\pm$0.141 & 2.711$\pm$0.158     & ---               & ---               & ---               & ---               \\
6584\ [N\ {\sc ii}]\                           & 0.102$\pm$0.079 & 0.086$\pm$0.079     & ---               & ---               & ---               & ---               \\
6717\ [S\ {\sc ii}]\                           & 0.171$\pm$0.087 & 0.144$\pm$0.087     & ---               & ---               & ---               & ---               \\
6731\ [S\ {\sc ii}]\                           & 0.080$\pm$0.078 & 0.068$\pm$0.078     & ---               & ---               & ---               & ---               \\
  & & \\
C(H$\beta$)\ dex          & \MC {2}{c}{0.00$\pm$0.06} & \MC {2}{c}{0.08$\pm$0.04} & \MC {2}{c}{0.15$\pm$0.04} \\
EW(abs)\ \AA\             & \MC {2}{c}{3.65$\pm$0.34} & \MC {2}{c}{2.50$\pm$1.00} & \MC {2}{c}{4.50$\pm$1.25} \\
F(H$\beta$)$^a$\          & \MC {2}{c}{1.09$\pm$0.03} & \MC {2}{c}{7.19$\pm$0.27} & \MC {2}{c}{0.48$\pm$0.04} \\
EW(H$\beta$)\ \AA\        & \MC {2}{c}{  20$\pm$ 1}   & \MC {2}{c}{  16$\pm$ 1}   & \MC {2}{c}{   9$\pm$ 1}   \\
& \\
$T_{\rm e}$(OIII)(K)\                & \MC {2}{c}{22319$\pm$1187~~}     & \MC {2}{c}{19673$\pm$1086~~}     & \MC {2}{c}{19793$\pm$2972~~}     \\
$T_{\rm e}$(OII)(K)\                 & \MC {2}{c}{16271$\pm$248 ~~}     & \MC {2}{c}{14947$\pm$67  ~~}     & \MC {2}{c}{14938$\pm$228 ~~}     \\
$N_{\rm e}$(SII)(cm$^{-3}$)\         & \MC {2}{c}{ 10$\pm$10 ~~}         & \MC {2}{c}{ 10$\pm$10 ~~}         & \MC {2}{c}{ 10$\pm$10 ~~}         \\
O$^{+}$/H$^{+}$($\times$10$^5$)\     & \MC {2}{c}{0.551$\pm$0.081~~}     & \MC {2}{c}{2.133$\pm$0.158~~}     & \MC {2}{c}{1.094$\pm$0.781~~}     \\
O$^{++}$/H$^{+}$($\times$10$^5$)\    & \MC {2}{c}{0.329$\pm$0.038~~}     & \MC {2}{c}{0.496$\pm$0.062~~}     & \MC {2}{c}{0.638$\pm$0.236~~}     \\
O/H($\times$10$^5$)\                 & \MC {2}{c}{0.880$\pm$0.090~~}     & \MC {2}{c}{2.629$\pm$0.170~~}     & \MC {2}{c}{1.732$\pm$0.816~~}     \\
12+log(O/H)(mse,c)\                  & \MC {2}{c}{~6.96$\pm$0.10~~}      & \MC {2}{c}{~7.43$\pm$0.10~~}      & \MC {2}{c}{~7.25$\pm$0.20~~}      \\
12+log(O/H)(se,c)\                   & \MC {2}{c}{~6.95$\pm$0.10~~}      & \MC {2}{c}{~7.42$\pm$0.10~~}      & \MC {2}{c}{~7.18$\pm$0.20~~}      \\
12+log(O/H)(s,c)\                    & \MC {2}{c}{~7.00$\pm$0.05~~}      & \MC {2}{c}{~7.31$\pm$0.05~~}      & \MC {2}{c}{~7.19$\pm$0.12~~}      \\
 & \\
\hline
\end{tabular}
\end{table*}

\begin{table*} 
\caption{Measured and corrected line intensities, and derived oxygen abundances}
\label{t:Intens3}
\begin{tabular}{lcccccc} \hline
\rule{0pt}{10pt}
& \MC{2}{c}{J0823+1758}  & \MC{2}{c}{AGC239144=J1349+3544}  & \MC{2}{c}{AGC249590=J1440+3416}  \\ \hline
\rule{0pt}{10pt}
$\lambda_{0}$(\AA) Ion                  & F($\lambda$)/F(H$\beta$)&I($\lambda$)/I(H$\beta$) & F($\lambda$)/F(H$\beta$)&I($\lambda$)/I(H$\beta$) & F($\lambda$)/F(H$\beta$)&I($\lambda$)/I(H$\beta$) \\ \hline
3727\ [O\ {\sc ii}]\                           & 1.807$\pm$0.149 & 1.908$\pm$0.173     & 2.575$\pm$0.384 & 2.180$\pm$0.397     & 1.471$\pm$0.136 & 1.478$\pm$0.150     \\
4101\ H$\delta$\                               & ---               & ---               & ---               & ---               & 0.143$\pm$0.020 & 0.245$\pm$0.043     \\
4340\ H$\gamma$\                               & 0.381$\pm$0.025 & 0.478$\pm$0.041     & 0.181$\pm$0.038 & 0.332$\pm$0.099     & 0.397$\pm$0.027 & 0.481$\pm$0.039     \\
4861\ H$\beta$\                                & 1.000$\pm$0.039 & 1.000$\pm$0.045     & 1.000$\pm$0.066 & 1.000$\pm$0.082     & 1.000$\pm$0.036 & 1.000$\pm$0.040     \\
4959\ [O\ {\sc iii}]\                          & 0.454$\pm$0.027 & 0.415$\pm$0.026     & 0.323$\pm$0.044 & 0.273$\pm$0.044     & 0.357$\pm$0.021 & 0.328$\pm$0.021     \\
5007\ [O\ {\sc iii}]\                          & 1.514$\pm$0.053 & 1.376$\pm$0.053     & 0.857$\pm$0.062 & 0.726$\pm$0.062     & 1.178$\pm$0.038 & 1.077$\pm$0.038     \\
6548\ [N\ {\sc ii}]\                           & ---               & ---               & 0.084$\pm$0.205 & 0.071$\pm$0.205     & ---               & ---               \\
6563\ H$\alpha$\                               & 3.387$\pm$0.103 & 2.760$\pm$0.100     & 3.177$\pm$0.295 & 2.787$\pm$0.332     & 3.201$\pm$0.110 & 2.741$\pm$0.111     \\
6584\ [N\ {\sc ii}]\                           & ---               & ---               & 0.253$\pm$0.217 & 0.214$\pm$0.217     & ---               & ---               \\
  & & \\
C(H$\beta$)\ dex          & \MC {2}{c}{0.19$\pm$0.04} & \MC {2}{c}{0.00$\pm$0.12} & \MC {2}{c}{0.12$\pm$0.04} \\
EW(abs)\ \AA\             & \MC {2}{c}{2.00$\pm$0.39} & \MC {2}{c}{2.00$\pm$0.28} & \MC {2}{c}{2.80$\pm$0.34} \\
F(H$\beta$)$^a$\          & \MC {2}{c}{2.40$\pm$0.10} & \MC {2}{c}{0.34$\pm$0.02} & \MC {2}{c}{0.46$\pm$0.01} \\
EW(H$\beta$)\ \AA\        & \MC {2}{c}{  24$\pm$ 1}   & \MC {2}{c}{  11$\pm$ 1}   & \MC {2}{c}{  33$\pm$ 1}   \\
& \\
$T_{\rm e}$(OIII)(K)\                & \MC {2}{c}{18066$\pm$1079~~}     & \MC {2}{c}{15179$\pm$1374~~}     & \MC {2}{c}{19536$\pm$1089~~}     \\
$T_{\rm e}$(OII)(K)\                 & \MC {2}{c}{14889$\pm$145 ~~}     & \MC {2}{c}{13994$\pm$668 ~~}     & \MC {2}{c}{14954$\pm$49  ~~}     \\
$N_{\rm e}$(SII)(cm$^{-3}$)\         & \MC {2}{c}{ 10$\pm$10 ~~}         & \MC {2}{c}{ 10$\pm$10 ~~}         & \MC {2}{c}{ 10$\pm$10 ~~}         \\
O$^{+}$/H$^{+}$($\times$10$^5$)\     & \MC {2}{c}{1.770$\pm$0.169~~}     & \MC {2}{c}{2.464$\pm$0.590~~}     & \MC {2}{c}{1.352$\pm$0.138~~}     \\
O$^{++}$/H$^{+}$($\times$10$^5$)\    & \MC {2}{c}{0.953$\pm$0.129~~}     & \MC {2}{c}{0.799$\pm$0.191~~}     & \MC {2}{c}{0.632$\pm$0.076~~}     \\
O/H($\times$10$^5$)\                 & \MC {2}{c}{2.722$\pm$0.213~~}     & \MC {2}{c}{3.263$\pm$0.620~~}     & \MC {2}{c}{1.984$\pm$0.158~~}     \\
12+log(O/H)(mse,c)\                  & \MC {2}{c}{~7.44$\pm$0.10~~}      & \MC {2}{c}{~7.39$\pm$0.13~~}      & \MC {2}{c}{~7.31$\pm$0.10~~}      \\
12+log(O/H)(se,c)\                   & \MC {2}{c}{~7.42$\pm$0.10~~}      & \MC {2}{c}{~7.37$\pm$0.12~~}      & \MC {2}{c}{~7.28$\pm$0.10~~}      \\
12+log(O/H)(s,c)\                    & \MC {2}{c}{~7.33$\pm$0.05~~}      & \MC {2}{c}{~7.25$\pm$0.07~~}      & \MC {2}{c}{~7.23$\pm$0.05~~}      \\
 & \\
\hline
\end{tabular}
\end{table*}

\begin{table*}
\caption{Measured and corrected line intensities, and derived oxygen abundances}
\label{t:Intens4}
\begin{tabular}{lcccccc} \hline
\rule{0pt}{10pt}
 & \MC{2}{c}{J1444+4242a}  & \MC{2}{c}{J1444+4242b}  & \MC{2}{c}{PGC2081790=J1447+3630}  \\ \hline
\rule{0pt}{10pt}
$\lambda_{0}$(\AA) Ion                  & F($\lambda$)/F(H$\beta$)&I($\lambda$)/I(H$\beta$) & F($\lambda$)/F(H$\beta$)&I($\lambda$)/I(H$\beta$) & F($\lambda$)/F(H$\beta$)&I($\lambda$)/I(H$\beta$) \\ \hline
3727\ [O\ {\sc ii}]\                           & 1.162$\pm$0.088 & 1.070$\pm$0.092     & 1.233$\pm$0.115 & 1.161$\pm$0.120     & 1.868$\pm$0.061 & 1.709$\pm$0.063     \\
3967\ [Ne\ {\sc iii}]\ +\ H7\                  & ---               & ---               & ---               & ---               & 0.109$\pm$0.017 & 0.226$\pm$0.046     \\
4101\ H$\delta$\                               & 0.132$\pm$0.017 & 0.273$\pm$0.048     & 0.110$\pm$0.016 & 0.212$\pm$0.043     & 0.195$\pm$0.018 & 0.299$\pm$0.034     \\
4340\ H$\gamma$\                               & 0.317$\pm$0.025 & 0.441$\pm$0.045     & 0.416$\pm$0.024 & 0.477$\pm$0.035     & 0.338$\pm$0.018 & 0.428$\pm$0.029     \\
4861\ H$\beta$\                                & 1.000$\pm$0.032 & 1.000$\pm$0.038     & 1.000$\pm$0.035 & 1.000$\pm$0.042     & 1.000$\pm$0.024 & 1.000$\pm$0.028     \\
4959\ [O\ {\sc iii}]\                          & 0.471$\pm$0.024 & 0.419$\pm$0.024     & 0.278$\pm$0.026 & 0.253$\pm$0.026     & 0.530$\pm$0.013 & 0.485$\pm$0.013     \\
5007\ [O\ {\sc iii}]\                          & 1.403$\pm$0.045 & 1.248$\pm$0.045     & 0.877$\pm$0.036 & 0.797$\pm$0.036     & 1.616$\pm$0.036 & 1.478$\pm$0.036     \\
6563\ H$\alpha$\                               & ---               & ---               & ---               & ---               & 2.865$\pm$0.057 & 2.756$\pm$0.066     \\
  & & \\
C(H$\beta$)\ dex          & \MC {2}{c}{0.04$\pm$0.04} & \MC {2}{c}{0.04$\pm$0.05} & \MC {2}{c}{0.00$\pm$0.03} \\
EW(abs)\ \AA\             & \MC {2}{c}{2.50$\pm$0.25} & \MC {2}{c}{1.25$\pm$0.22} & \MC {2}{c}{4.70$\pm$0.41} \\
F(H$\beta$)$^a$\          & \MC {2}{c}{0.77$\pm$0.02} & \MC {2}{c}{0.56$\pm$0.02} & \MC {2}{c}{1.78$\pm$0.03}     \\
EW(H$\beta$)\ \AA\        & \MC {2}{c}{  21$\pm$0.4}  & \MC {2}{c}{  13$\pm$0.3}   & \MC {2}{c}{  50$\pm$ 1}   \\
& \\
$T_{\rm e}$(OIII)(K)\                & \MC {2}{c}{19443$\pm$1060~~}     & \MC {2}{c}{21189$\pm$1113~~}     & \MC {2}{c}{17909$\pm$1016~~}     \\
$T_{\rm e}$(OII)(K)\                 & \MC {2}{c}{14958$\pm$36  ~~}     & \MC {2}{c}{16011$\pm$279 ~~}     & \MC {2}{c}{14867$\pm$156 ~~}     \\
$N_{\rm e}$(SII)(cm$^{-3}$)\         & \MC {2}{c}{ 10$\pm$10 ~~}         & \MC {2}{c}{ 10$\pm$10 ~~}         & \MC {2}{c}{ 10$\pm$10 ~~}         \\
O$^{+}$/H$^{+}$($\times$10$^5$)\     & \MC {2}{c}{0.978$\pm$0.085~~}     & \MC {2}{c}{0.864$\pm$0.100~~}     & \MC {2}{c}{1.592$\pm$0.078~~}     \\
O$^{++}$/H$^{+}$($\times$10$^5$)\    & \MC {2}{c}{0.757$\pm$0.090~~}     & \MC {2}{c}{0.400$\pm$0.045~~}     & \MC {2}{c}{1.065$\pm$0.135~~}     \\
O/H($\times$10$^5$)\                 & \MC {2}{c}{1.736$\pm$0.123~~}     & \MC {2}{c}{1.265$\pm$0.109~~}     & \MC {2}{c}{2.657$\pm$0.156~~}     \\
12+log(O/H)(mse,c)\                  & \MC {2}{c}{~7.25$\pm$0.10~~}      & \MC {2}{c}{~7.12$\pm$0.10~~}      & \MC {2}{c}{~7.43$\pm$0.10~~}      \\
12+log(O/H)(se,c)\                   & \MC {2}{c}{~7.18$\pm$0.10~~}      & \MC {2}{c}{~7.08$\pm$0.10~~}      & \MC {2}{c}{~7.40$\pm$0.11~~}      \\
12+log(O/H)(s,c)\                    & \MC {2}{c}{~7.20$\pm$0.05~~}      & \MC {2}{c}{~7.11$\pm$0.05~~}      & \MC {2}{c}{~7.32$\pm$0.04~~}      \\
 & \\
\hline
\end{tabular}
\end{table*}

\begin{table*}
\caption{Measured and corrected line intensities, and derived oxygen abundances}
\label{t:Intens5}
\begin{tabular}{lcccccc} \hline
\rule{0pt}{10pt}
& \MC{2}{c}{J1522+4201}  & \MC{2}{c}{J2103--0049}  & \MC{2}{c}{AGC321307=J2214+2540}   \\ \hline
\rule{0pt}{10pt}
$\lambda_{0}$(\AA) Ion                  & F($\lambda$)/F(H$\beta$)&I($\lambda$)/I(H$\beta$) & F($\lambda$)/F(H$\beta$)&I($\lambda$)/I(H$\beta$) & F($\lambda$)/F(H$\beta$)&I($\lambda$)/I(H$\beta$) \\ \hline
3727\ [O\ {\sc ii}]\                           & 2.481$\pm$0.169 & 2.191$\pm$0.171     & 1.621$\pm$0.084 & 1.594$\pm$0.085     & 2.764$\pm$0.384 & 2.729$\pm$0.396     \\
4340\ H$\gamma$\                               & 0.208$\pm$0.026 & 0.314$\pm$0.065      & ---               & ---               & ---               & ---              \\
4861\ H$\beta$\                                & 1.000$\pm$0.037 & 1.000$\pm$0.057     & 1.000$\pm$0.028 & 1.000$\pm$0.039     & 1.000$\pm$0.103 & 1.000$\pm$0.224     \\
4959\ [O\ {\sc iii}]\                          & 0.390$\pm$0.028 & 0.344$\pm$0.028     & 0.311$\pm$0.011 & 0.305$\pm$0.011     & 0.764$\pm$0.091 & 0.751$\pm$0.092     \\
5007\ [O\ {\sc iii}]\                          & 1.026$\pm$0.037 & 0.906$\pm$0.037     & 0.913$\pm$0.026 & 0.897$\pm$0.026     & 2.800$\pm$0.223 & 2.754$\pm$0.223     \\
6563\ H$\alpha$\                               & 3.000$\pm$0.087 & 2.736$\pm$0.102     & 2.767$\pm$0.066 & 2.733$\pm$0.075     & 2.790$\pm$0.227 & 2.817$\pm$0.337     \\
  & & \\
C(H$\beta$)\ dex          & \MC {2}{c}{0.00$\pm$0.04} & \MC {2}{c}{0.00$\pm$0.03} & \MC {2}{c}{0.01$\pm$0.10} \\
EW(abs)\ \AA\             & \MC {2}{c}{2.30$\pm$0.68} & \MC {2}{c}{0.10$\pm$0.16} & \MC {2}{c}{0.05$\pm$0.61} \\
F(H$\beta$)$^a$\          & \MC {2}{c}{0.77$\pm$0.02} & \MC {2}{c}{0.50$\pm$0.01} & \MC {2}{c}{0.55$\pm$0.04} \\
EW(H$\beta$)\ \AA\        & \MC {2}{c}{17.4$\pm$0.55} & \MC {2}{c}{2.04$\pm$0.04} & \MC {2}{c}{3.09$\pm$0.25} \\
& \\
$T_{\rm e}$(OIII)(K)\                & \MC {2}{c}{19435$\pm$1080~~}     & \MC {2}{c}{20085$\pm$1034~~}     & \MC {2}{c}{13512$\pm$1212~~}     \\
$T_{\rm e}$(OII)(K)\                 & \MC {2}{c}{14958$\pm$36  ~~}     & \MC {2}{c}{14911$\pm$116 ~~}     & \MC {2}{c}{13014$\pm$836 ~~}     \\
$N_{\rm e}$(SII)(cm$^{-3}$)\         & \MC {2}{c}{ 10$\pm$10 ~~}         & \MC {2}{c}{ 10$\pm$10 ~~}         & \MC {2}{c}{ 10$\pm$10 ~~}         \\
O$^{+}$/H$^{+}$($\times$10$^5$)\     & \MC {2}{c}{2.003$\pm$0.157~~}     & \MC {2}{c}{1.472$\pm$0.086~~}     & \MC {2}{c}{3.939$\pm$1.048~~}     \\
O$^{++}$/H$^{+}$($\times$10$^5$)\    & \MC {2}{c}{0.569$\pm$0.069~~}     & \MC {2}{c}{0.511$\pm$0.055~~}     & \MC {2}{c}{3.800$\pm$0.965~~}     \\
O/H($\times$10$^5$)\                 & \MC {2}{c}{2.572$\pm$0.172~~}     & \MC {2}{c}{1.983$\pm$0.103~~}     & \MC {2}{c}{7.739$\pm$1.424~~}     \\
12+log(O/H)(mse,c)\                  & \MC {2}{c}{~7.42$\pm$0.10~~}      & \MC {2}{c}{~7.31$\pm$0.09~~}      & \MC {2}{c}{~7.89$\pm$0.10~~}      \\
12+log(O/H)(se,c)\                   & \MC {2}{c}{~7.41$\pm$0.11~~}      & \MC {2}{c}{~7.30$\pm$0.10~~}      & \MC {2}{c}{~7.75$\pm$0.10~~}      \\
12+log(O/H)(s,c)\                    & \MC {2}{c}{~7.30$\pm$0.05~~}      & \MC {2}{c}{~7.21$\pm$0.04~~}      & \MC {2}{c}{---             }      \\
\hline
\end{tabular}
\end{table*}

\begin{table*}
\caption{Measured and corrected line intensities, and derived oxygen and neon abundances}
\label{t:Intens6}
\begin{tabular}{lcccc} \hline
\rule{0pt}{10pt}
& \MC{2}{c}{AGC332939=J2308+3154}   & \MC{2}{c}{AGC334513=J2348+2335}  \\ \hline
\rule{0pt}{10pt}
$\lambda_{0}$(\AA) Ion                  & F($\lambda$)/F(H$\beta$)&I($\lambda$)/I(H$\beta$) & F($\lambda$)/F(H$\beta$)&I($\lambda$)/I(H$\beta$) \\ \hline
3727\ [O\ {\sc ii}]\                           & 0.720$\pm$0.024 & 0.714$\pm$0.026     & 1.784$\pm$0.143 & 1.750$\pm$0.165     \\
4101\ H$\delta$\                               & 0.190$\pm$0.011 & 0.200$\pm$0.015     & ---               & ---               \\
4340\ H$\gamma$\                               & 0.363$\pm$0.015 & 0.369$\pm$0.017     & 0.324$\pm$0.028 & 0.471$\pm$0.105     \\
4363\ [O\ {\sc iii}]\                          & 0.076$\pm$0.010 & 0.076$\pm$0.010     & ---               & ---               \\
4471\ He\ {\sc i}\                             & 0.033$\pm$0.003 & 0.033$\pm$0.003     & ---               & ---               \\
4861\ H$\beta$\                                & 1.000$\pm$0.031 & 1.000$\pm$0.032     & 1.000$\pm$0.038 & 1.000$\pm$0.078     \\
4959\ [O\ {\sc iii}]\                          & 1.414$\pm$0.049 & 1.402$\pm$0.049     & 0.351$\pm$0.029 & 0.306$\pm$0.029     \\
5007\ [O\ {\sc iii}]\                          & 3.727$\pm$0.114 & 3.695$\pm$0.115     & 1.081$\pm$0.040 & 0.939$\pm$0.040     \\
6548\ [N\ {\sc ii}]\                           & 0.017$\pm$0.014 & 0.016$\pm$0.014     & 0.000$\pm$0.054 & 0.000$\pm$0.054     \\
6563\ H$\alpha$\                               & 2.811$\pm$0.086 & 2.791$\pm$0.094     & 3.405$\pm$0.164 & 2.740$\pm$0.168     \\
6584\ [N\ {\sc ii}]\                           & 0.021$\pm$0.021 & 0.021$\pm$0.021     & 0.081$\pm$0.081 & 0.063$\pm$0.072     \\
6717\ [S\ {\sc ii}]\                           & 0.093$\pm$0.018 & 0.093$\pm$0.018     & 0.243$\pm$0.028 & 0.188$\pm$0.025     \\
6731\ [S\ {\sc ii}]\                           & 0.069$\pm$0.018 & 0.068$\pm$0.018     & 0.108$\pm$0.027 & 0.084$\pm$0.024     \\
  & & \\
C(H$\beta$)\ dex          & \MC {2}{c}{0.00$\pm$0.04} & \MC {2}{c}{0.15$\pm$0.06} \\
EW(abs)\ \AA\             & \MC {2}{c}{0.75$\pm$0.62} & \MC {2}{c}{2.55$\pm$1.19} \\
F(H$\beta$)$^a$\          & \MC {2}{c}{32.1$\pm$0.62} & \MC {2}{c}{3.72$\pm$0.06} \\
EW(H$\beta$)\ \AA\        & \MC {2}{c}{  89$\pm$ 2}   & \MC {2}{c}{  19$\pm$ 1}   \\
& \\
$T_{\rm e}$(OIII)(K)\                & \MC {2}{c}{15172$\pm$977 ~~}     & \MC {2}{c}{19757$\pm$1102~~}     \\
$T_{\rm e}$(OII)(K)\                 & \MC {2}{c}{13990$\pm$476 ~~}     & \MC {2}{c}{14941$\pm$80  ~~}     \\
$N_{\rm e}$(SII)(cm$^{-3}$)\         & \MC {2}{c}{ 45$\pm$455~~}         & \MC {2}{c}{ 10$\pm$10 ~~}         \\
O$^{+}$/H$^{+}$($\times$10$^5$)\     & \MC {2}{c}{0.811$\pm$0.104~~}     & \MC {2}{c}{1.606$\pm$0.154~~}     \\
O$^{++}$/H$^{+}$($\times$10$^5$)\    & \MC {2}{c}{4.084$\pm$0.665~~}     & \MC {2}{c}{0.547$\pm$0.067~~}     \\
O/H($\times$10$^5$)\                 & \MC {2}{c}{4.895$\pm$0.673~~}     & \MC {2}{c}{2.154$\pm$0.168~~}     \\
12+log(O/H)(dir)\                    & \MC {2}{c}{~7.69$\pm$0.06~~}      & \MC {2}{c}{---             }      \\
12+log(O/H)(mse,c)\                  & \MC {2}{c}{---             }      & \MC {2}{c}{~7.34$\pm$0.10~~}      \\
12+log(O/H)(se,c)\                   & \MC {2}{c}{---             }      & \MC {2}{c}{~7.32$\pm$0.11~~}      \\
12+log(O/H)(s,c)\                    & \MC {2}{c}{---             }      & \MC {2}{c}{~7.24$\pm$0.05~~}      \\
\hline
\end{tabular}
\end{table*}

\begin{table*}
\caption{Measured and corrected line intensities, and derived oxygen and neon abundances}
\label{t:Intens7}
\begin{tabular}{lcccc} \hline
\rule{0pt}{10pt}
& \MC{2}{c}{AGC124609=J0249+3444}   & \MC{2}{c}{AGC189201=J0823+1754}  \\ \hline
\rule{0pt}{10pt}
$\lambda_{0}$(\AA) Ion                  & F($\lambda$)/F(H$\beta$)&I($\lambda$)/I(H$\beta$) & F($\lambda$)/F(H$\beta$)&I($\lambda$)/I(H$\beta$) \\ \hline
3727\ [O\ {\sc ii}]\                           & 0.733$\pm$0.031 & 0.720$\pm$0.032     & 0.760$\pm$0.046 & 0.880$\pm$0.056     \\
3835\ H9\                                      & 0.046$\pm$0.018 & 0.068$\pm$0.029     & ---               & ---               \\
3967\ [Ne\ {\sc iii}]\ +\ H7\                  & 0.266$\pm$0.008 & 0.284$\pm$0.011     & 0.178$\pm$0.009 & 0.251$\pm$0.015     \\
4101\ H$\delta$\                               & 0.237$\pm$0.007 & 0.256$\pm$0.010     & 0.223$\pm$0.006 & 0.290$\pm$0.011     \\
4340\ H$\gamma$\                               & 0.463$\pm$0.012 & 0.476$\pm$0.014     & 0.404$\pm$0.010 & 0.466$\pm$0.013     \\
4363\ [O\ {\sc iii}]\                          & 0.120$\pm$0.005 & 0.118$\pm$0.005     & 0.051$\pm$0.006 & 0.053$\pm$0.006     \\
4388\ He\ {\sc i}\                             & ---               & ---               & 0.009$\pm$0.003 & 0.009$\pm$0.003     \\
4438\ He\ {\sc i}\                             & ---               & ---               & 0.009$\pm$0.003 & 0.009$\pm$0.003     \\
4471\ He\ {\sc i}\                             & ---               & ---               & 0.023$\pm$0.006 & 0.023$\pm$0.006     \\
4686\ He\ {\sc ii}\                            & 0.040$\pm$0.006 & 0.039$\pm$0.006     & ---               & ---               \\
4712\ [Ar\ {\sc iv]}\ +\ He\ {\sc i}\          & 0.021$\pm$0.004 & 0.021$\pm$0.004     & ---               & ---               \\
4861\ H$\beta$\                                & 1.000$\pm$0.023 & 1.000$\pm$0.024     & 1.000$\pm$0.020 & 1.000$\pm$0.021     \\
4959\ [O\ {\sc iii}]\                          & 2.034$\pm$0.046 & 1.998$\pm$0.046     & 0.630$\pm$0.012 & 0.603$\pm$0.012     \\
5007\ [O\ {\sc iii}]\                          & 6.056$\pm$0.124 & 5.951$\pm$0.124     & 1.901$\pm$0.032 & 1.808$\pm$0.031     \\
5869\ He\ {\sc ii}\                            & ---               & ---               & 0.127$\pm$0.006 & 0.109$\pm$0.005     \\
6548\ [N\ {\sc ii}]\                           & ---               & ---               & 0.006$\pm$0.011 & 0.004$\pm$0.009     \\
6563\ H$\alpha$\                               & ---               & ---               & 3.424$\pm$0.055 & 2.756$\pm$0.049     \\
6584\ [N\ {\sc ii}]\                           & ---               & ---               & 0.025$\pm$0.014 & 0.020$\pm$0.012     \\
6678\ He\ {\sc i}\                             & ---               & ---               & 0.037$\pm$0.006 & 0.029$\pm$0.005     \\
6717\ [S\ {\sc ii}]\                           & ---               & ---               & 0.096$\pm$0.003 & 0.076$\pm$0.003     \\
6731\ [S\ {\sc ii}]\                           & ---               & ---               & 0.079$\pm$0.006 & 0.062$\pm$0.005     \\
  & & \\
C(H$\beta$)\ dex          & \MC {2}{c}{0.00$\pm$0.03} & \MC {2}{c}{0.25$\pm$0.02} \\
EW(abs)\ \AA\             & \MC {2}{c}{1.50$\pm$0.42} & \MC {2}{c}{3.15$\pm$0.35} \\
F(H$\beta$)$^a$\          & \MC {2}{c}{14.77$\pm$0.19} & \MC {2}{c}{3.53$\pm$0.04} \\
EW(H$\beta$)\ \AA\        & \MC {2}{c}{  89$\pm$ 1.5}   & \MC {2}{c}{ 101$\pm$ 1}     \\
& \\
$T_{\rm e}$(OIII)(K)\                & \MC {2}{c}{15146$\pm$341 ~~}    & \MC {2}{c}{18464$\pm$1209~~}     \\
$T_{\rm e}$(OII)(K)\                 & \MC {2}{c}{13978$\pm$167 ~~}    & \MC {2}{c}{14933$\pm$103 ~~}     \\
$N_{\rm e}$(SII)(cm$^{-3}$)\         & \MC {2}{c}{ 10$\pm$10 ~~}        & \MC {2}{c}{235$\pm$161~~}         \\
O$^{+}$/H$^{+}$($\times$10$^5$)\     & \MC {2}{c}{0.820$\pm$0.048~~}    & \MC {2}{c}{0.829$\pm$0.057~~}     \\
O$^{++}$/H$^{+}$($\times$10$^5$)\    & \MC {2}{c}{6.415$\pm$0.378~~}    & \MC {2}{c}{1.223$\pm$0.175~~}     \\
O$^{+++}$/H$^{+}$($\times$10$^5$)\   & \MC {2}{c}{0.542$\pm$0.154~~}    & \MC {2}{c}{ ---             }     \\
O/H($\times$10$^5$)\                 & \MC {2}{c}{7.778$\pm$0.374~~}    & \MC {2}{c}{2.052$\pm$0.184~~}     \\
12+log(O/H)(dir)\                    & \MC {2}{c}{~7.89$\pm$0.02~~}     & \MC {2}{c}{~7.31$\pm$0.04~~}      \\
12+log(O/H)(se,c)\                   & \MC {2}{c}{~8.00$\pm$0.10~~}     & \MC {2}{c}{~7.29$\pm$0.09~~}     \\
12+log(O/H)(mse,c)\                  & \MC {2}{c}{~7.86$\pm$0.10~~}     & \MC {2}{c}{~7.33$\pm$0.09~~}     \\
12+log(O/H)(s,c)\                    & \MC {2}{c}{ ---            }     & \MC {2}{c}{~7.27$\pm$0.04~~}     \\
 & \\
Ne$^{++}$/H$^{+}$($\times$10$^5$)\   & \MC {2}{c}{1.032$\pm$0.074~~}    & \MC {2}{c}{0.235$\pm$0.041~~}      \\
ICF(Ne)\                             & \MC {2}{c}{1.073}                & \MC {2}{c}{1.176}                  \\
Ne/H($\times$10$^5$)\                & \MC {2}{c}{1.11$\pm$0.08~~}      & \MC {2}{c}{0.28$\pm$0.05~~}        \\
12+log(Ne/H)\                        & \MC {2}{c}{7.04$\pm$0.03~~}      & \MC {2}{c}{6.44$\pm$0.08~~}        \\
log(Ne/O)\                           & \MC {2}{c}{ -0.85$\pm$0.04~~}    & \MC {2}{c}{ -0.87$\pm$0.09~~}      \\
 & \\
\hline
\end{tabular}
\end{table*}

\clearpage

\section{On-line data. Finding charts with slit positions and plots
of 1D spectra}
\label{sec:online.data}

In Figures~S1--S2 
we provide  the slit
positions for each of our program galaxy overlaid on the galaxy images taken
either from  DECaLS \citep{DECaLS}, or PanSTARRS PS1 \citep{PS1-database},
or SDSS \citep{DR7, DR14} databases.
In Figure~S3 
we show the slit positions for three
observed galaxies which are not in the XMP void candidate list.

The plots of obtained 1D spectra are presented in Figures~S4--S5 
of this Appendix. Here on $X$ axis the wavelengths are
shown in \AA. In $Y$ axis the measured flux densities
are marked in units of 10$^{-16}$~erg~s$^{-1}$~cm$^{-2}$~\AA$^{-1}$. The names
of objects are inserted along with their adopted value of 12+log(O/H), when
available.

\begin{figure*}
\includegraphics[width=5.5cm,angle=0,clip=]{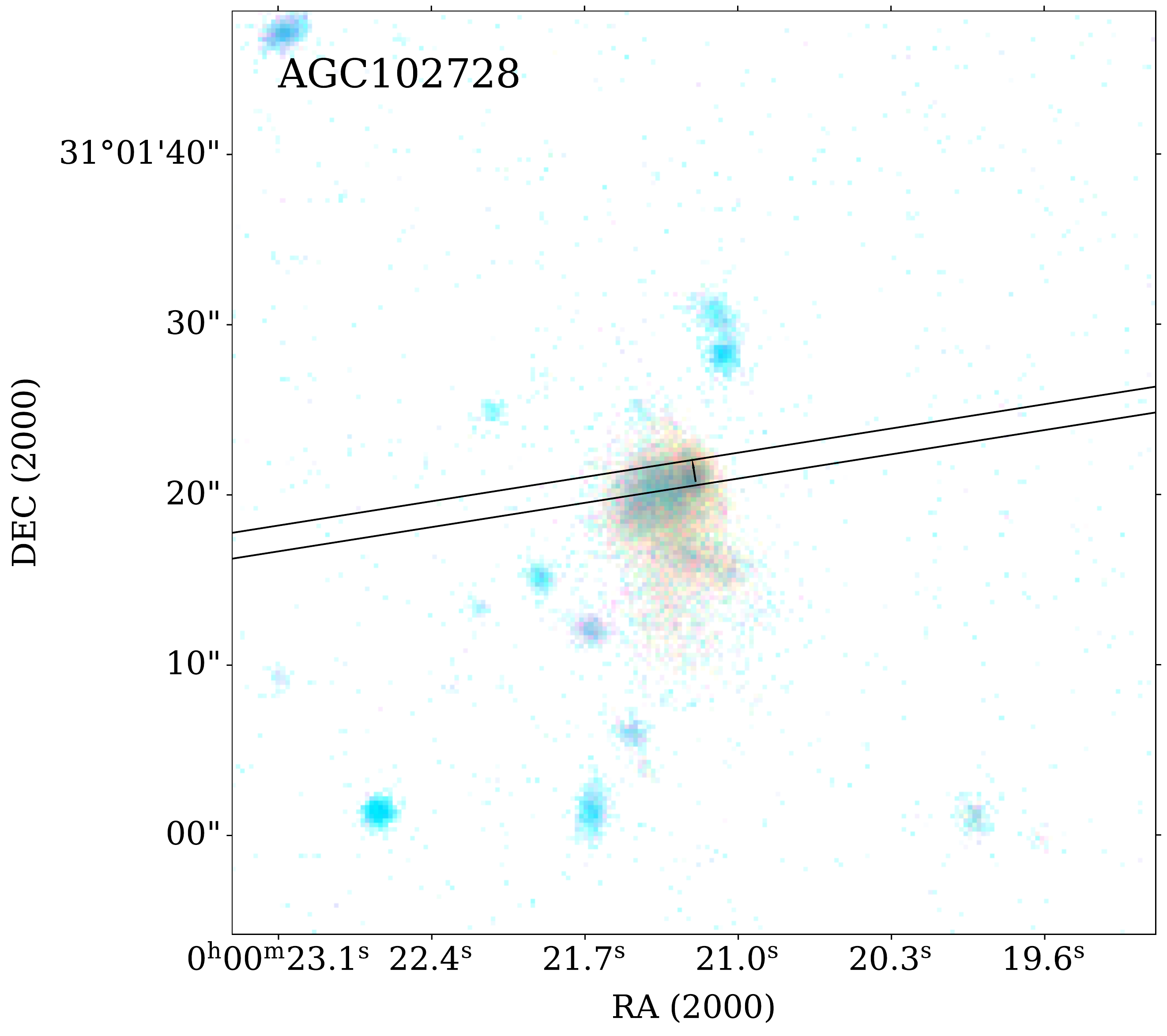}
\includegraphics[width=5.5cm,angle=0,clip=]{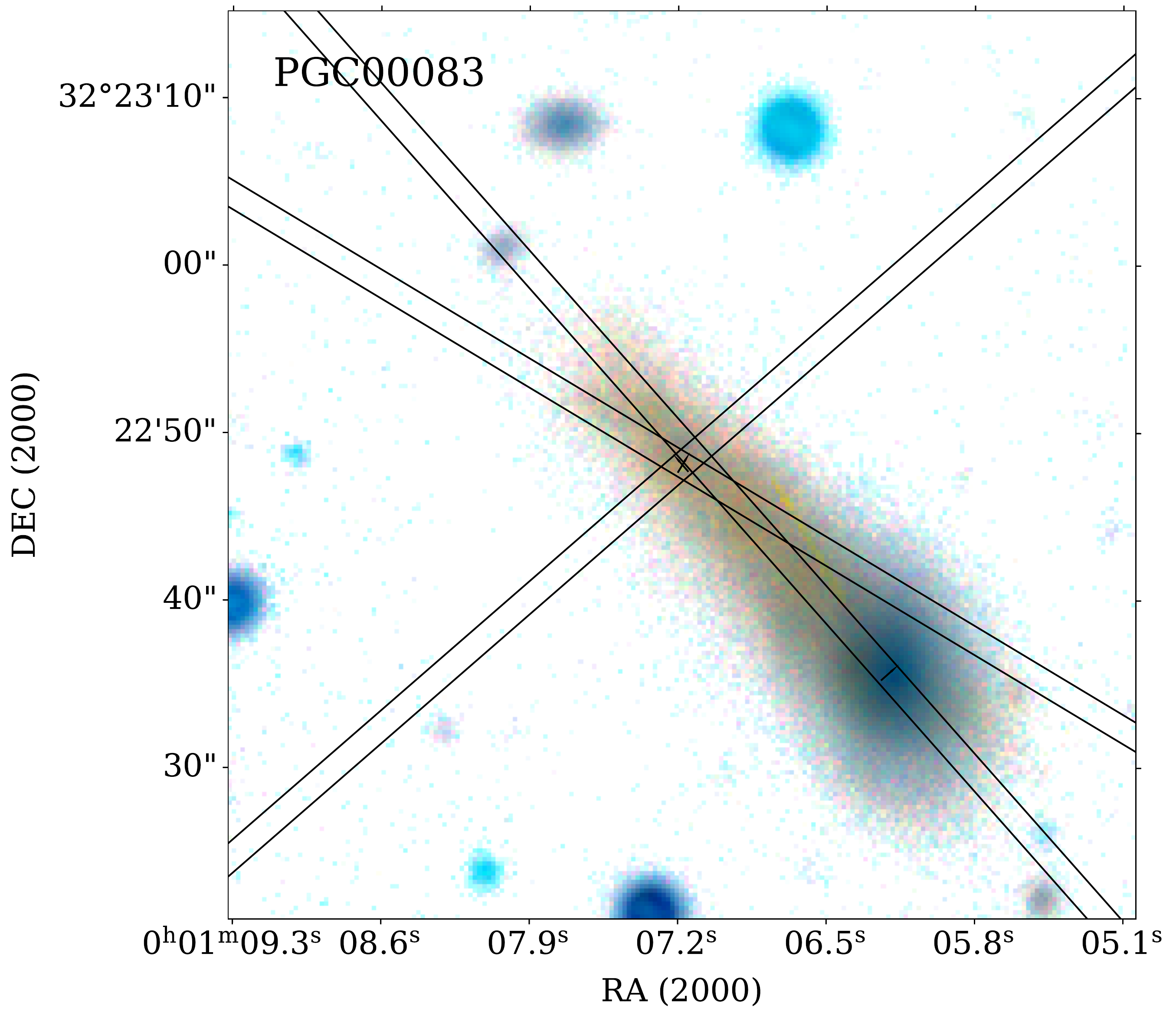}
\includegraphics[width=5.5cm,angle=0,clip=]{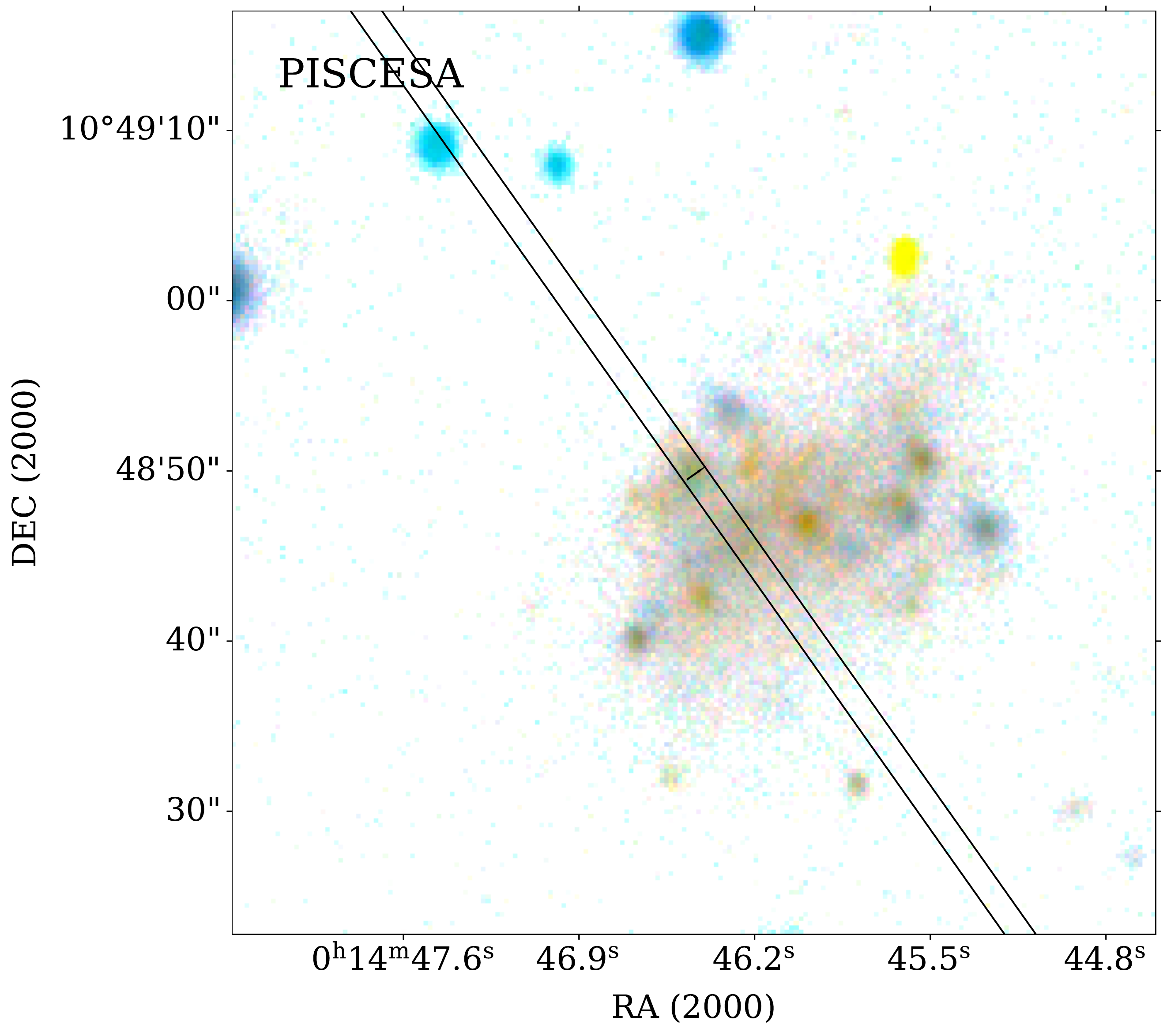}
\includegraphics[width=5.5cm,angle=0,clip=]{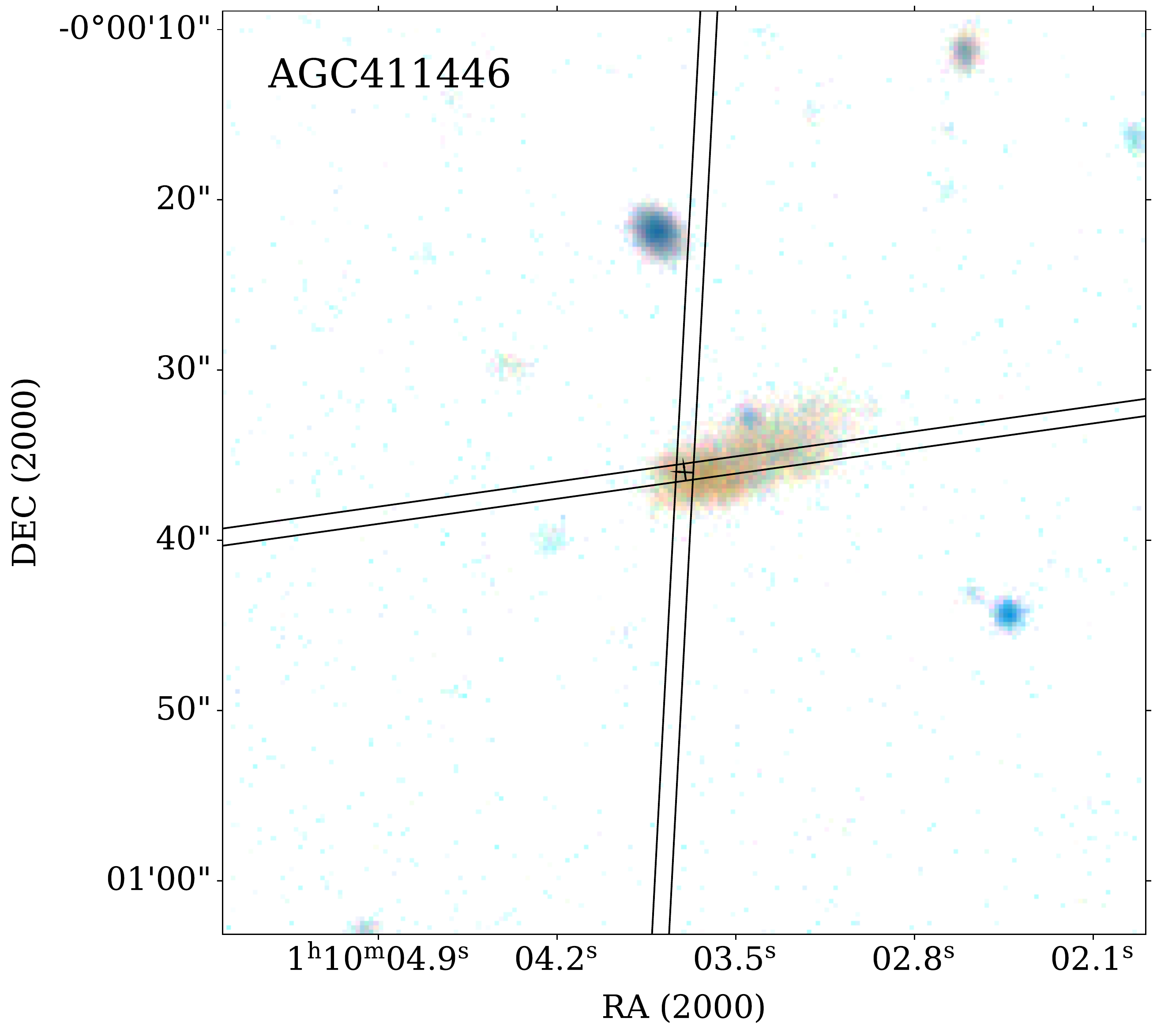}
\includegraphics[width=5.5cm,angle=0,clip=]{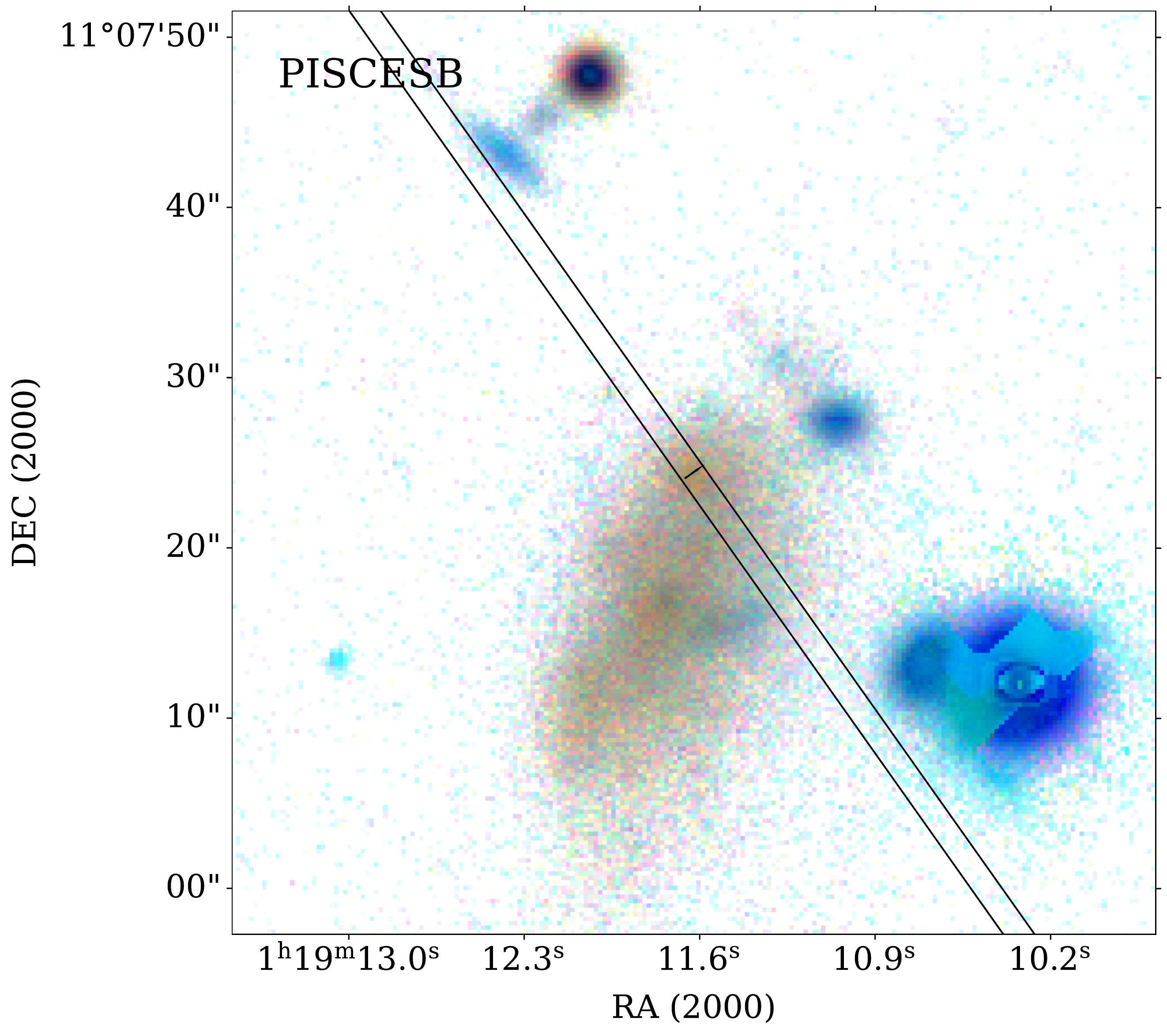}
\includegraphics[width=5.5cm,angle=0,clip=]{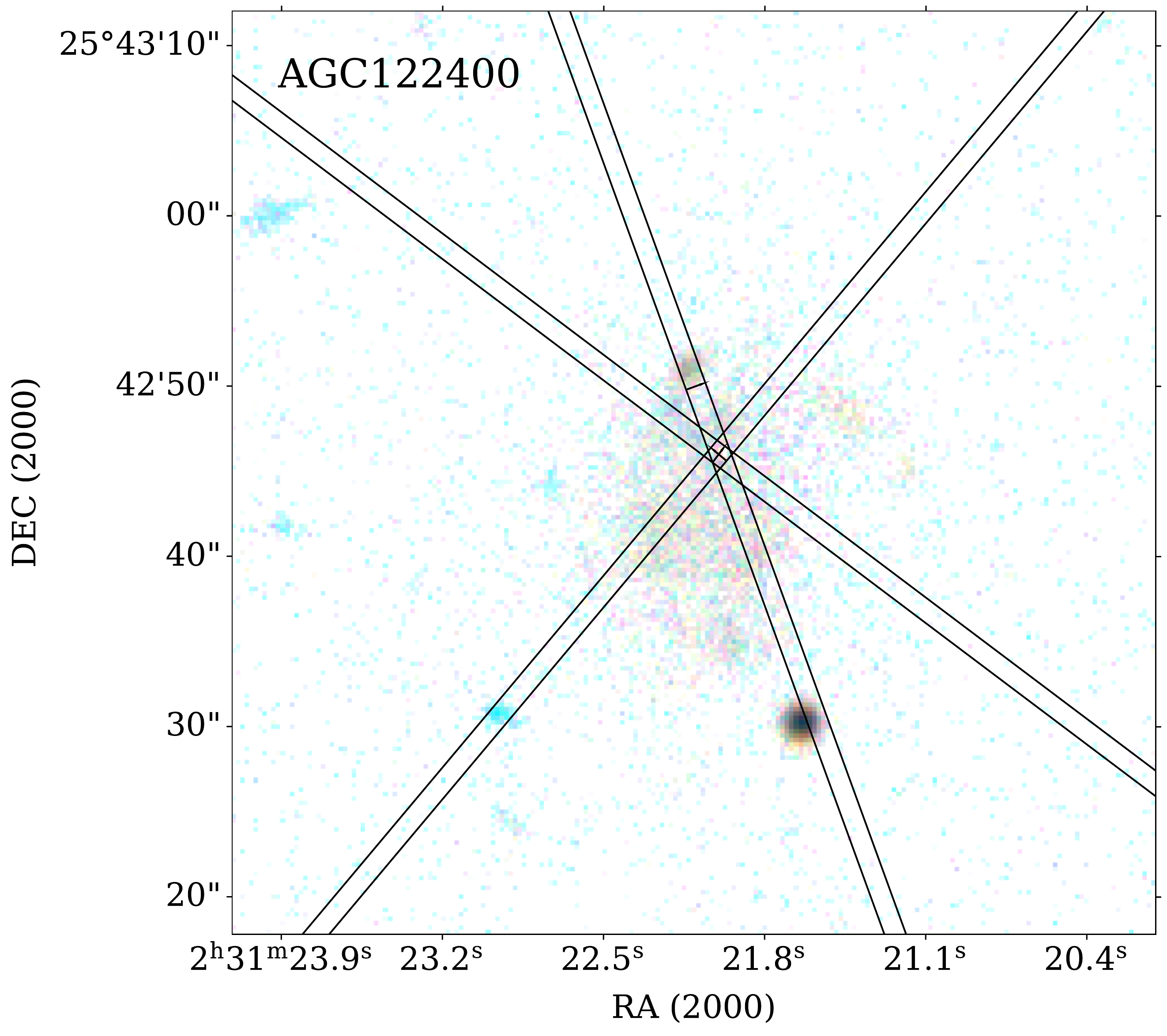}
\includegraphics[width=5.5cm,angle=0,clip=]{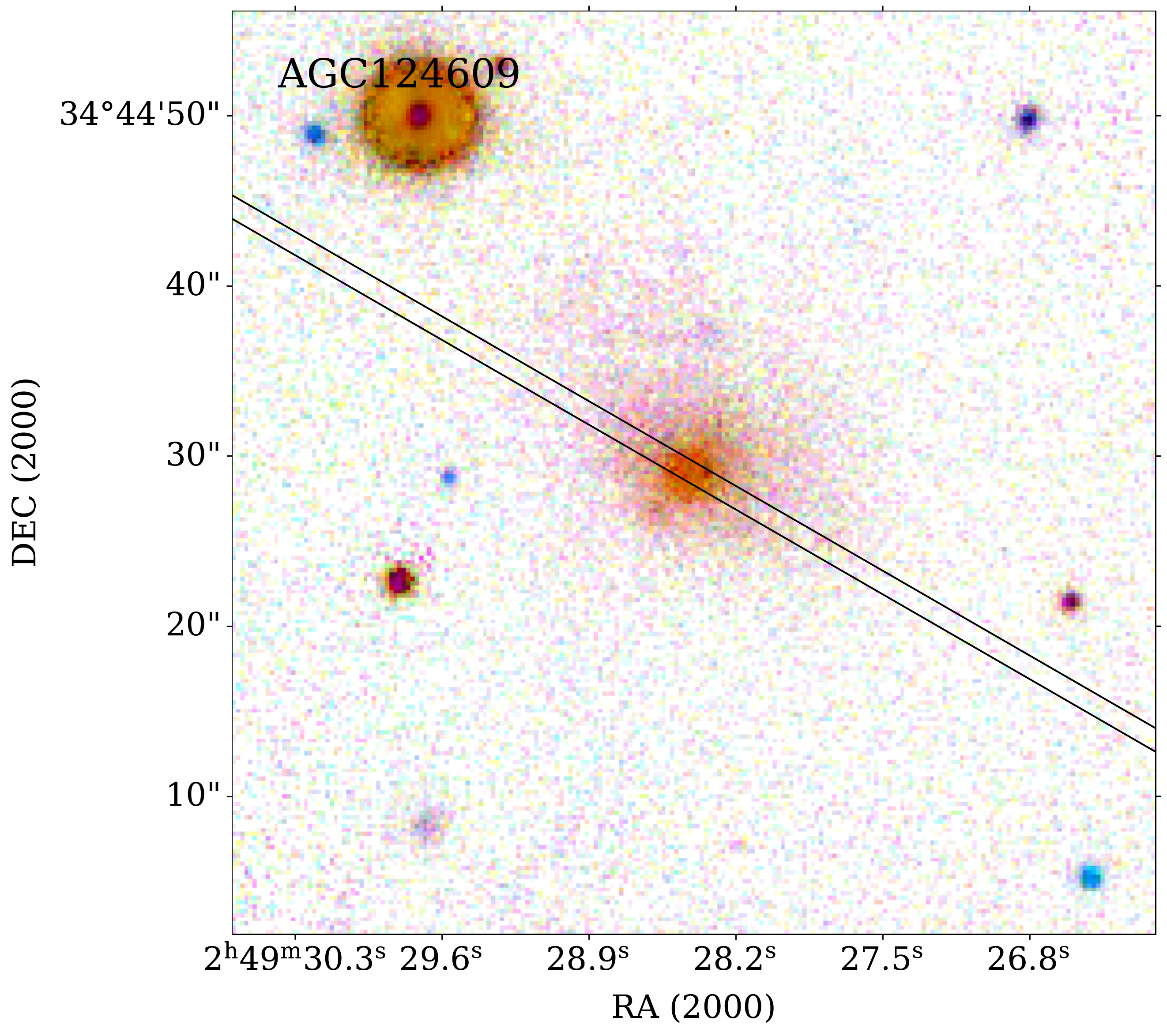}
\includegraphics[width=5.5cm,angle=0,clip=]{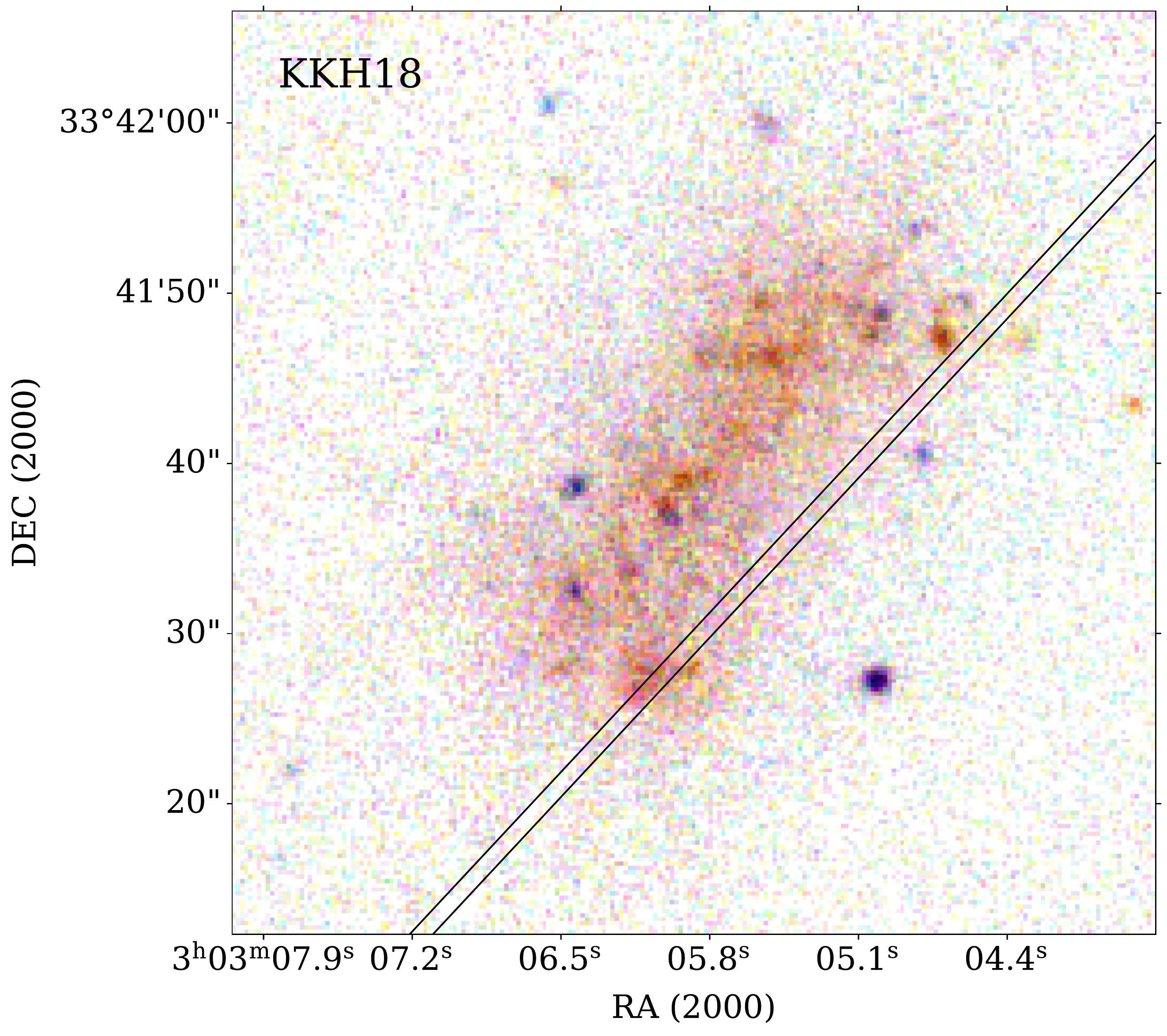}
\includegraphics[width=5.5cm,angle=0,clip=]{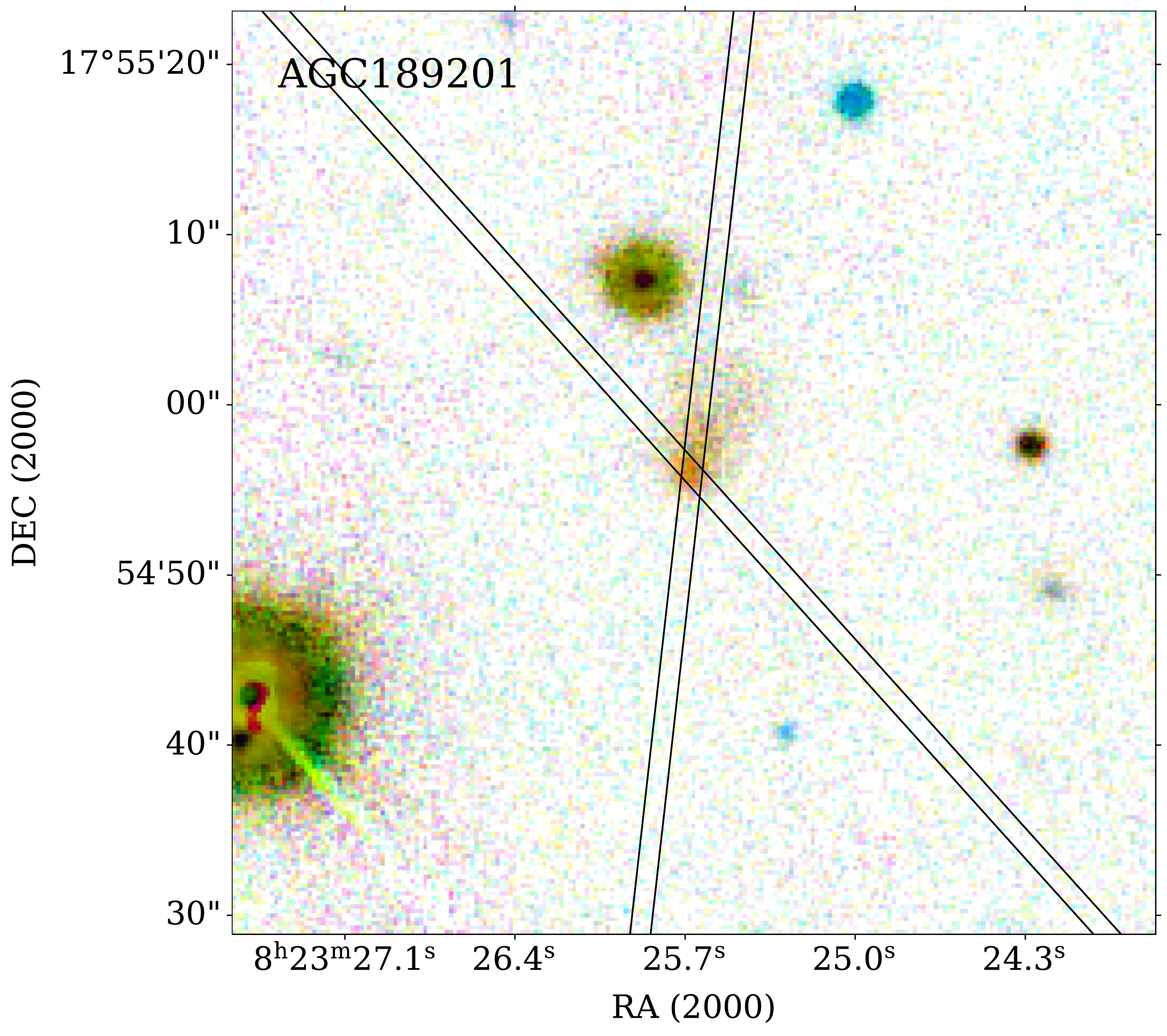}
\includegraphics[width=5.5cm,angle=0,clip=]{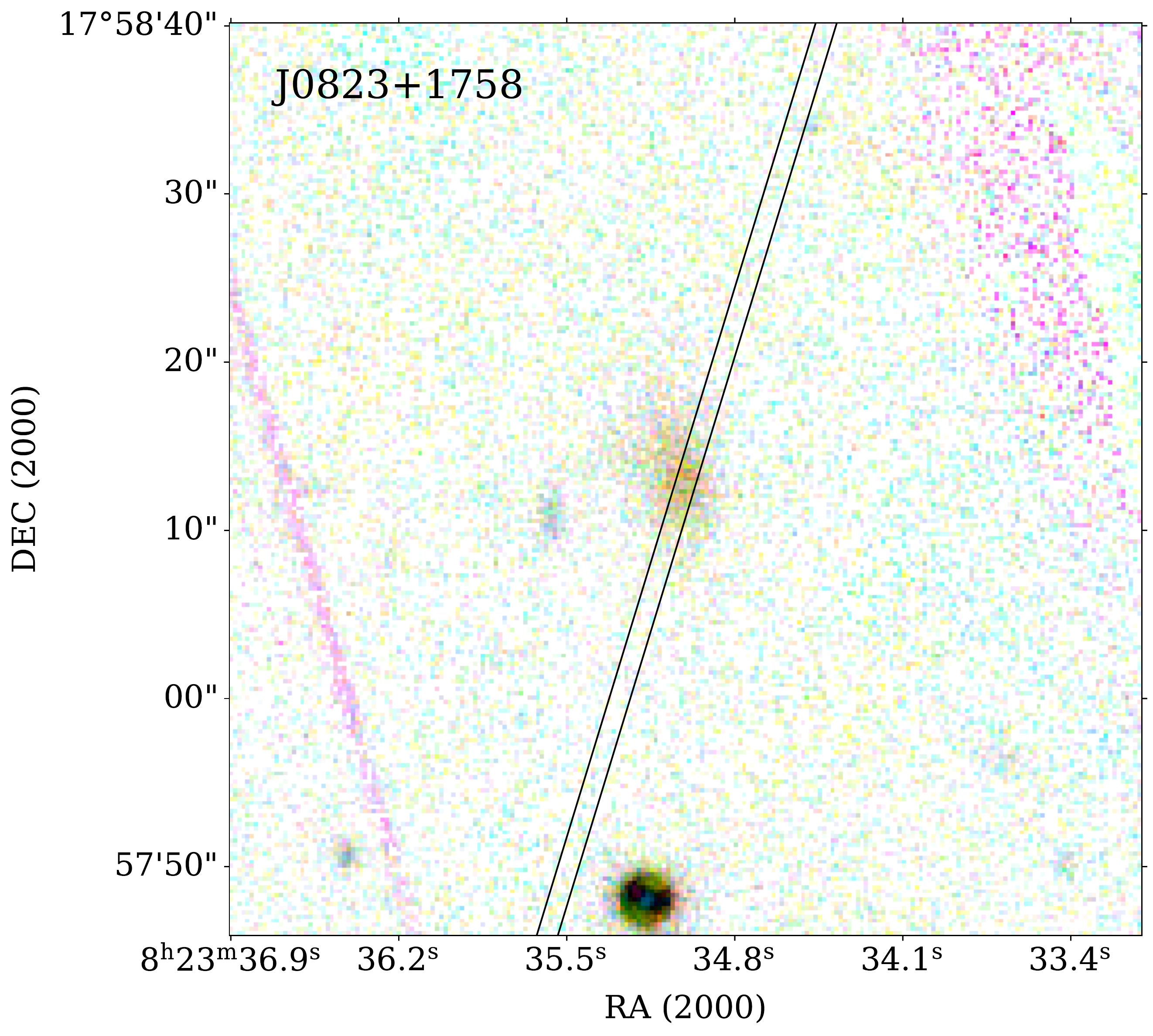}
\includegraphics[width=5.5cm,angle=0,clip=]{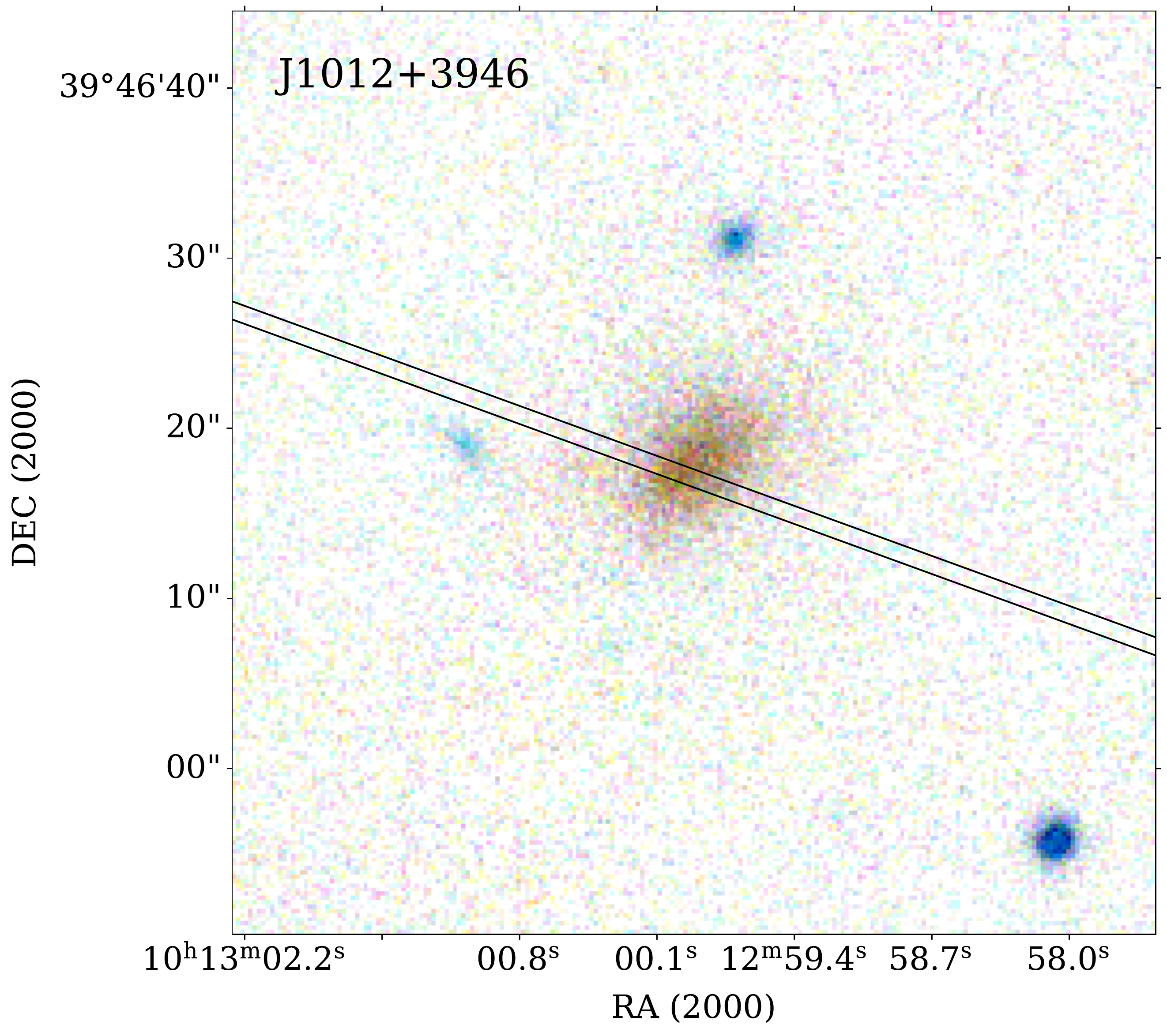}
\includegraphics[width=5.5cm,angle=0,clip=]{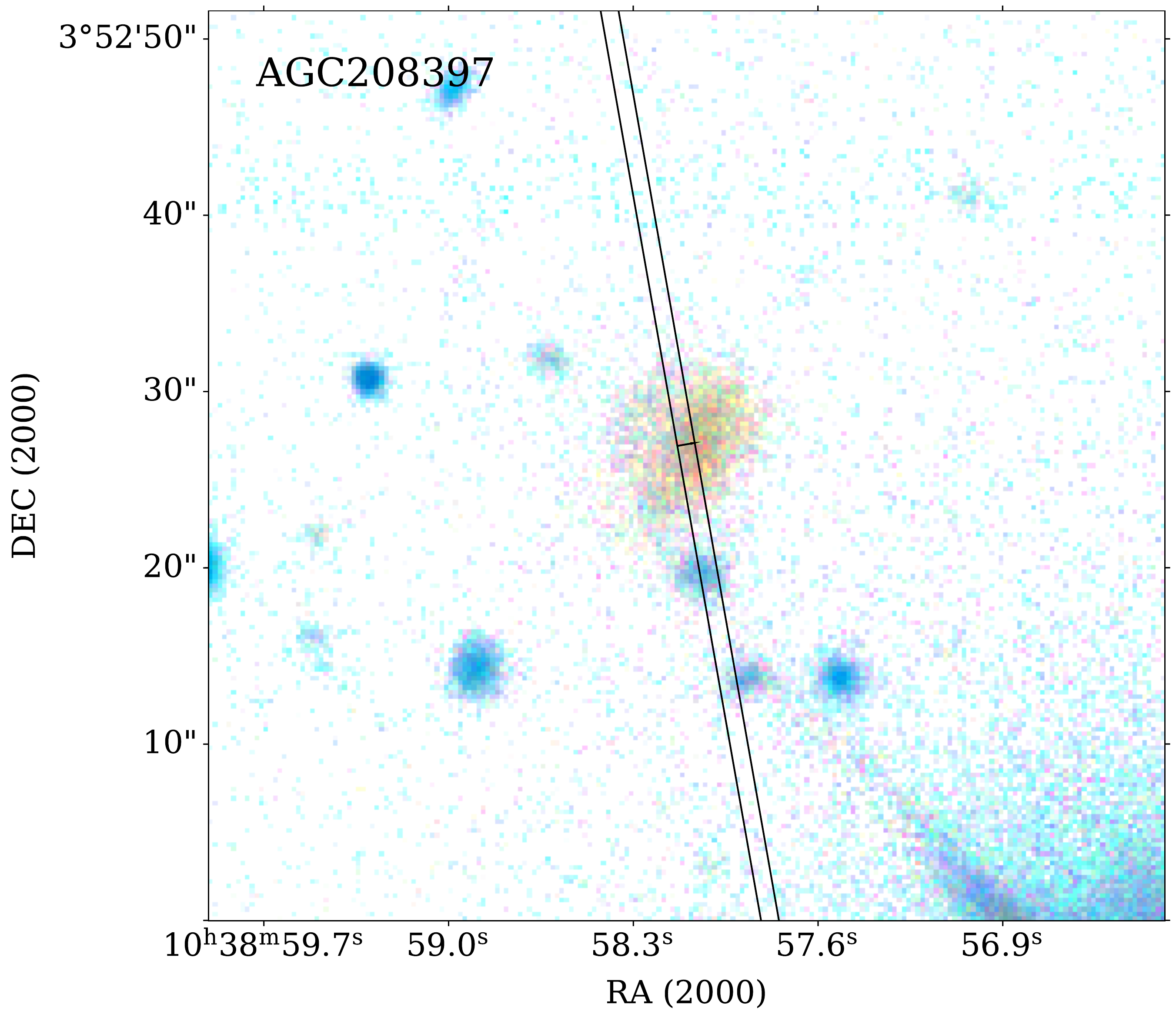}
\caption{Finding charts of void XMP candidates from BTA program with
slit positions superimposed. Inverted colours are used to underline the LSB
features of dwarf morphology. North is up, East is to the left.
{\bf Top row, from left to right:} AGC102728 = J0000+3101,
PGC00083 = J0001+3222; PISCESA = J0014+1048.
{\bf Second row, from left to right:}
AGC411446 = J0110--0000; PISCESB = J0119+1107; AGC122400 = J0231+2542;
{\bf Third row, from left to right:} AGC124609=J0249+3444; KKH18=J0303+3341;
AGC189201=J0823+1754,
{\bf Bottom row, from left to right:} J0823+1758;  J1012++3946; AGC208397 =
J1038+0352
}
\label{fig:slits1}
\end{figure*}

\begin{figure*}
\includegraphics[width=5.5cm,angle=0,clip=]{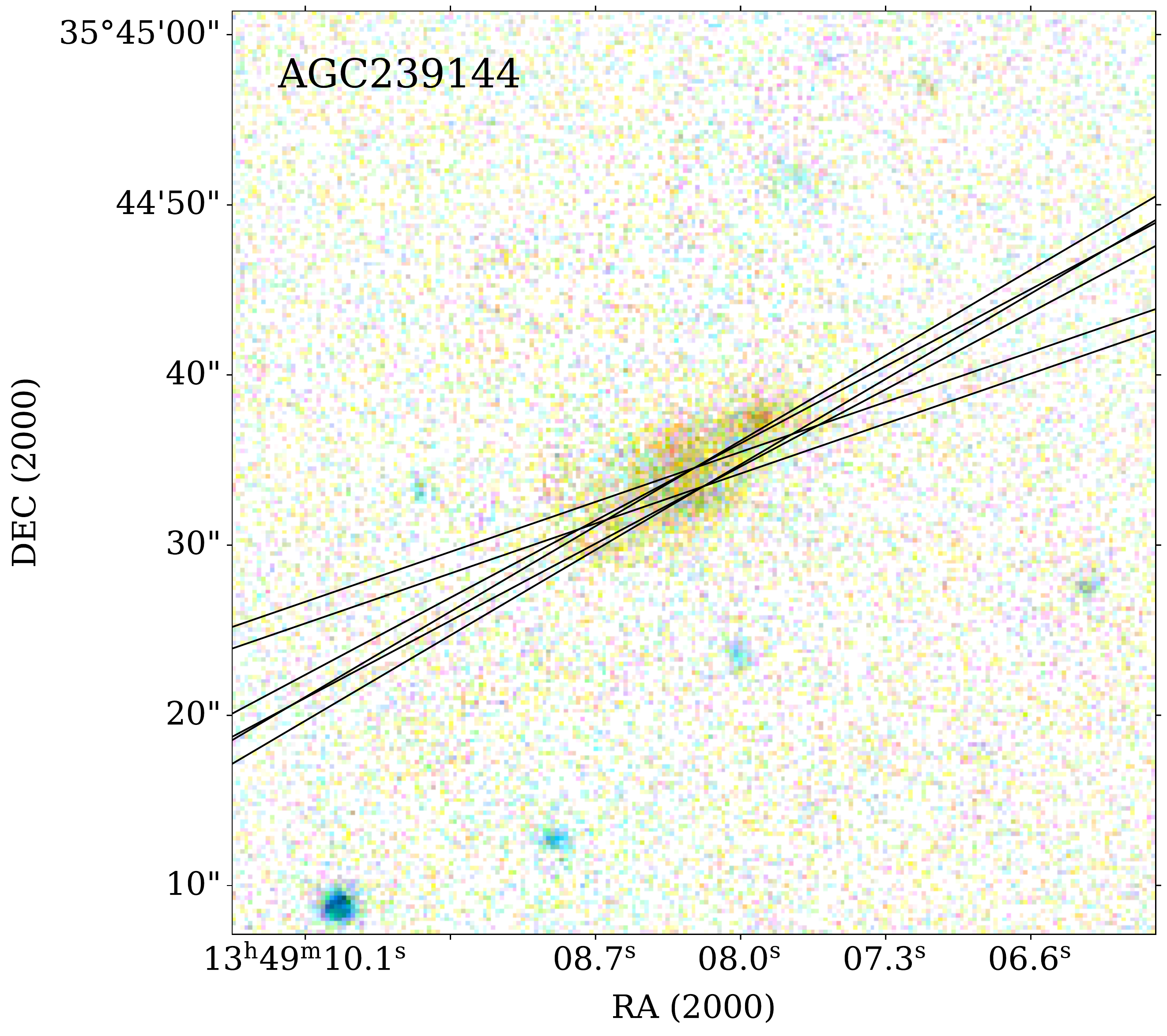}
\includegraphics[width=5.5cm,angle=0,clip=]{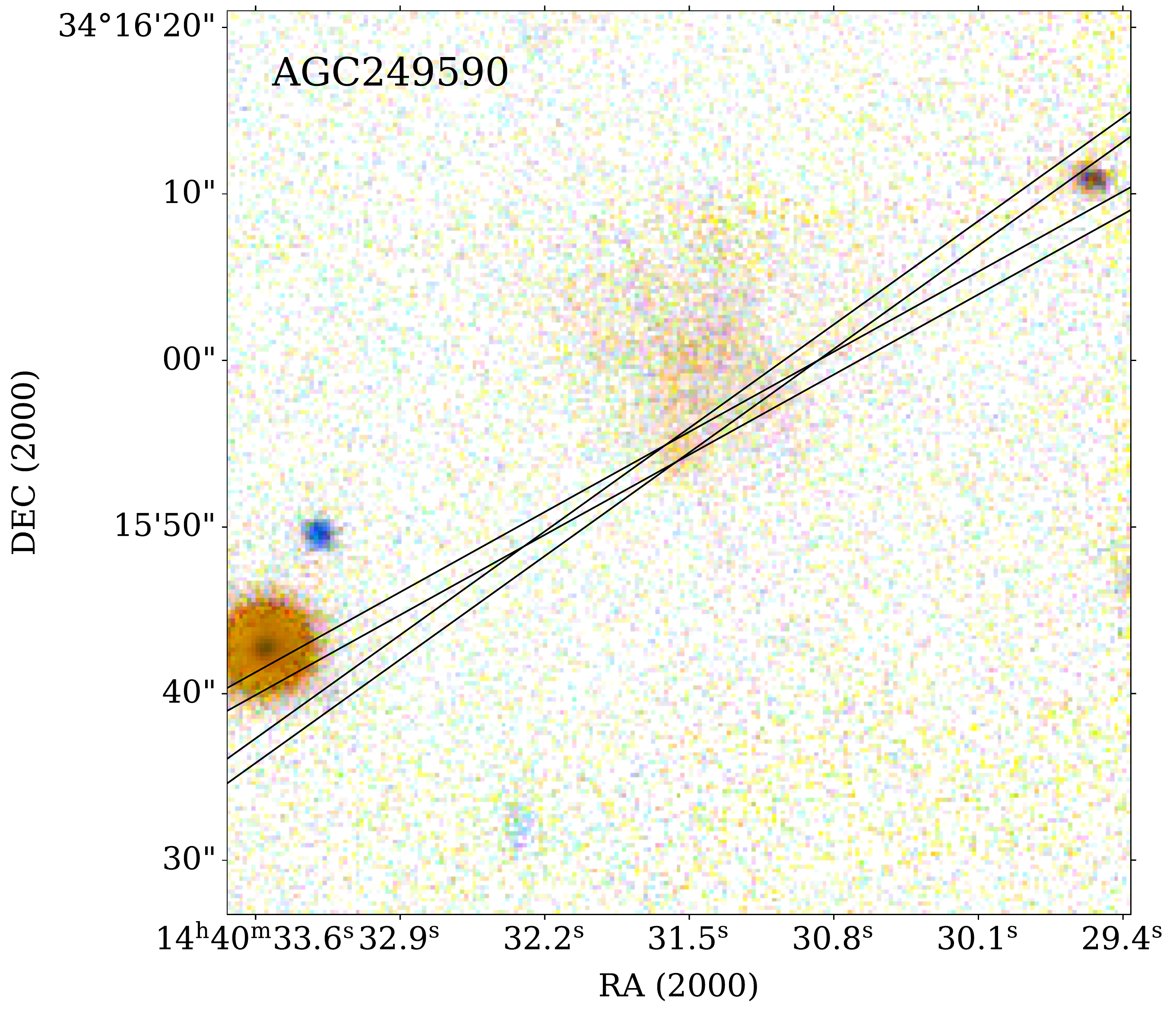}
\includegraphics[width=5.5cm,angle=0,clip=]{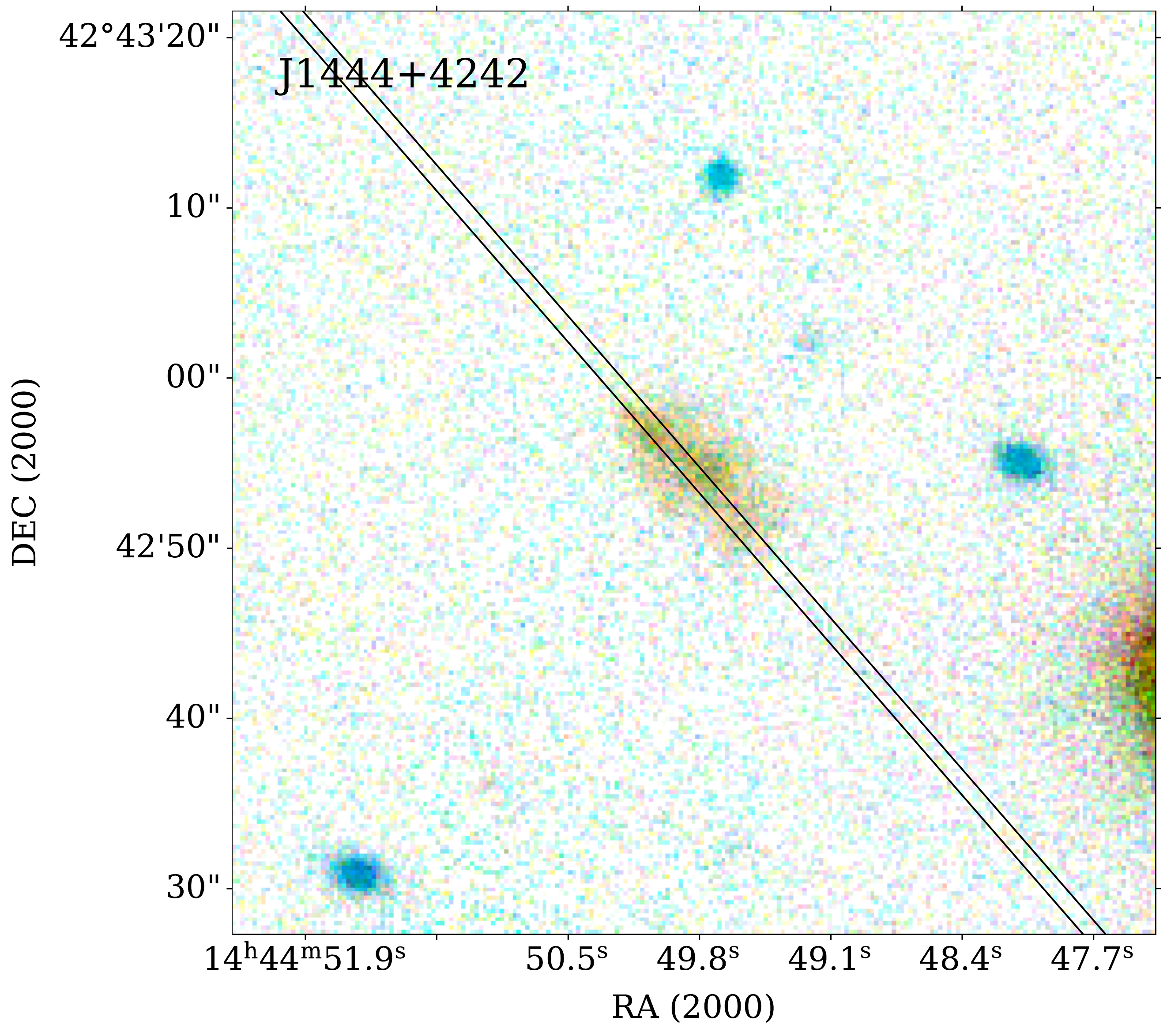}
\includegraphics[width=5.5cm,angle=0,clip=]{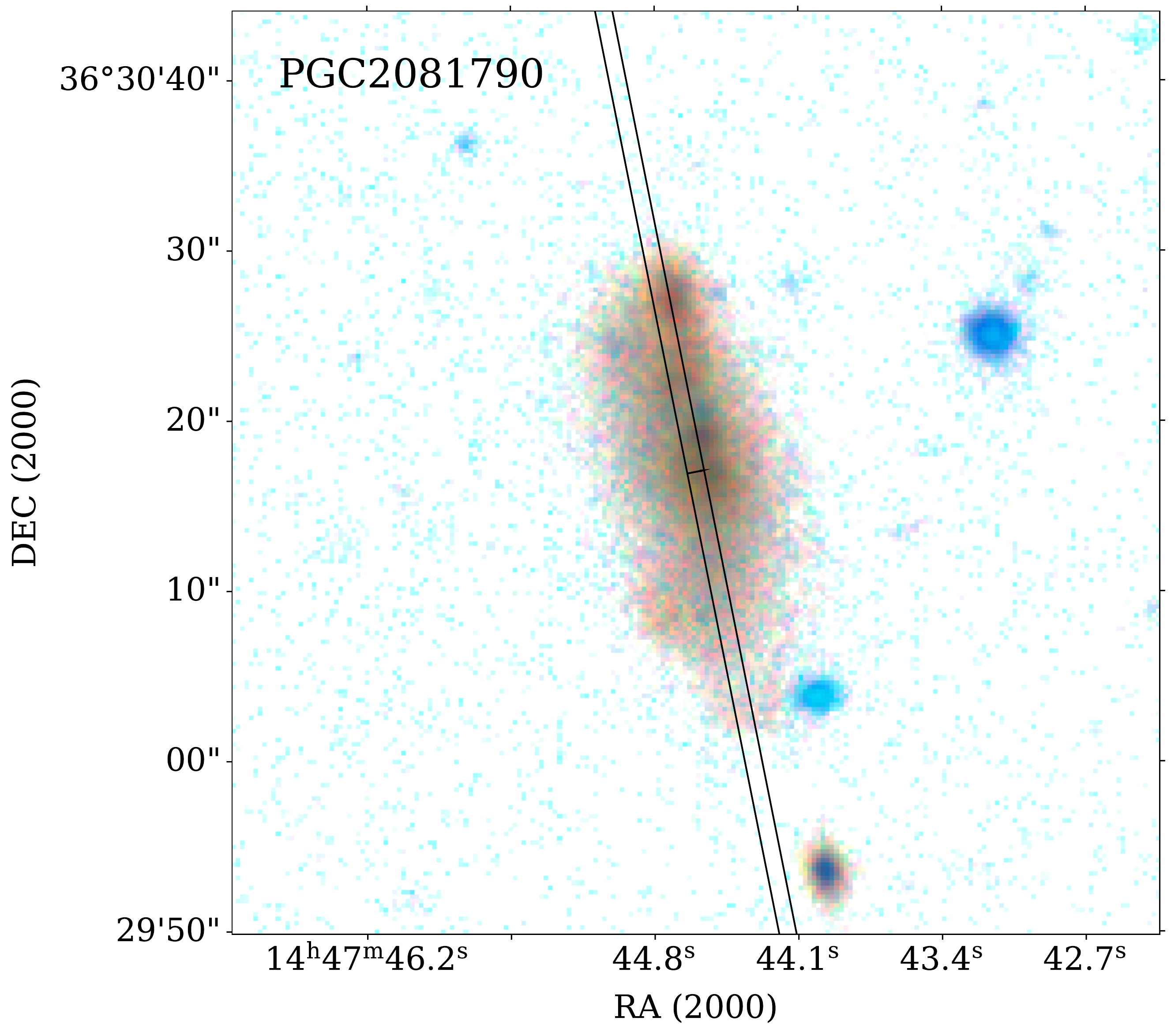}
\includegraphics[width=5.5cm,angle=0,clip=]{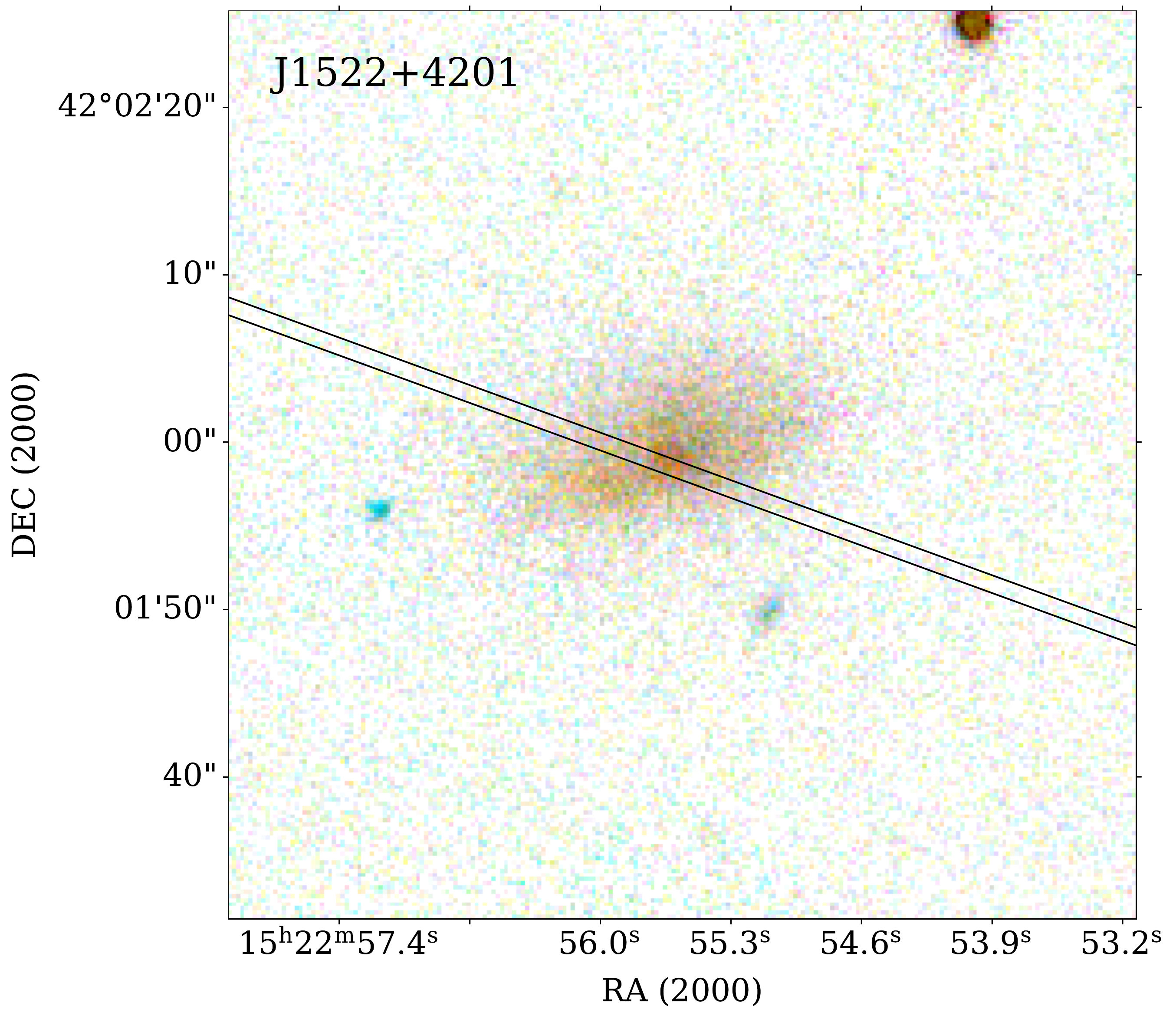}
\includegraphics[width=5.5cm,angle=0,clip=]{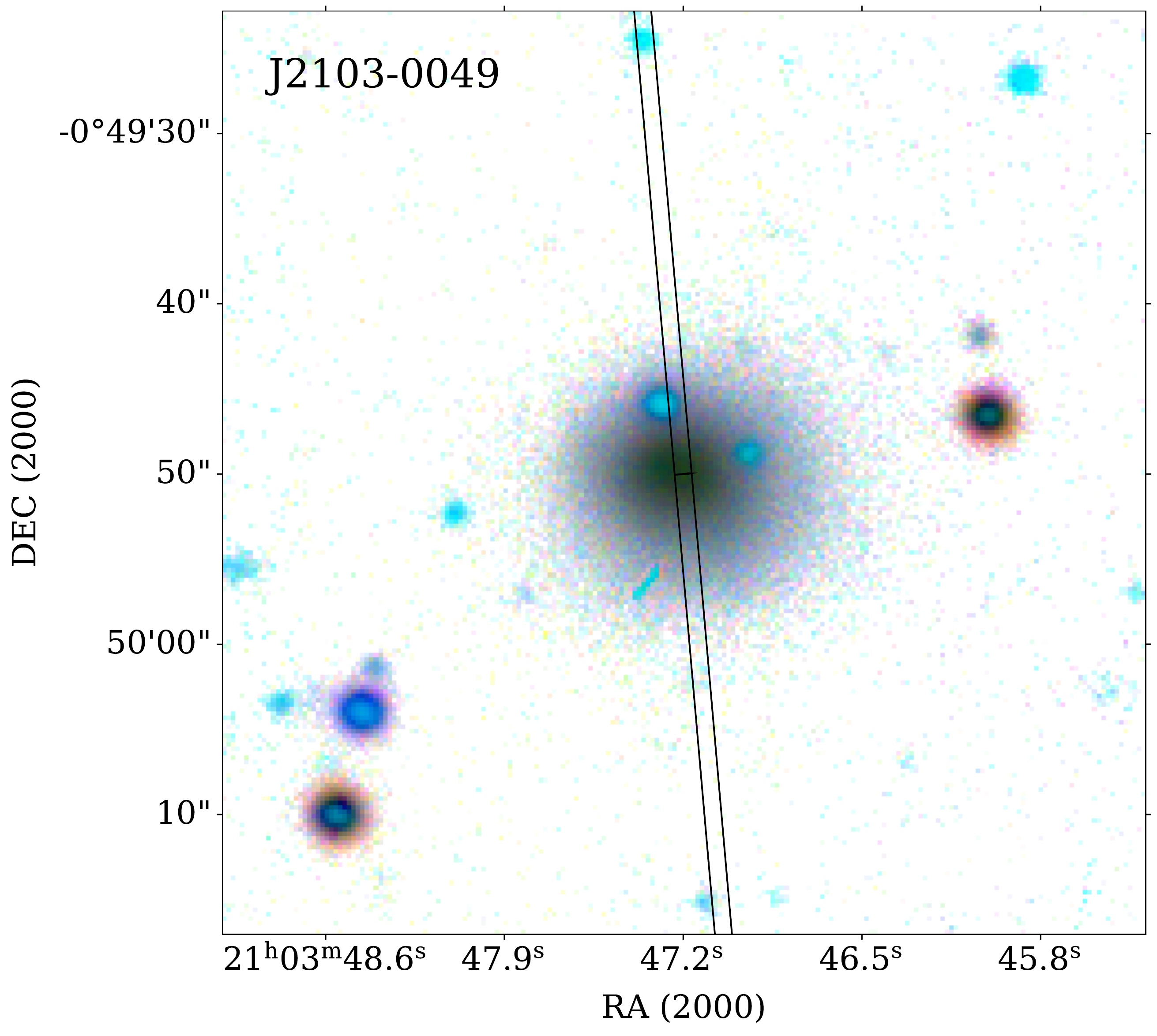}
\includegraphics[width=5.5cm,angle=0,clip=]{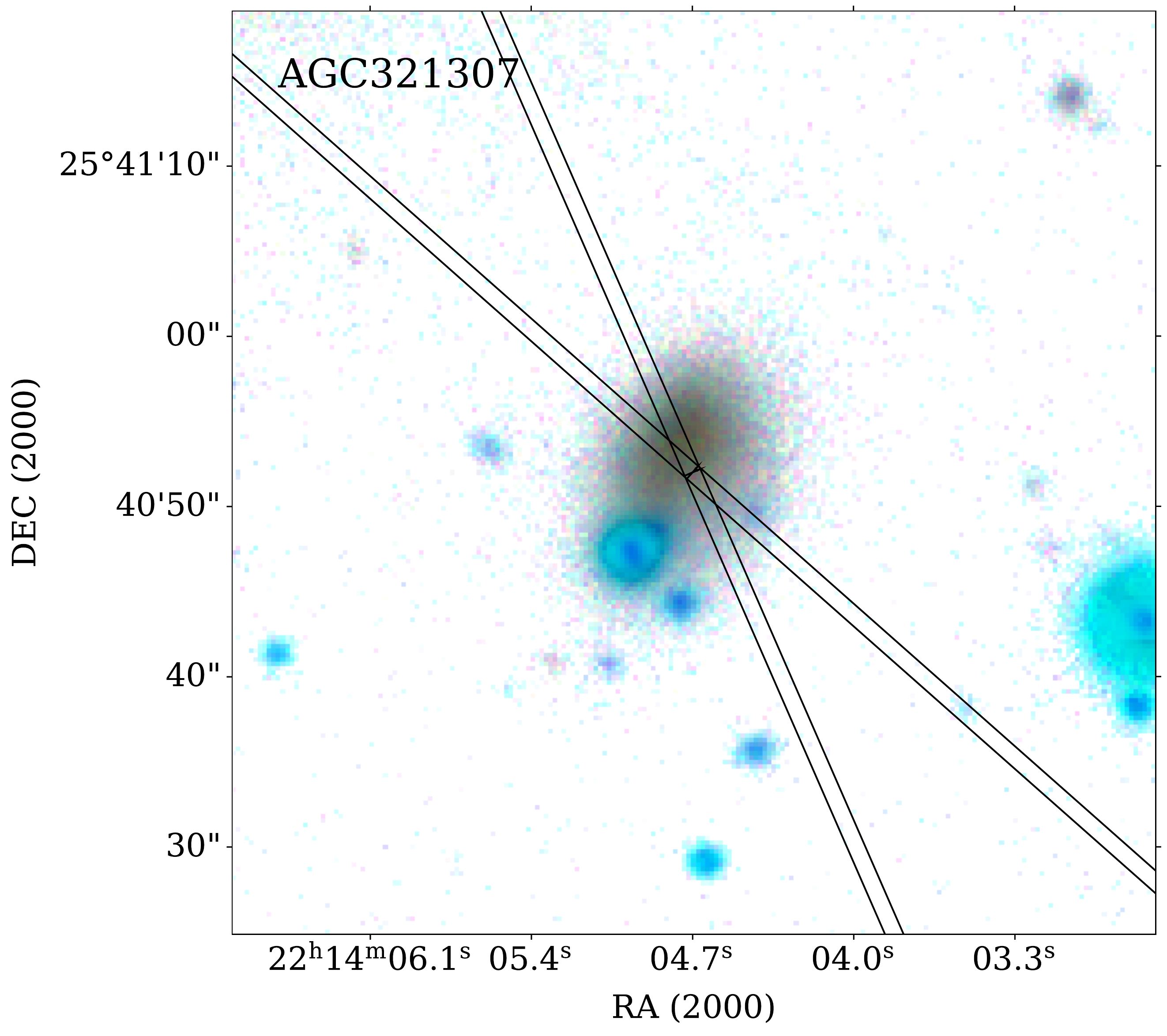}
\includegraphics[width=5.5cm,angle=0,clip=]{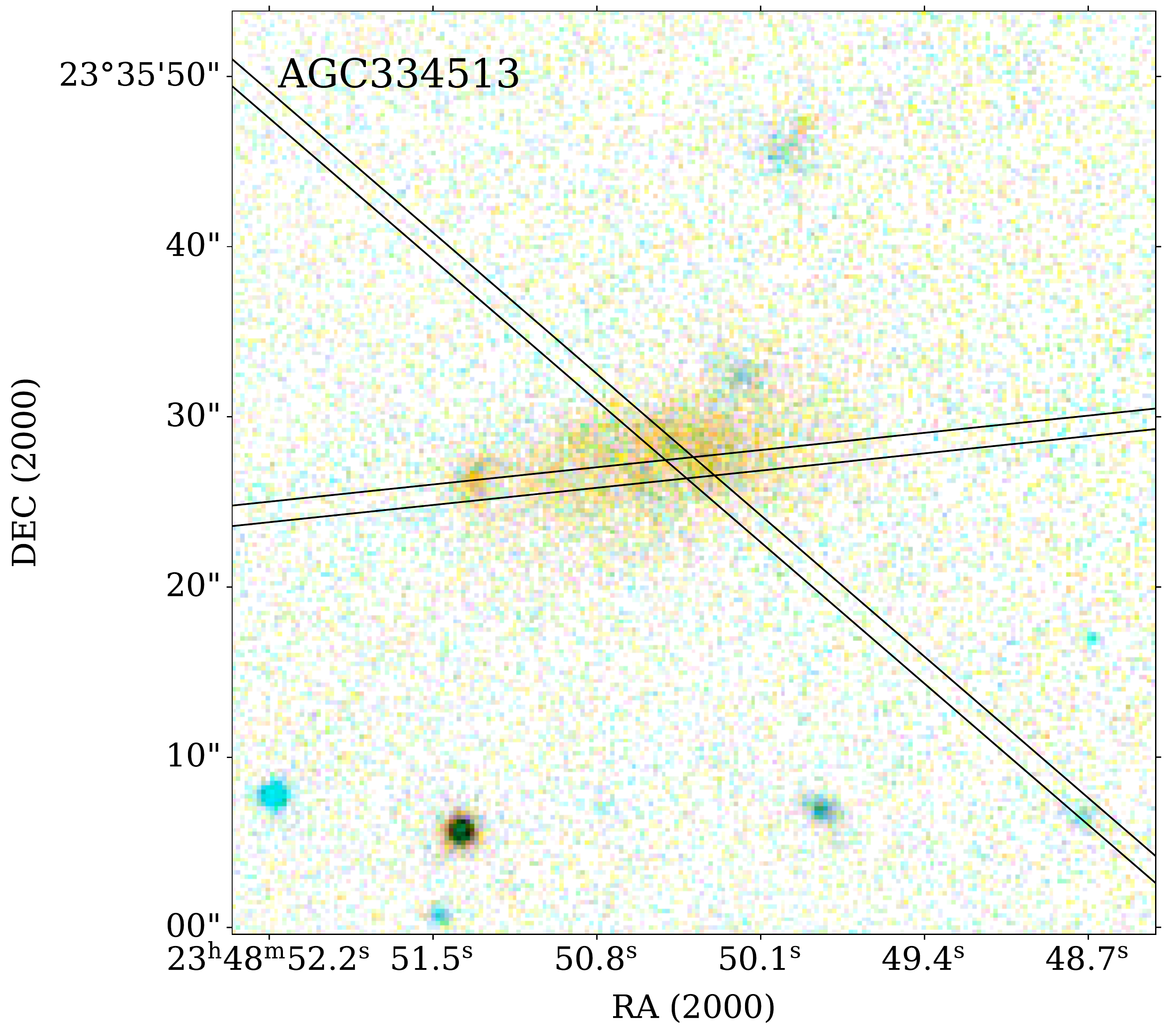}
\caption{Finding charts of void XMP candidates from BTA program with
slit positions superimposed. Inverted colours are used to underline the LSB
features of dwarf morphology. North is up, East is to the left.
{\bf Top row, from left to right:} AGC239144=J1349+3544; AGC249590=J1440+3416;
J1444+4242; {\bf Second row, from left to right:} PGC2081790 = J1447+3630,
 J1522+4201; J2103--0049; {\bf Bottom row, from left to right:}
AGC321307=J2214+2540; AGC334513=J2348+2335.
}
\label{fig:slits2}
\end{figure*}

\begin{figure*}
\includegraphics[width=5.5cm,angle=0,clip=]{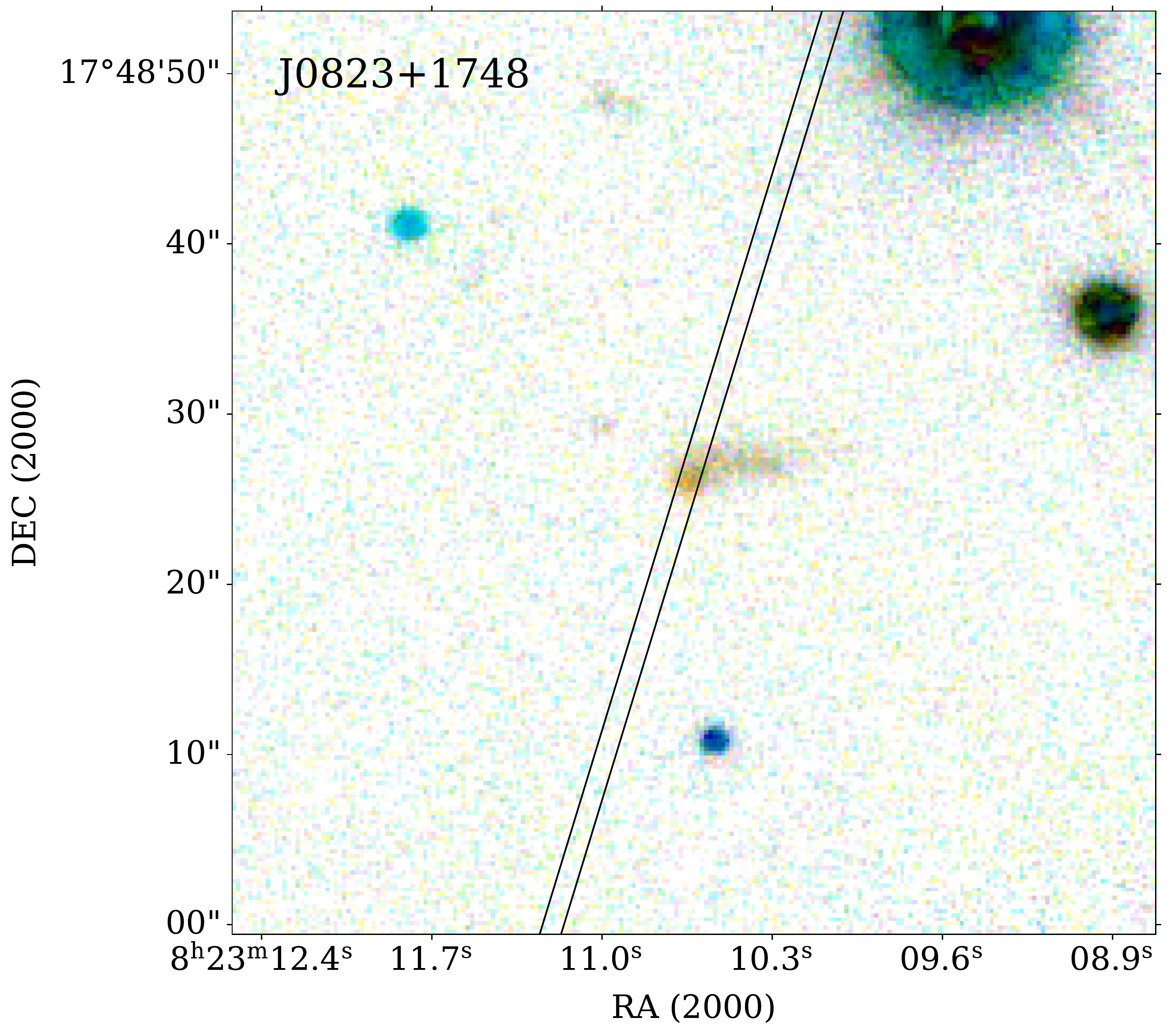}
\includegraphics[width=5.5cm,angle=0,clip=]{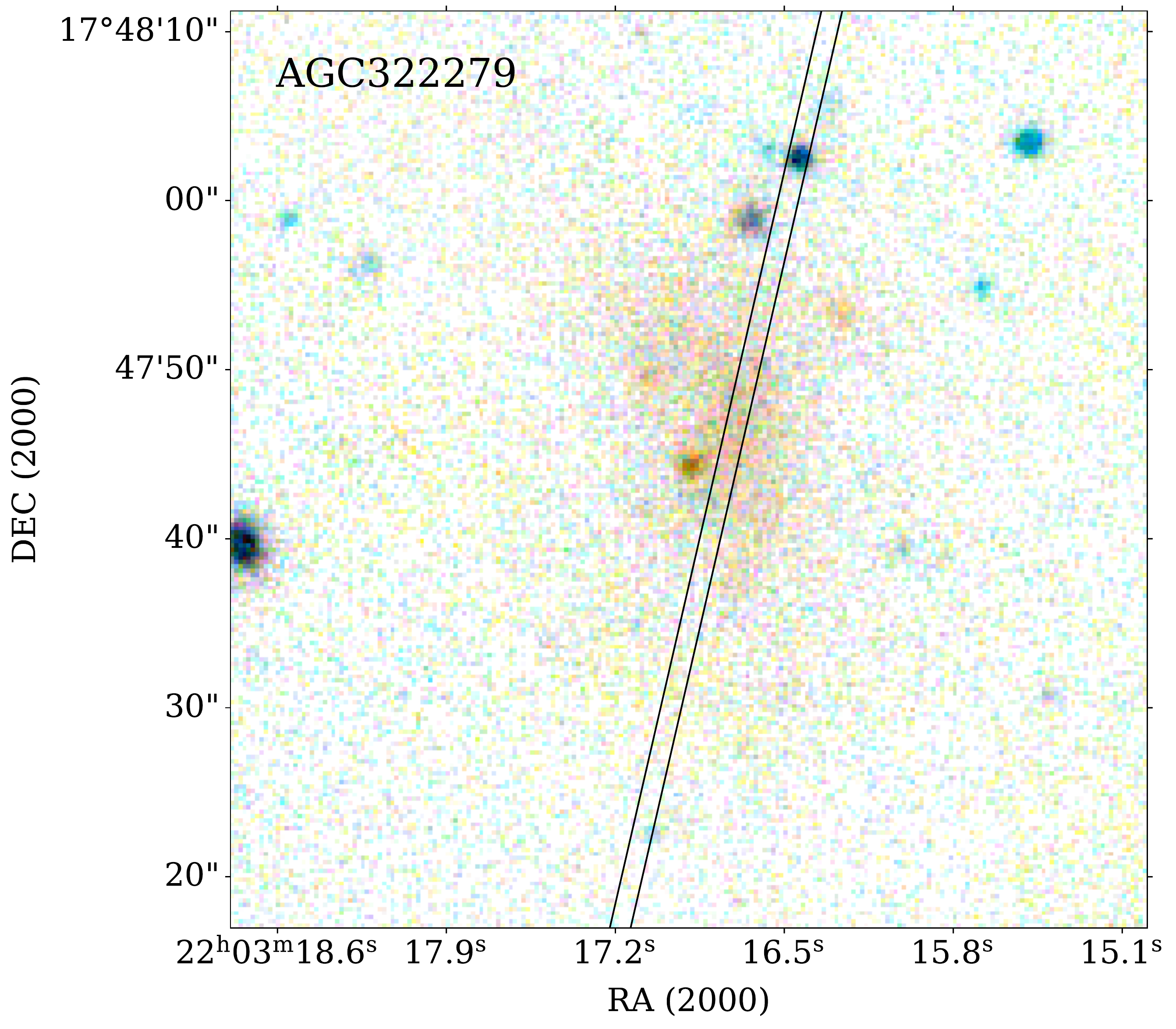}
\includegraphics[width=5.5cm,angle=0,clip=]{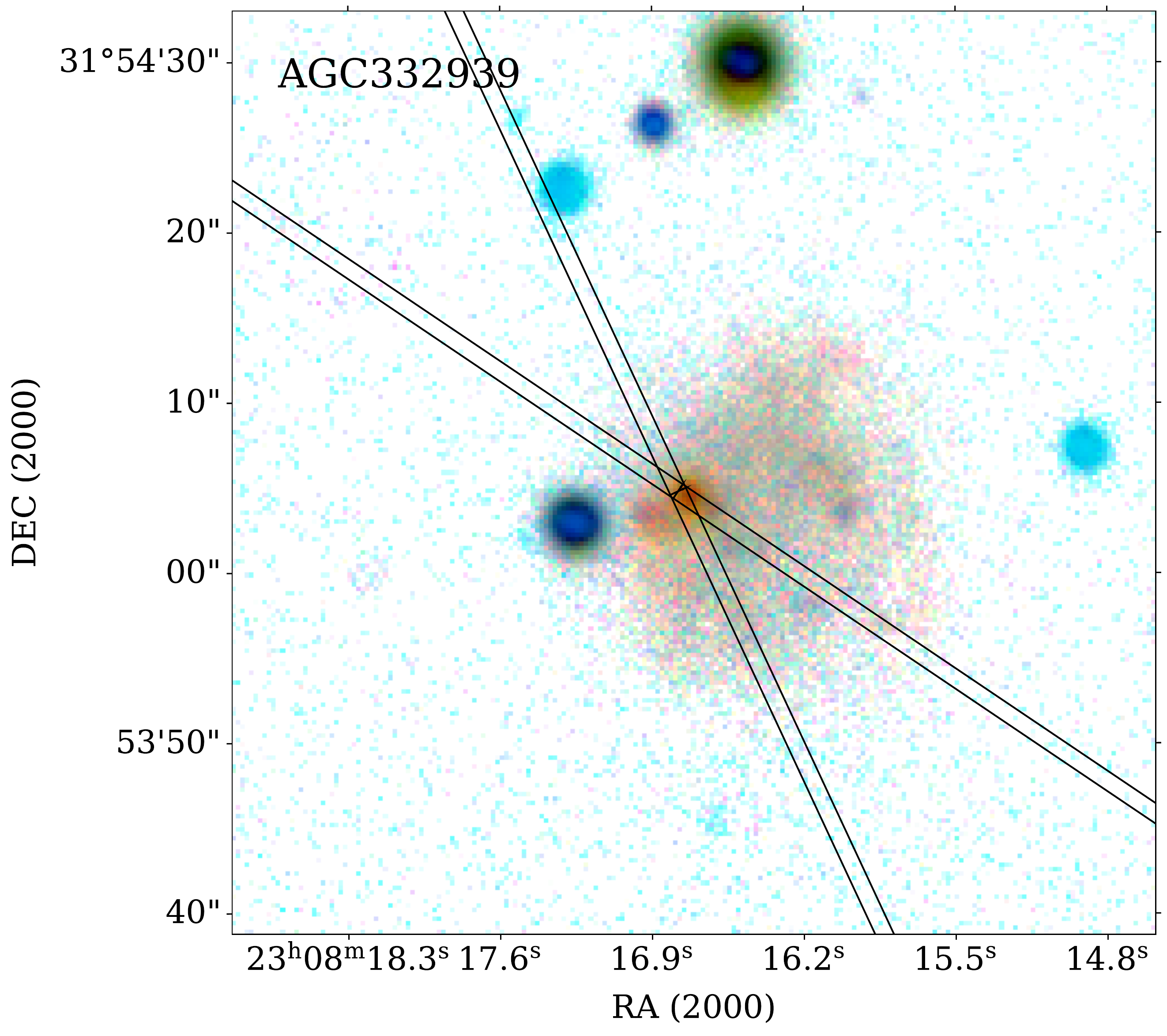}
\caption{Finding charts of the three not-NVG sample galaxies
observed at BTA, with slit positions superimposed.
Inverted colours are used to underline the LSB features. North is up,
East is to the left.
{\bf From left to right:} J0823+1748 -- a distant SF galaxy in  the
field of AGC189201; AGC322279=J2203+1747 -- non-void galaxy;
AGC332939 = J2308+3154 -- non-void galaxy.}
\label{fig:slits3}
\end{figure*}

\begin{figure*}
\includegraphics[width=4.0cm,angle=-90,clip=]{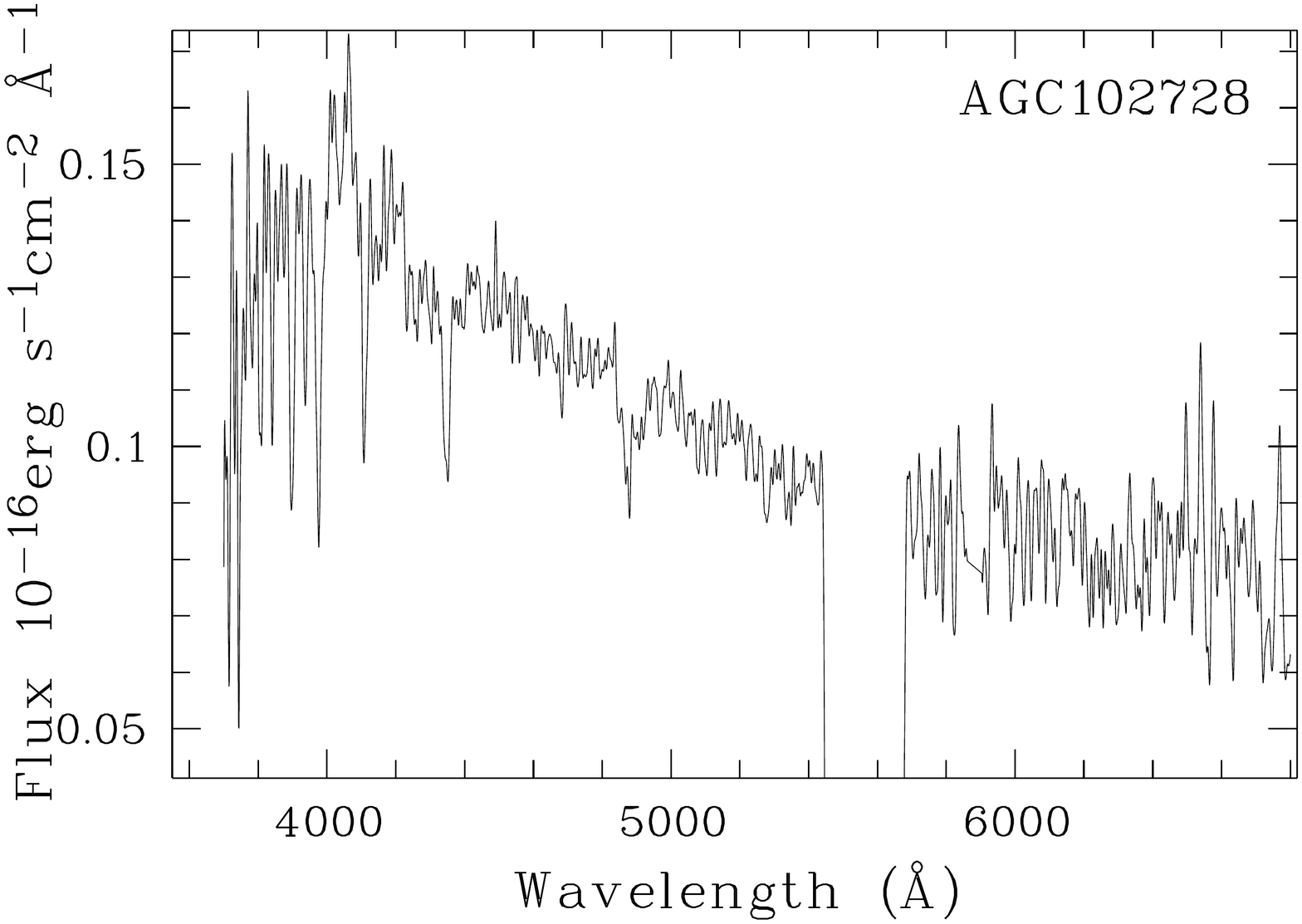}
\includegraphics[width=4.0cm,angle=-90,clip=]{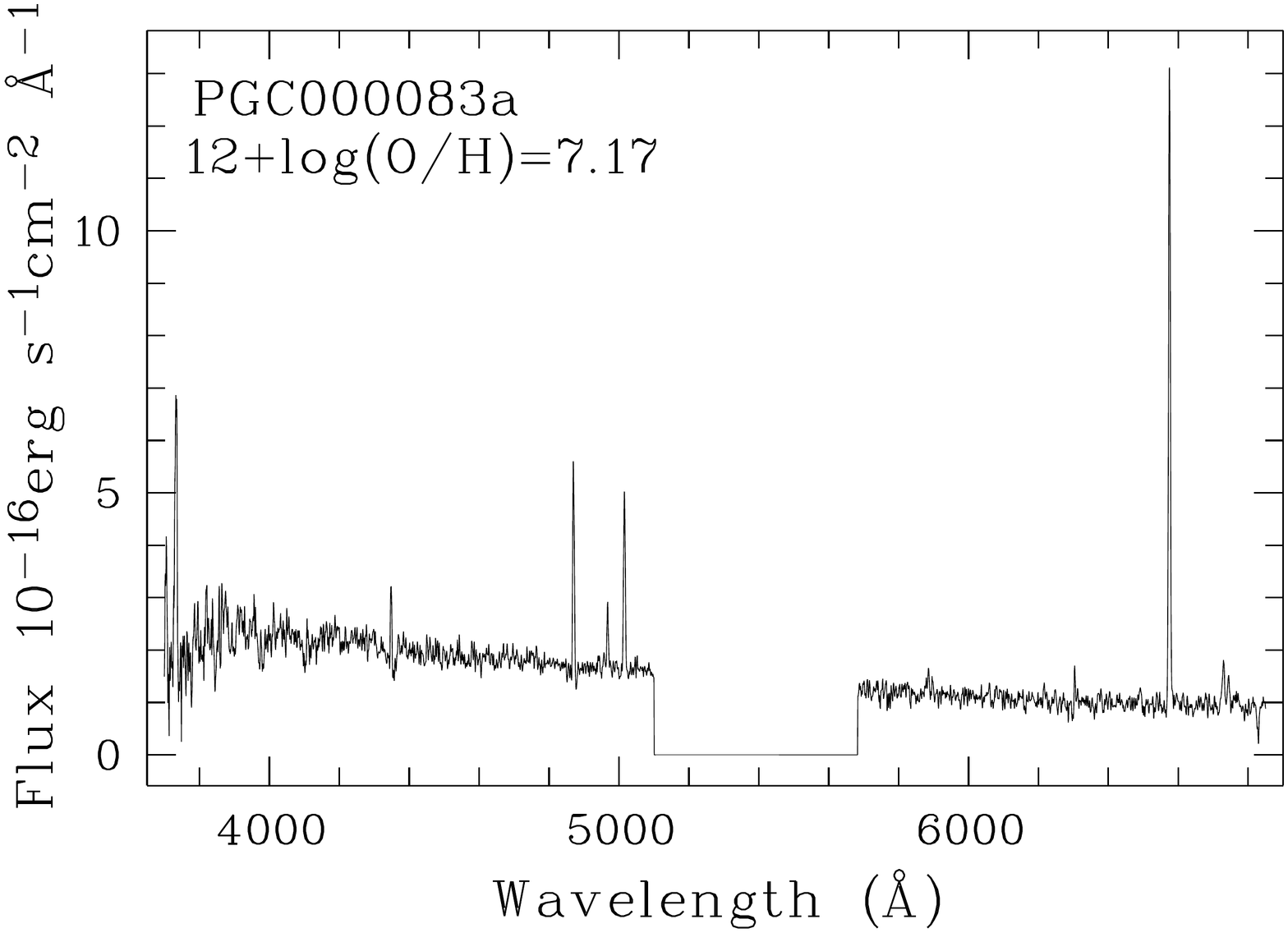}
\includegraphics[width=4.0cm,angle=-90,clip=]{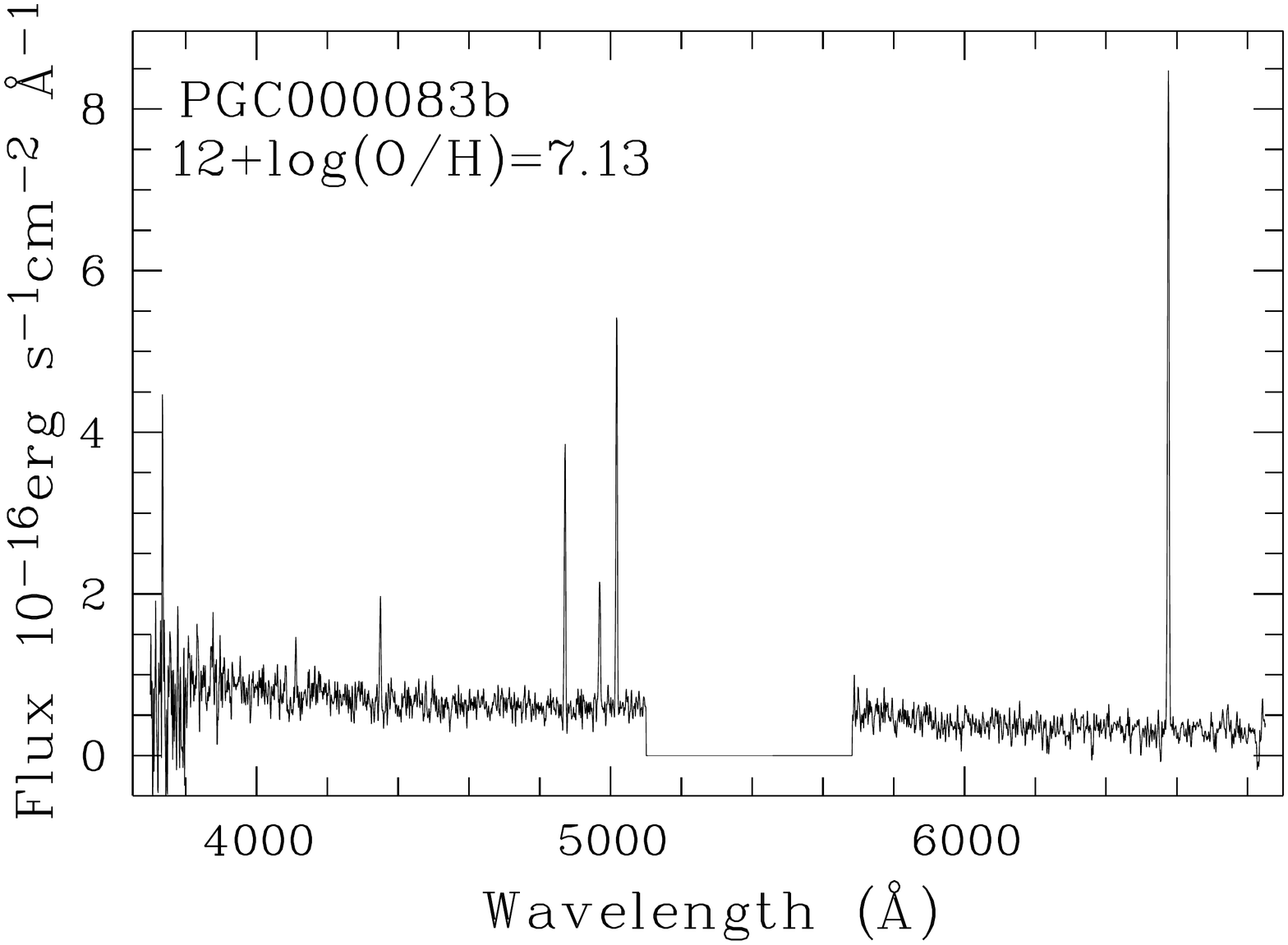}
\includegraphics[width=4.0cm,angle=-90,clip=]{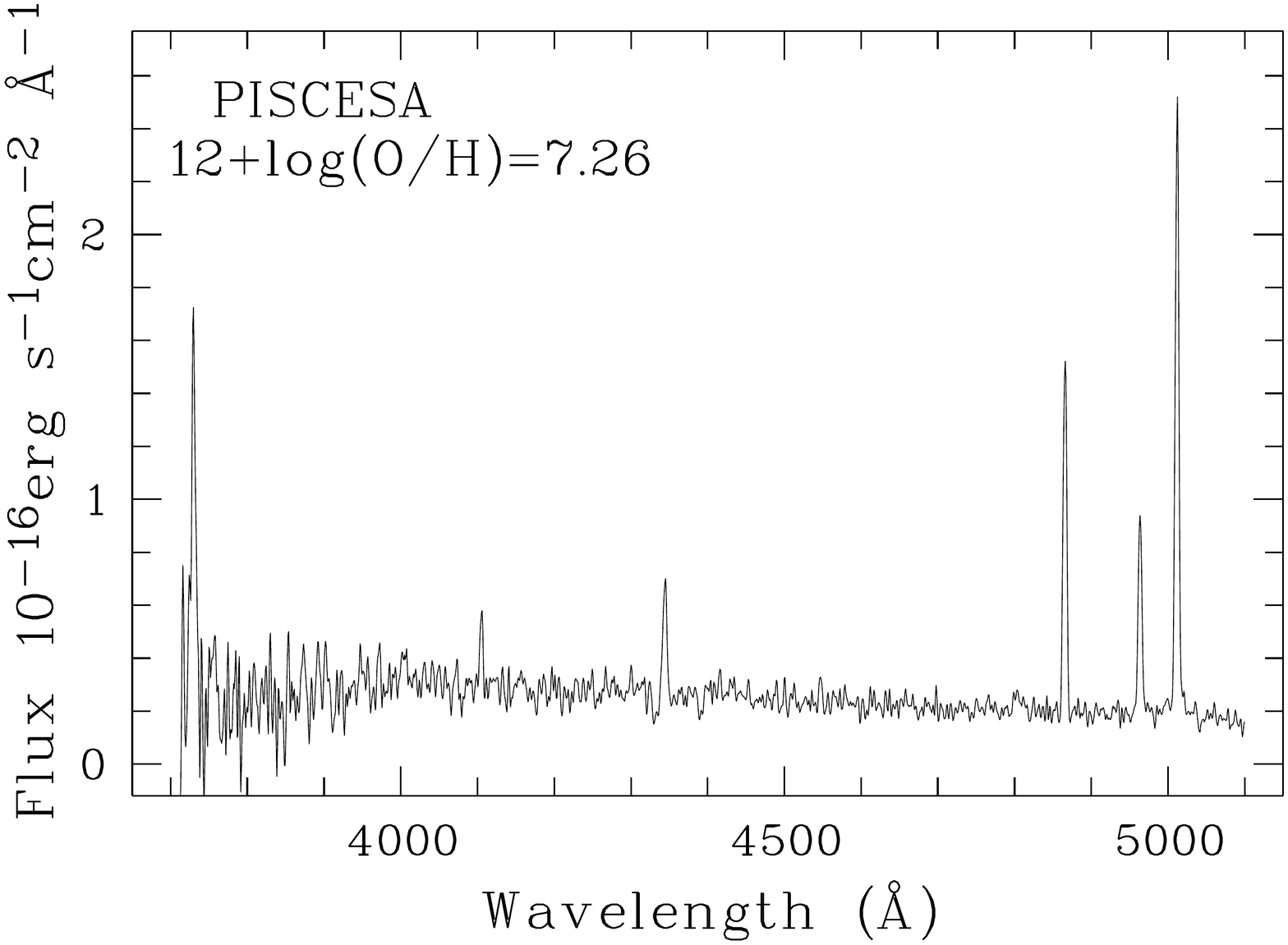}
\includegraphics[width=4.0cm,angle=-90,clip=]{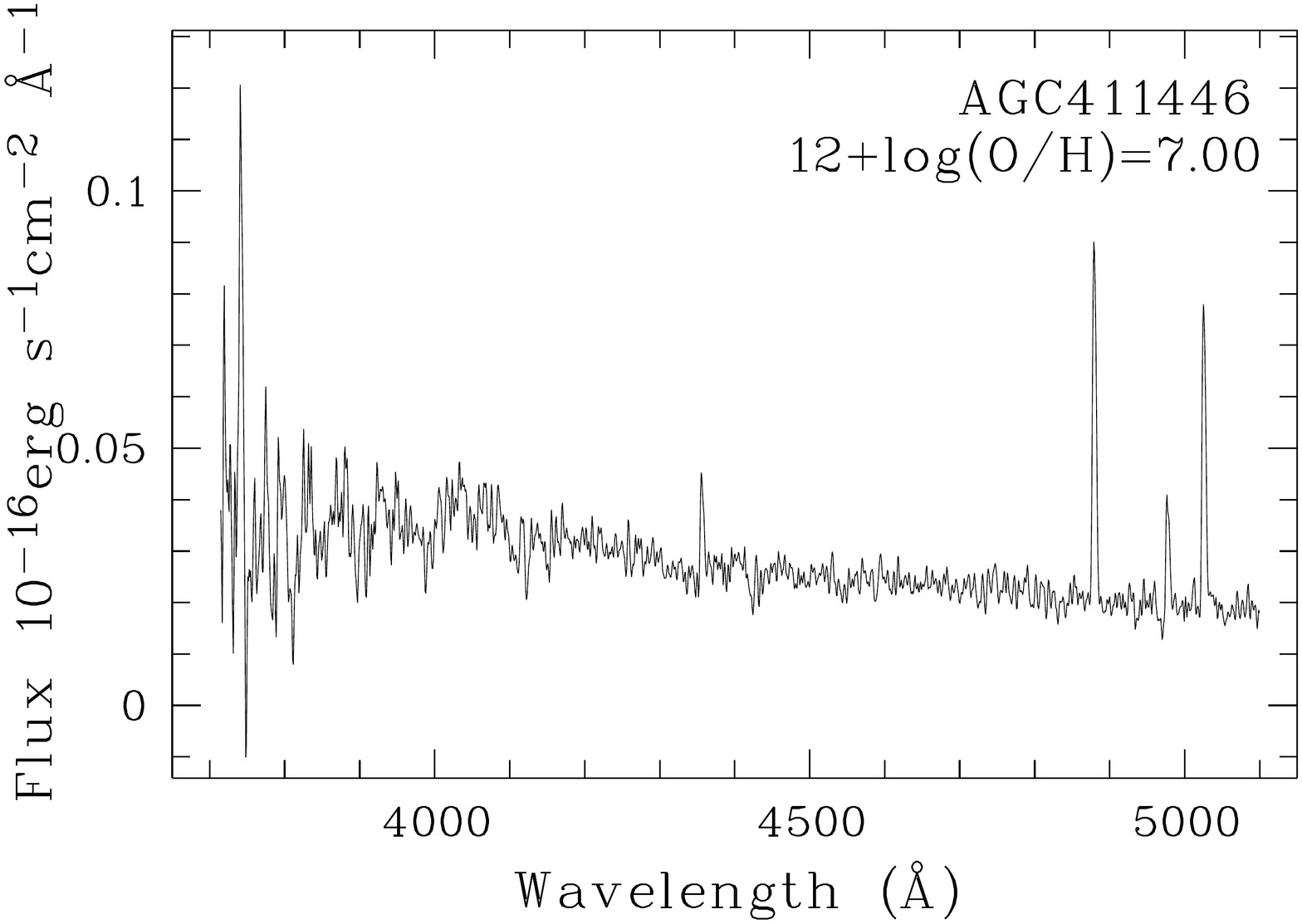}
\includegraphics[width=4.0cm,angle=-90,clip=]{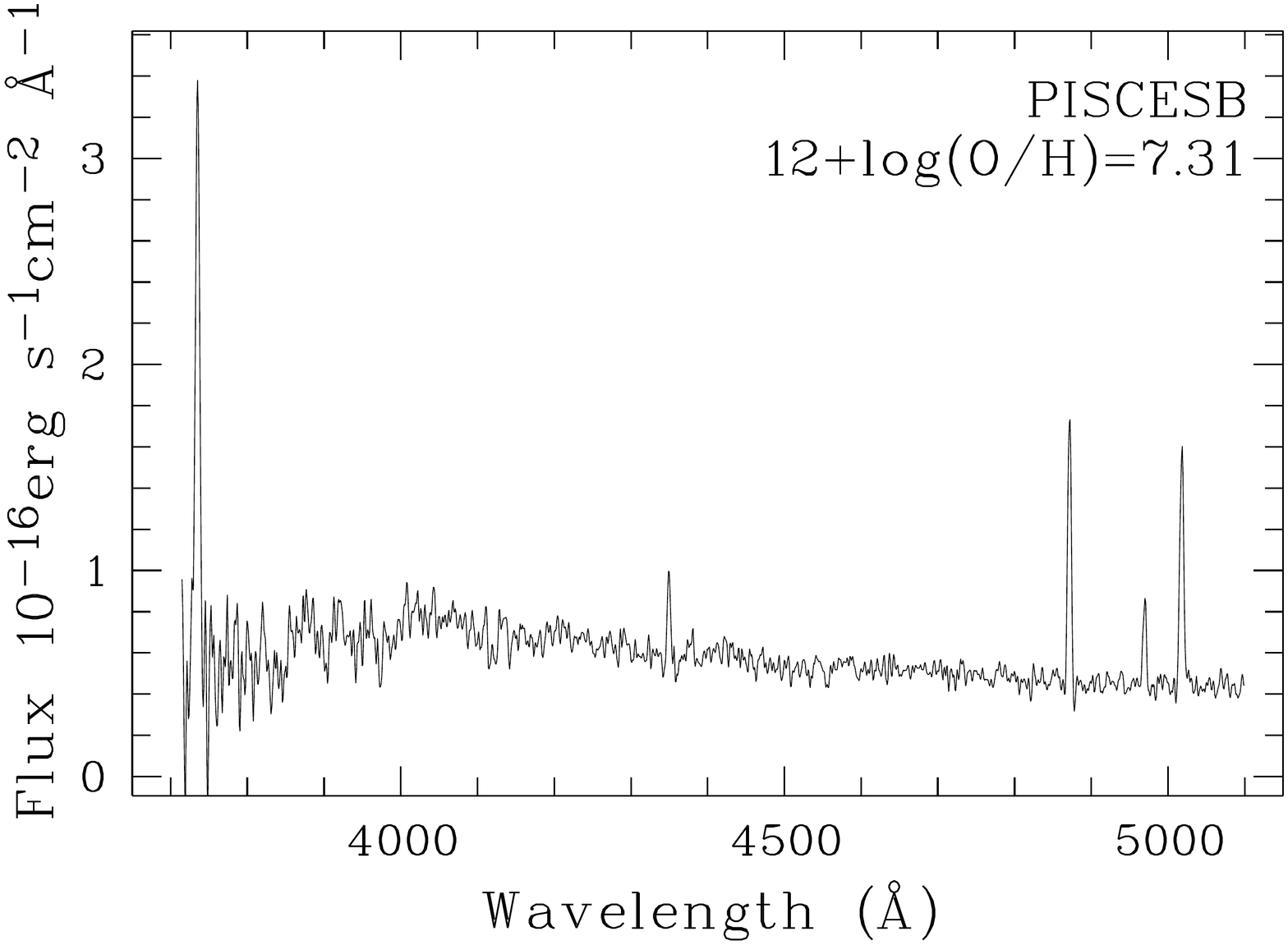}
\includegraphics[width=4.0cm,angle=-90,clip=]{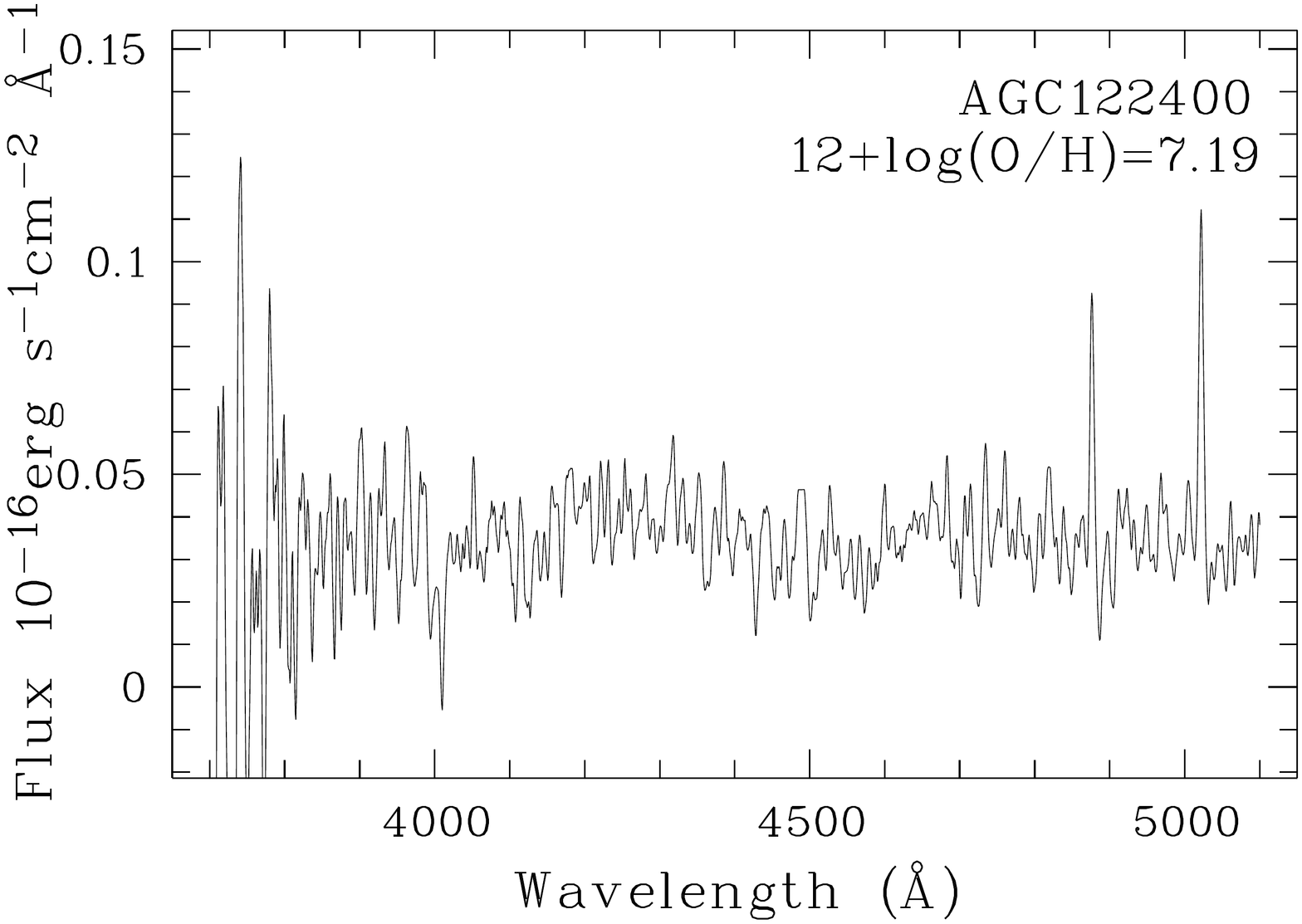}
\includegraphics[width=4.0cm,angle=-90,clip=]{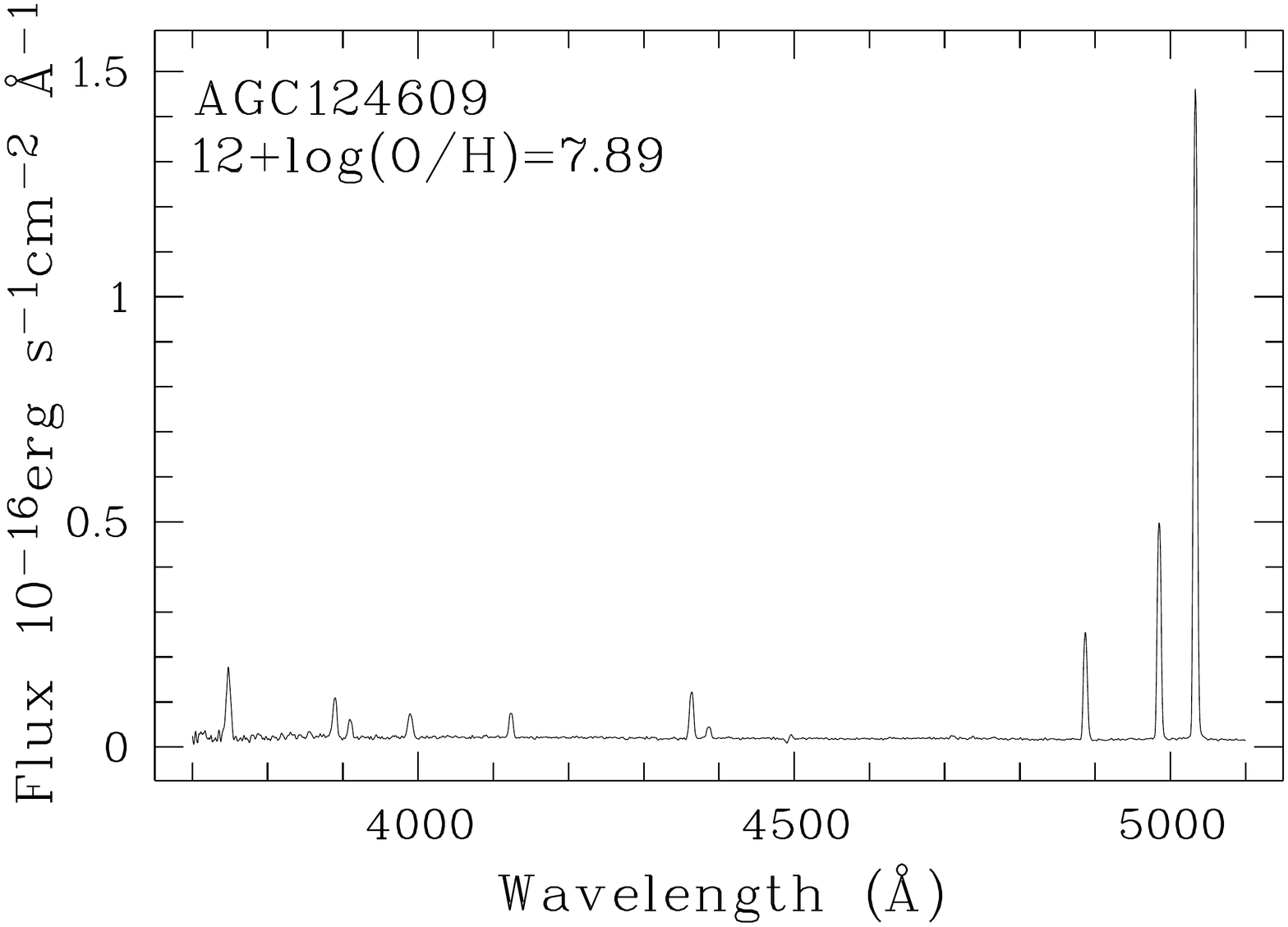}
\includegraphics[width=4.0cm,angle=-90,clip=]{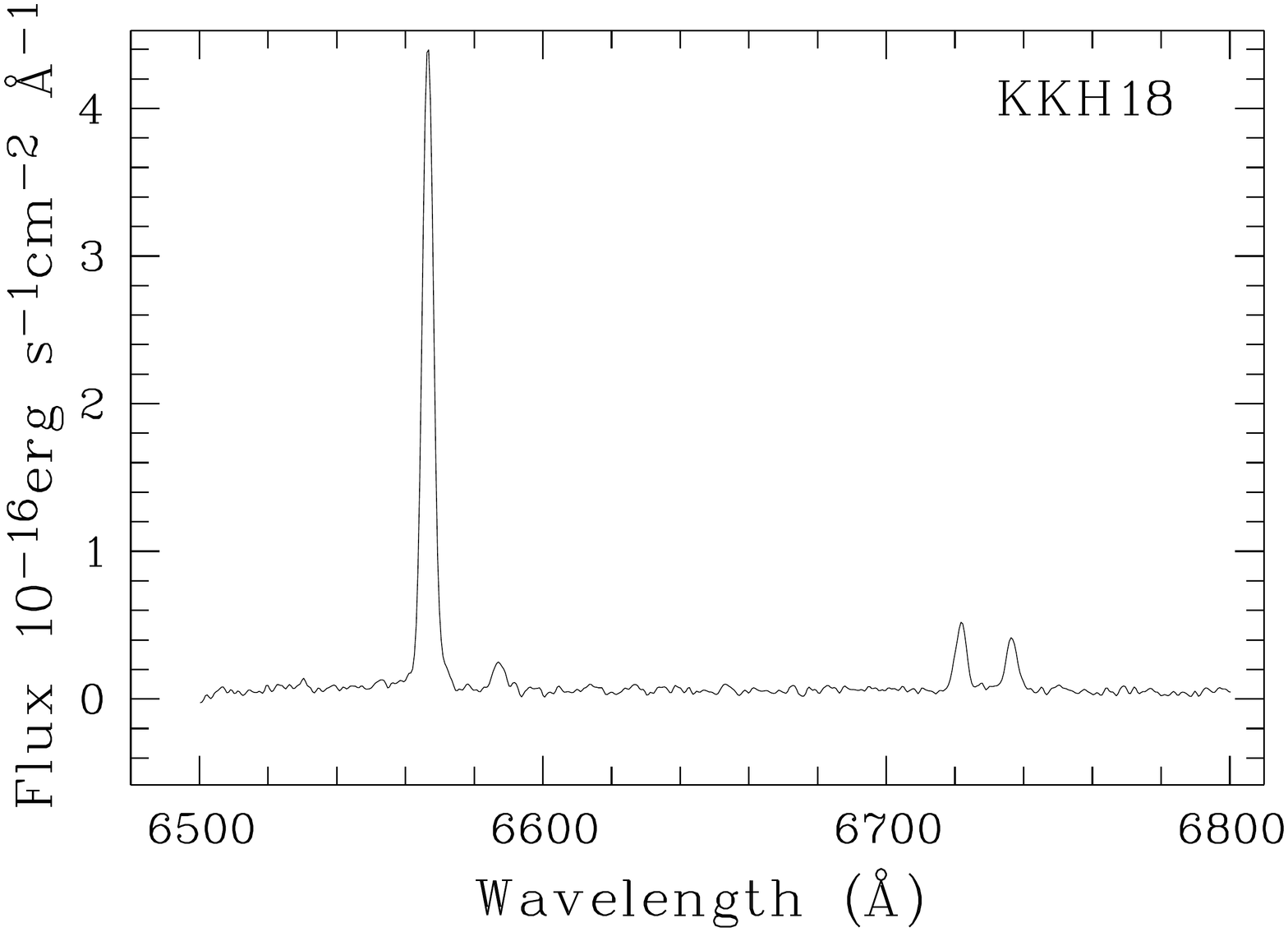}
\includegraphics[width=4.0cm,angle=-90,clip=]{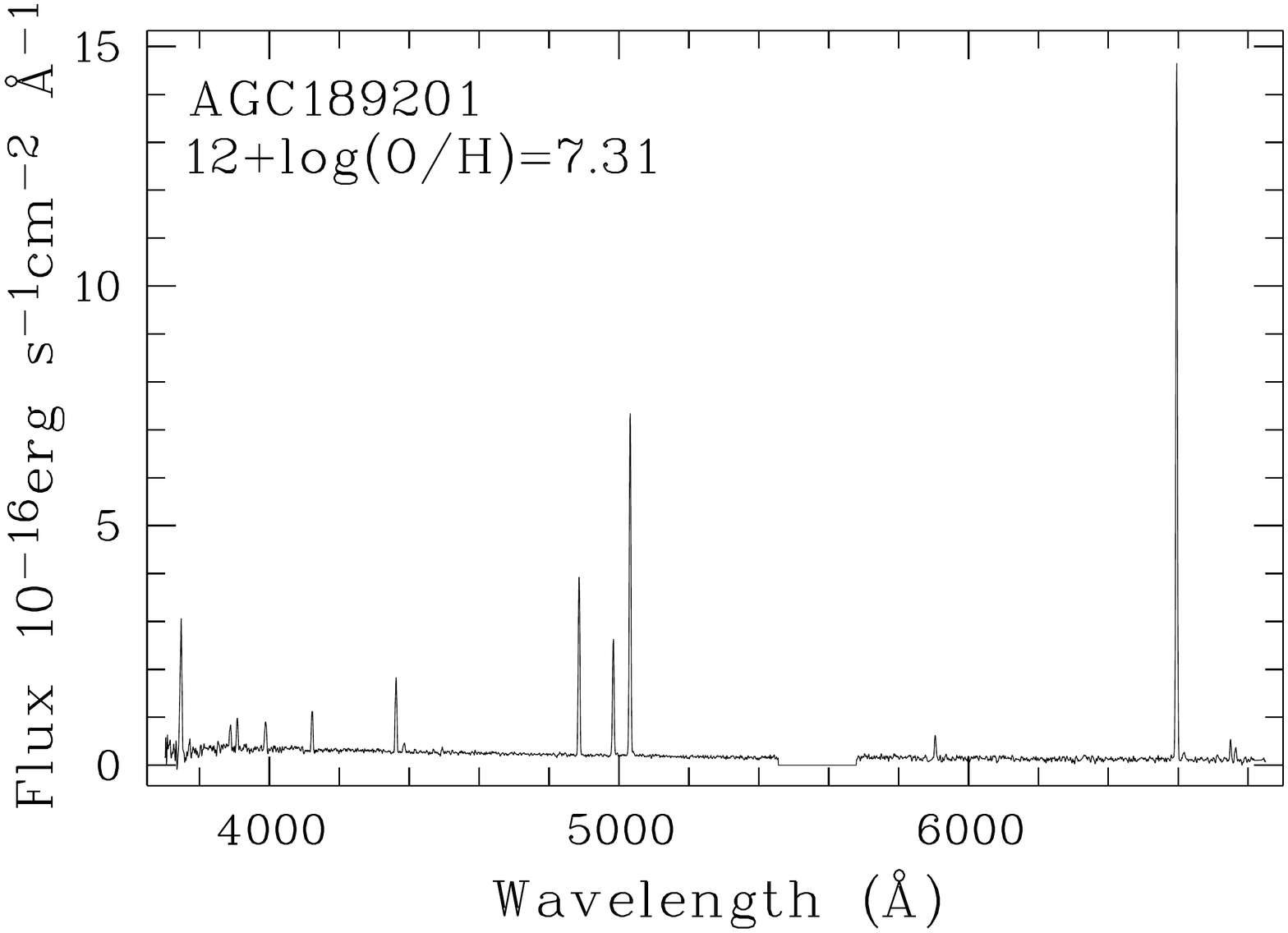}
\includegraphics[width=4.0cm,angle=-90,clip=]{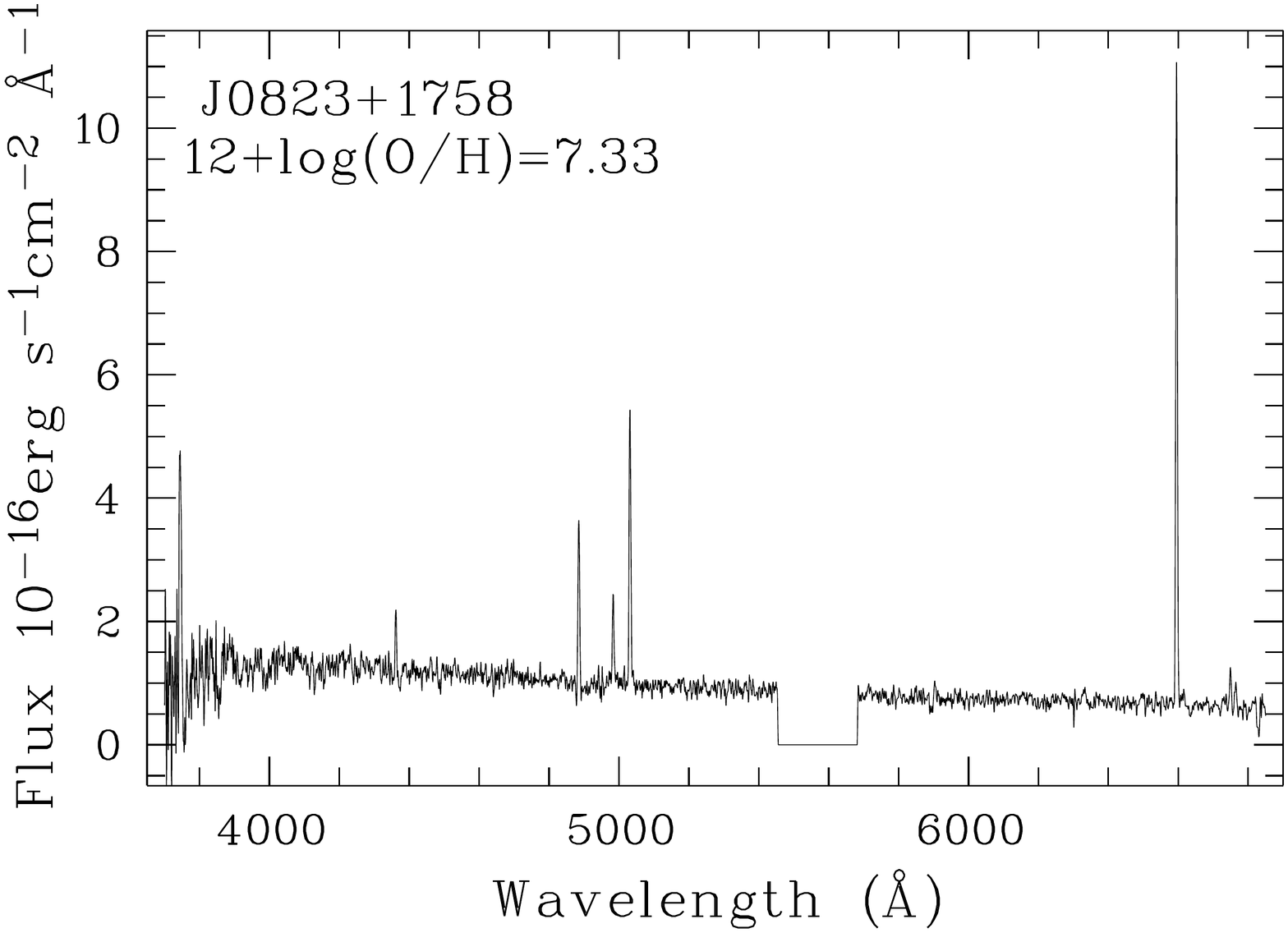}
\includegraphics[width=4.0cm,angle=-90,clip=]{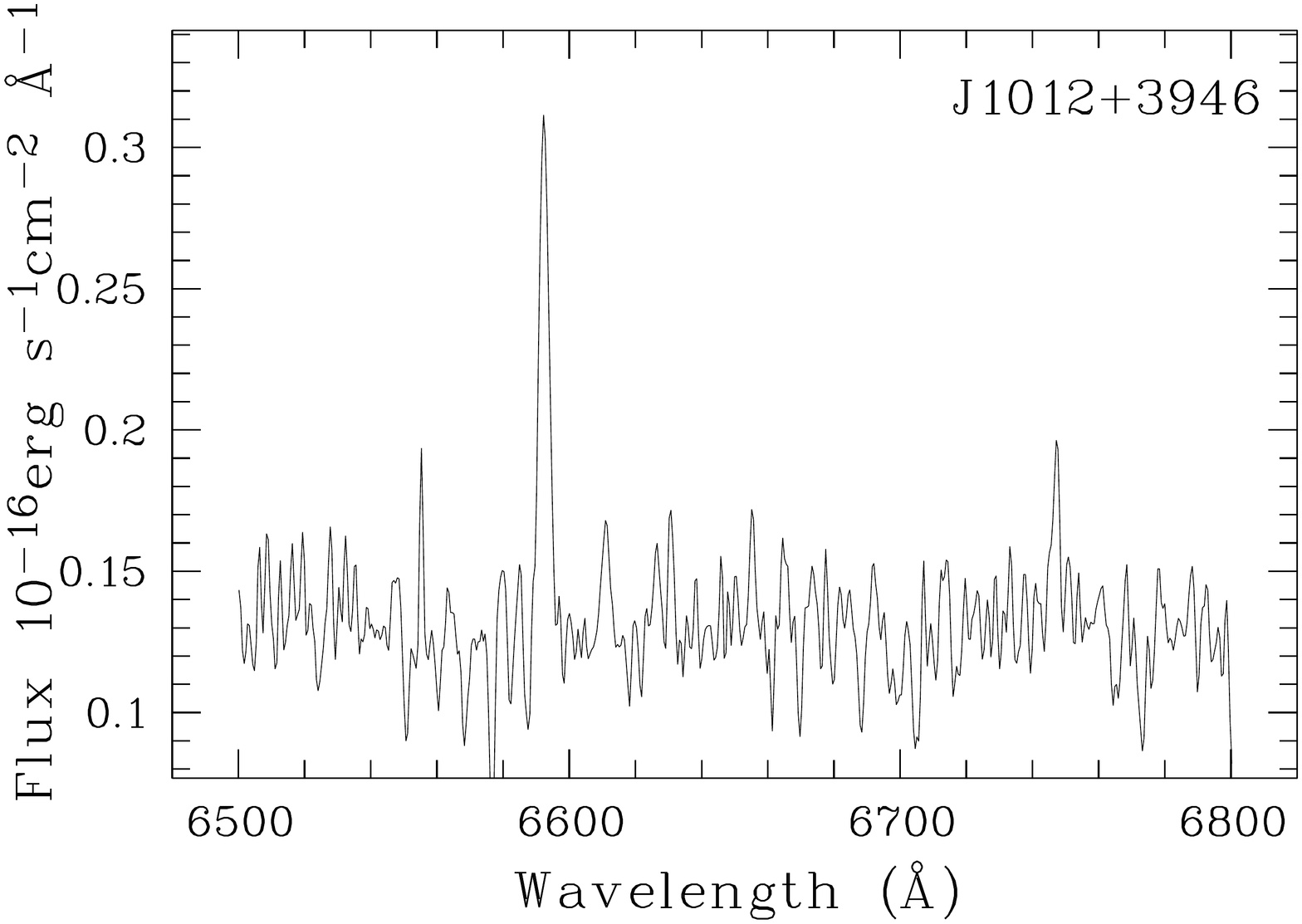}
\includegraphics[width=4.0cm,angle=-90,clip=]{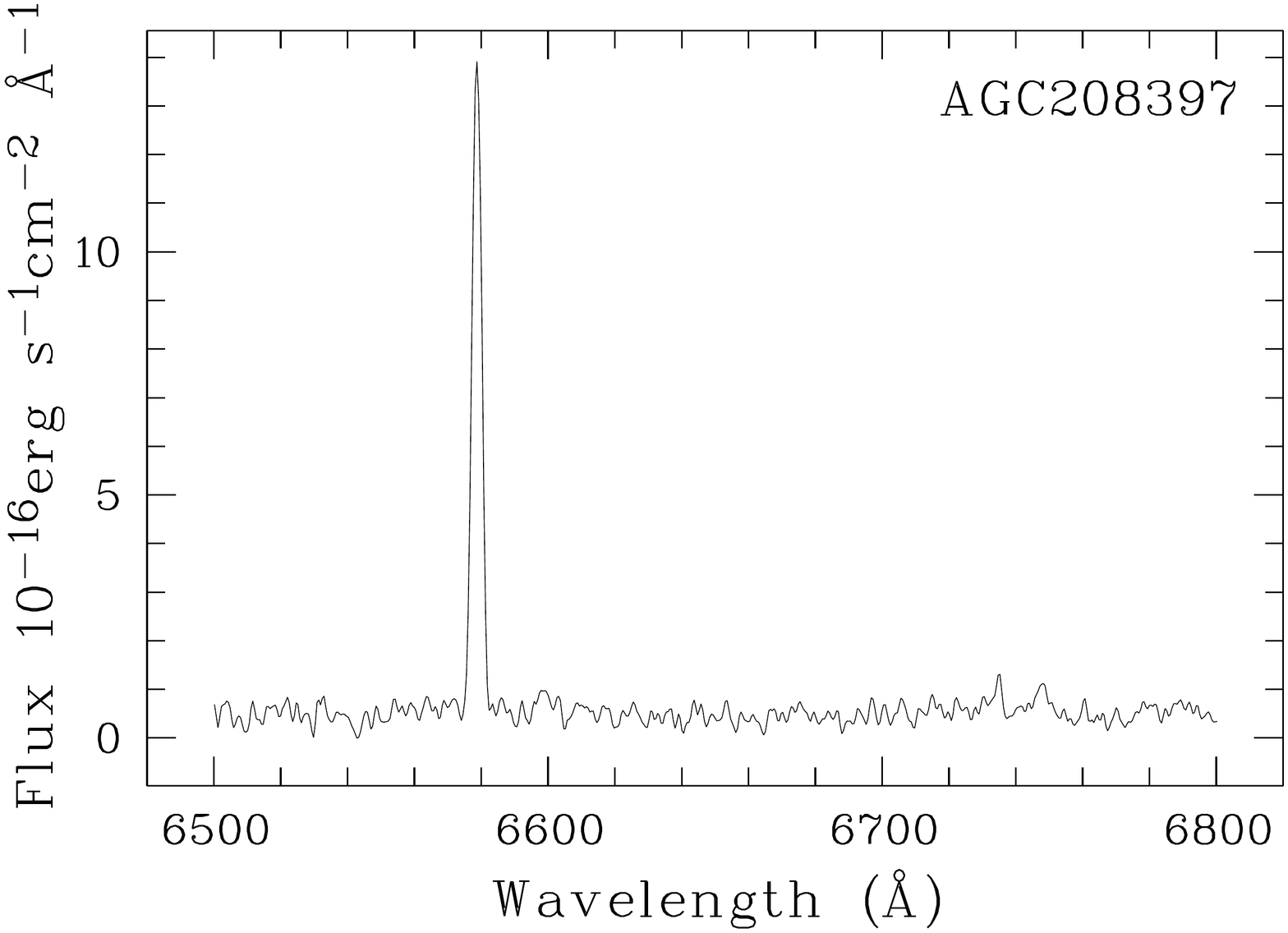}
\includegraphics[width=4.0cm,angle=-90,clip=]{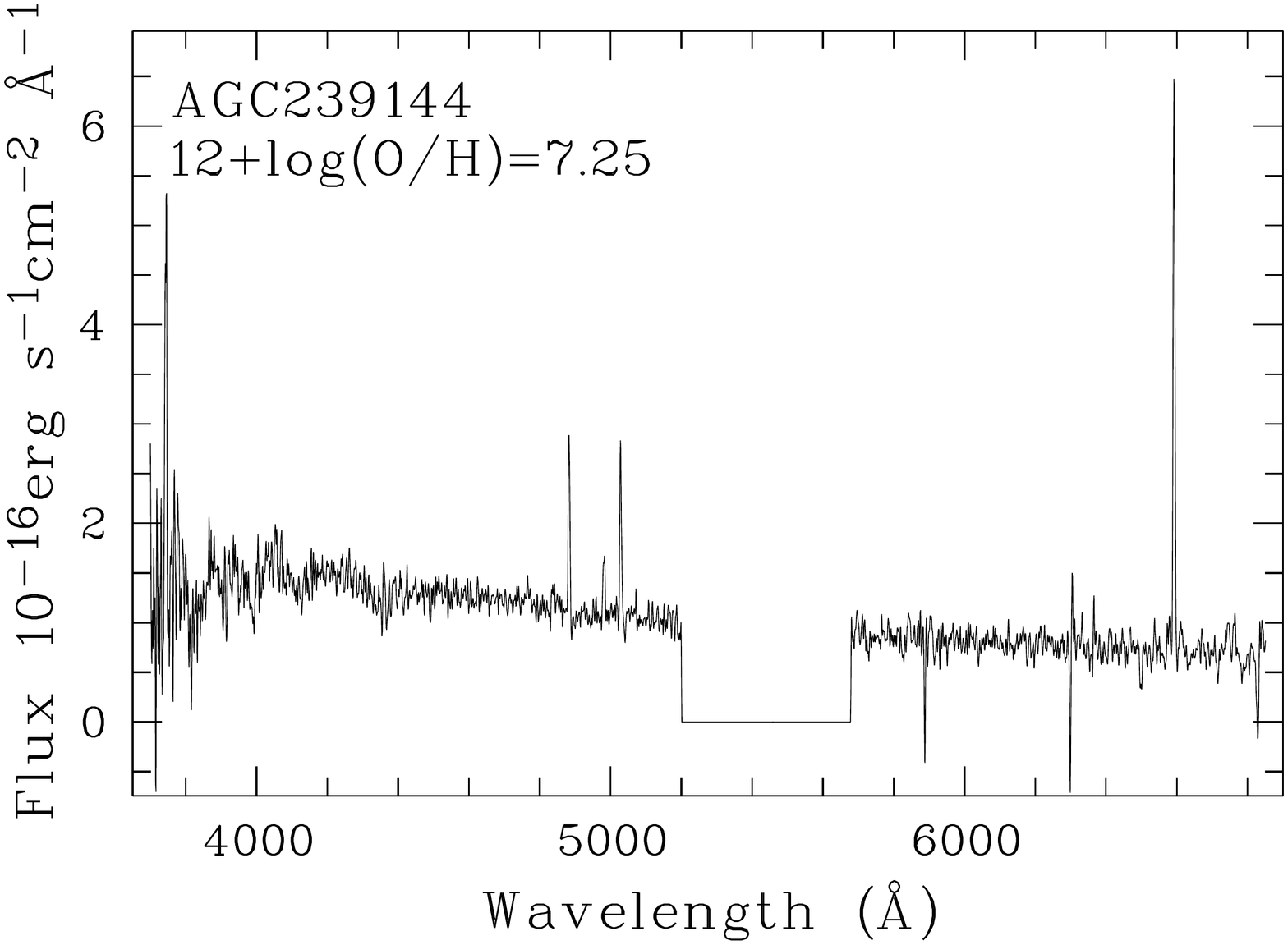}
\includegraphics[width=4.0cm,angle=-90,clip=]{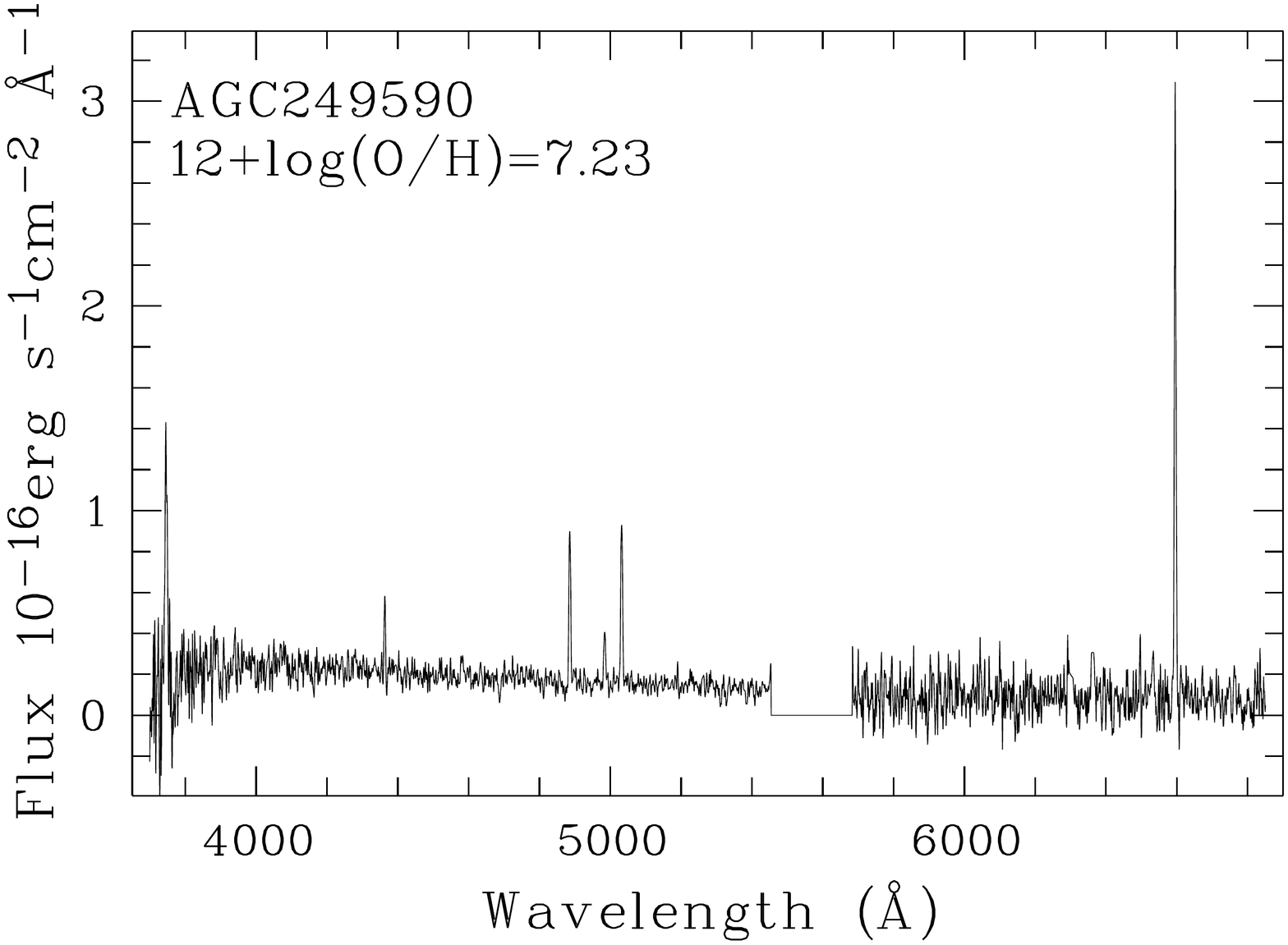}
\caption{1D spectra of void XMP candidates obtained with BTA.
The wavelengths on the $X$ axis are observed, not in the rest-frame. The flux
densities on the $Y$ axis are shown in units of
10$^{-16}$ ergs~s$^{-1}$cm$^{-2}$~\AA$^{-1}$.
The galaxy name and derived value of 12+log(O/H) are shown at the top
of each box.
{\bf Top panel, from left to right:} AGC102728=J0000+3101.
Only very faint emission H$\alpha$ is detected besides Balmer absorptions.
PGC00083=J0001+3222. Spectra of two knots are shown with somewhat different
O/H = 7.17 for "a" and 7.13 for "b".
{\bf Second panel, from left to right:} PiscesA=J0014+1048, O/H=7.26;
AGC411446=J011003--000036, O/H = 7.00;
PiscesB=J0119+1107, O/H = 7.31;
{\bf Third  panel, left to right:} AGC122400 = J0231+3542, O/H = 7.19;
AGC124609=J0249+3444,  O/H = 7.89; KKH18=J0303+3341, red spectrum.
{\bf Fourth panel, from left to right:}
AGC189201=J0823+1754, O/H = 7.31; J0823+1758, O/H = 7.33; J1012+3946, red spectrum.
{\bf Bottom panel, from left to right:}
back-up time red spectrum of
AGC208397  = J1038+0352;  AGC239144 = J1349+3544, O/H = 7.25;
AGC249590=J1440+3416, O/H = 7.23.
}
\label{fig:BTA_1Da}
\end{figure*}

\begin{figure*}
\includegraphics[width=4.0cm,angle=-90,clip=]{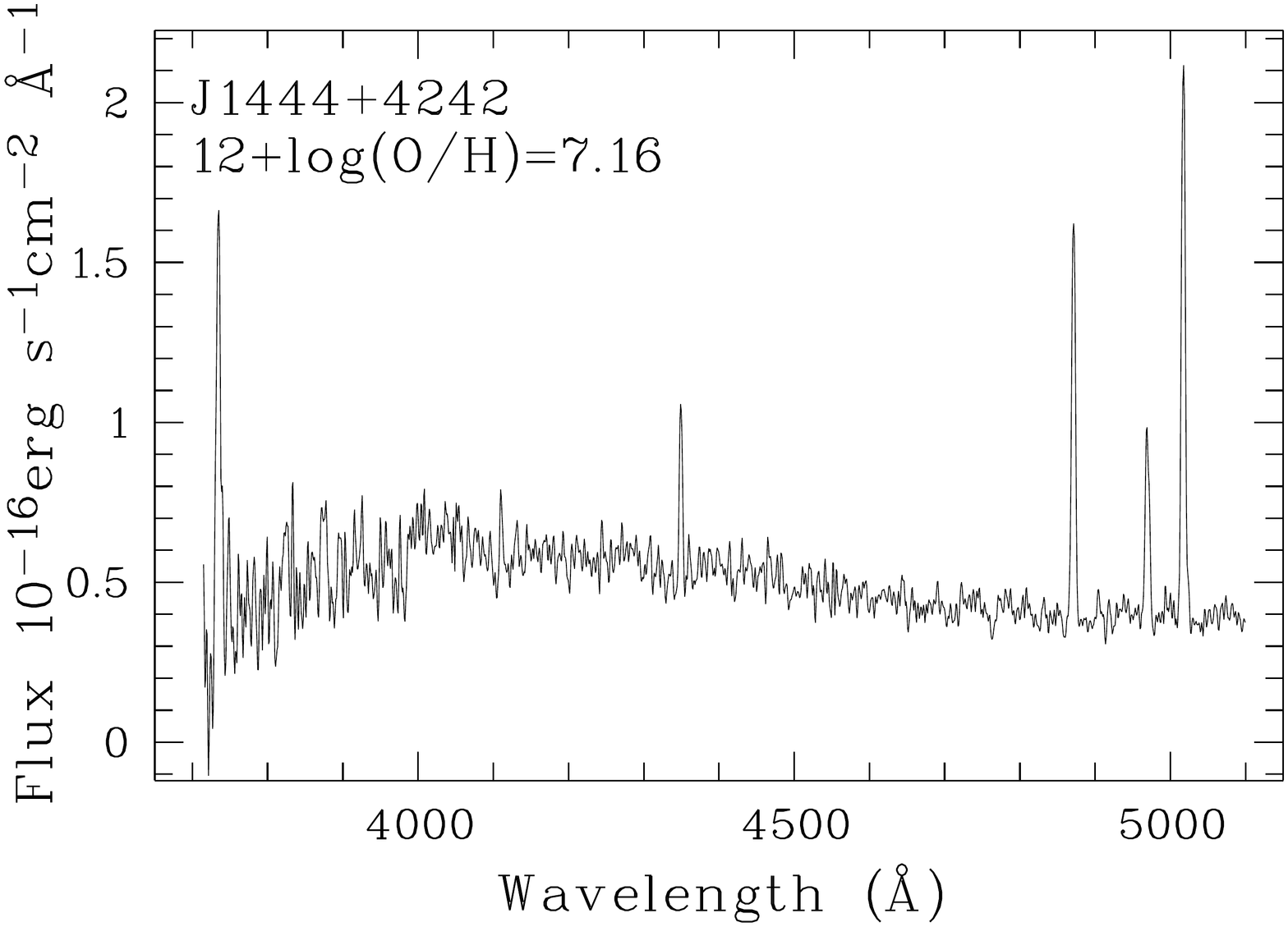}
\includegraphics[width=4.0cm,angle=-90,clip=]{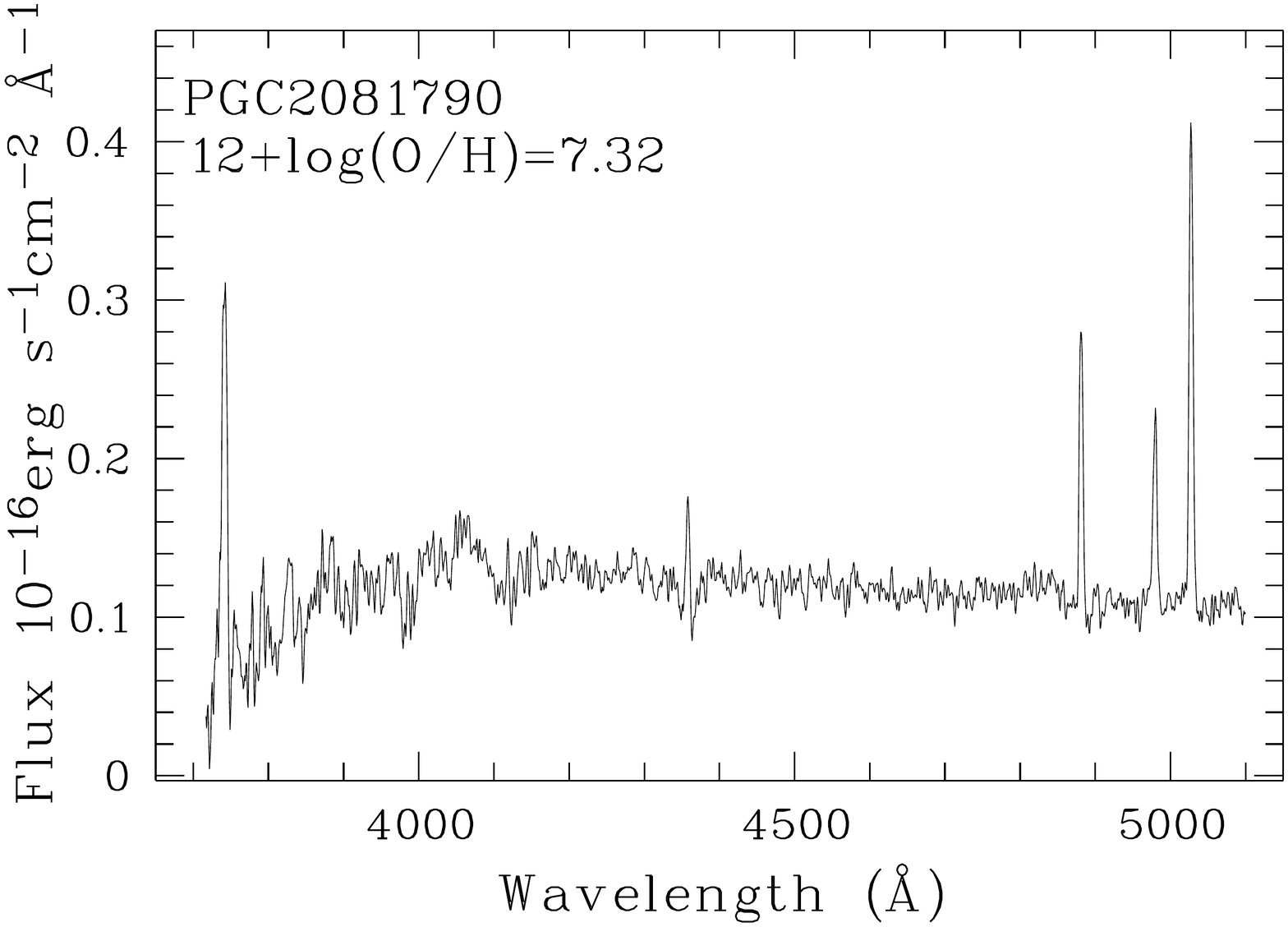}
\includegraphics[width=4.0cm,angle=-90,clip=]{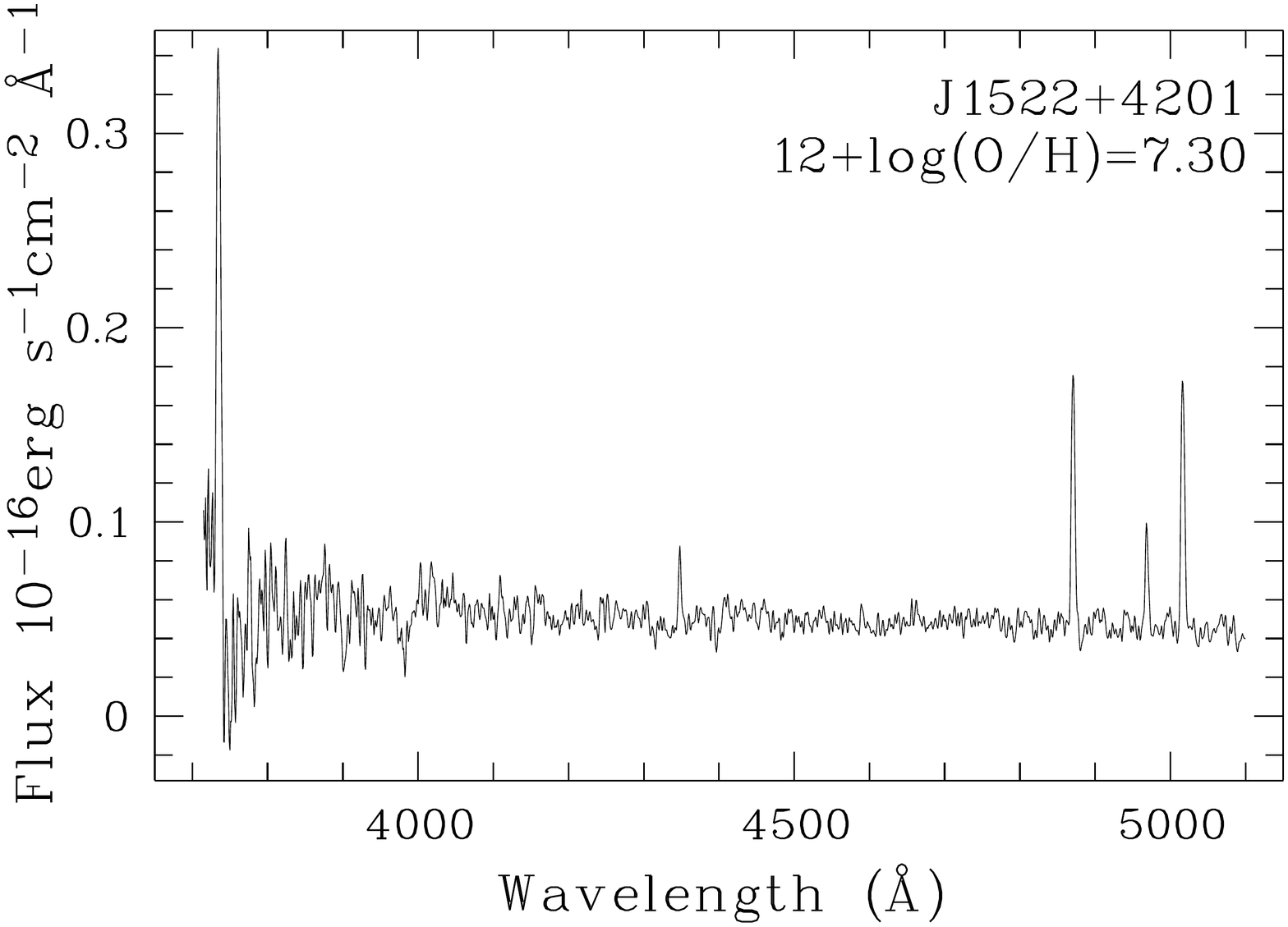}
\includegraphics[width=4.0cm,angle=-90,clip=]{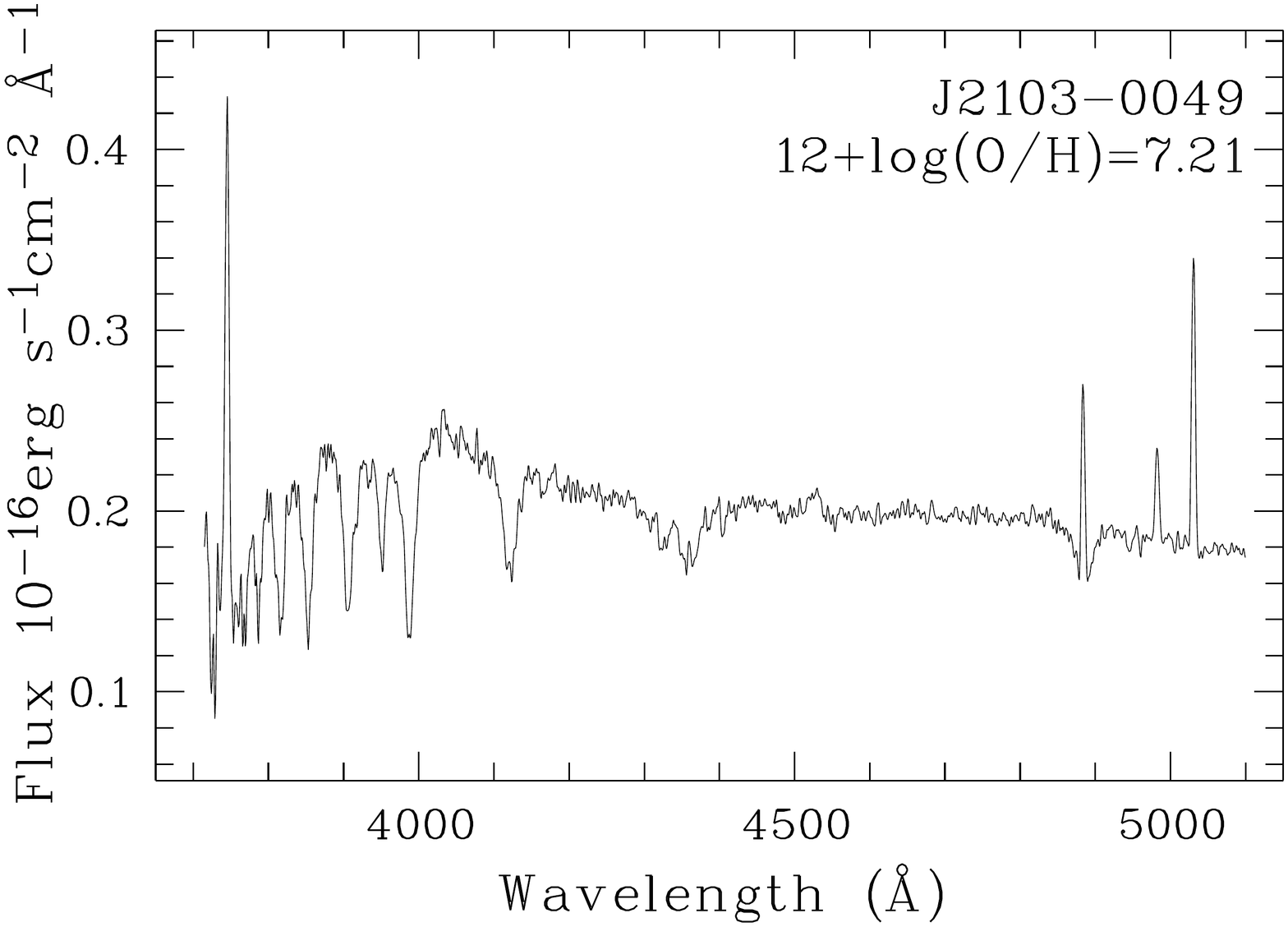}
\includegraphics[width=4.0cm,angle=-90,clip=]{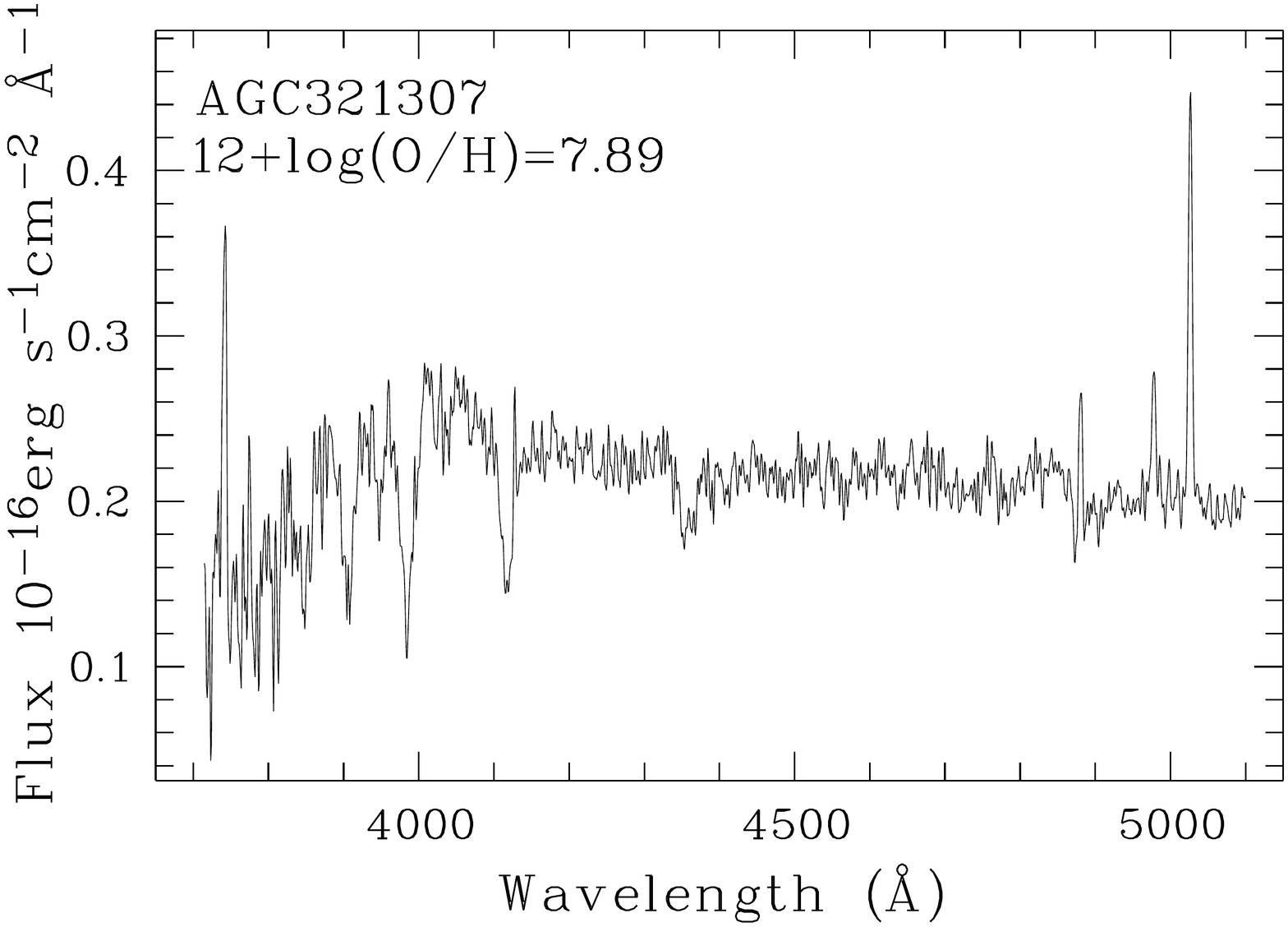}
\includegraphics[width=4.0cm,angle=-90,clip=]{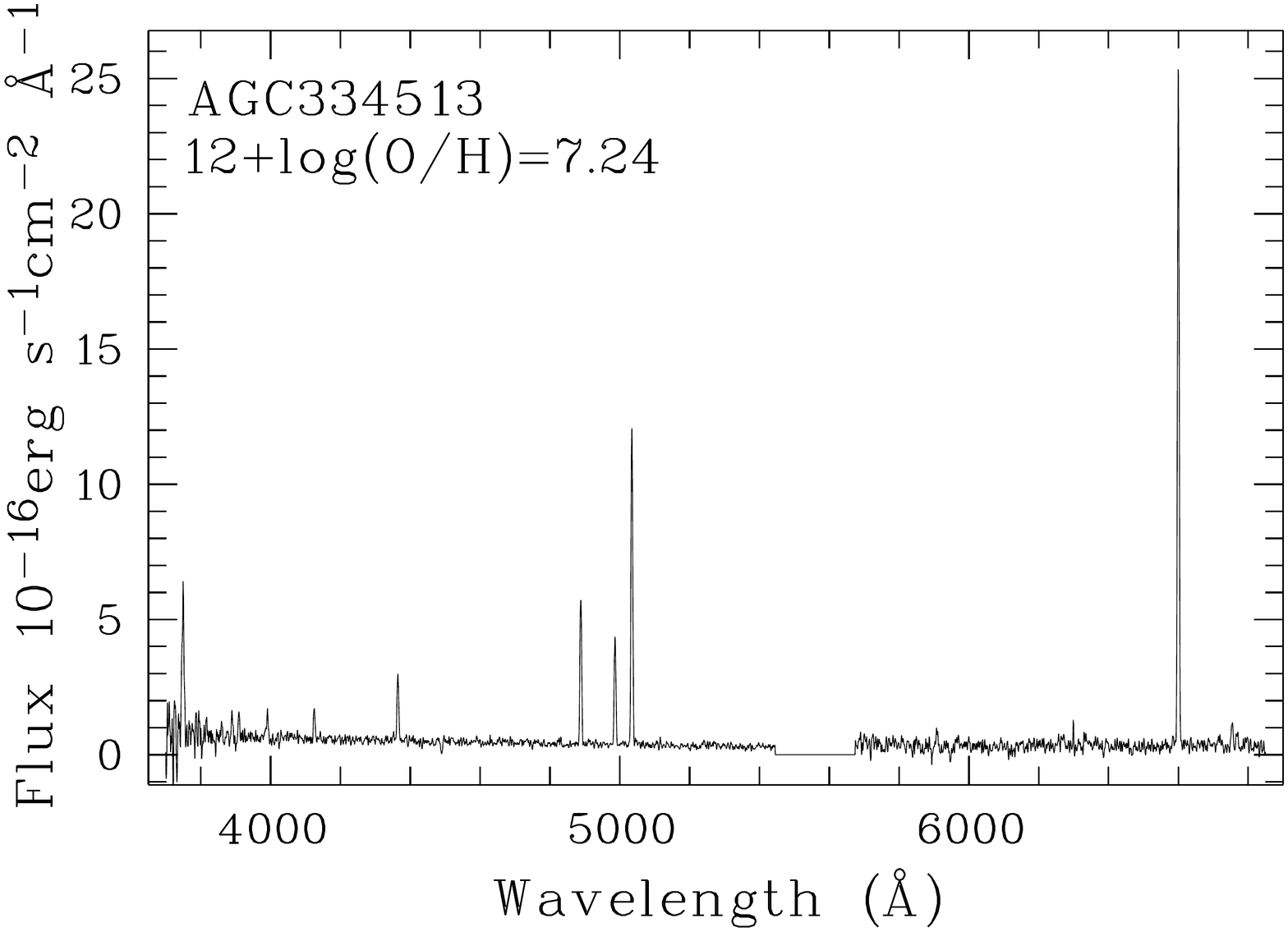}
\includegraphics[width=4.0cm,angle=-90,clip=]{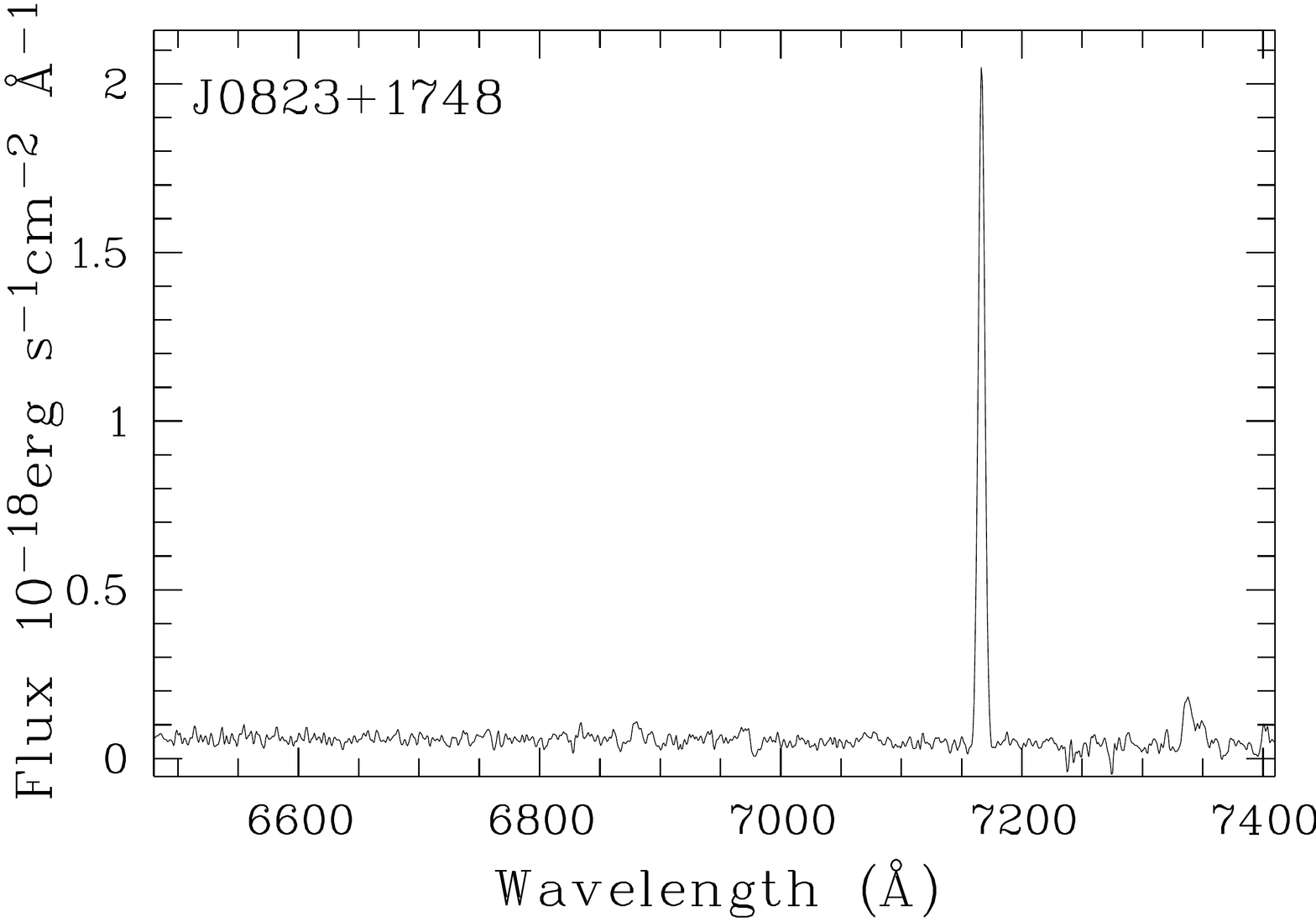}
\includegraphics[width=4.0cm,angle=-90,clip=]{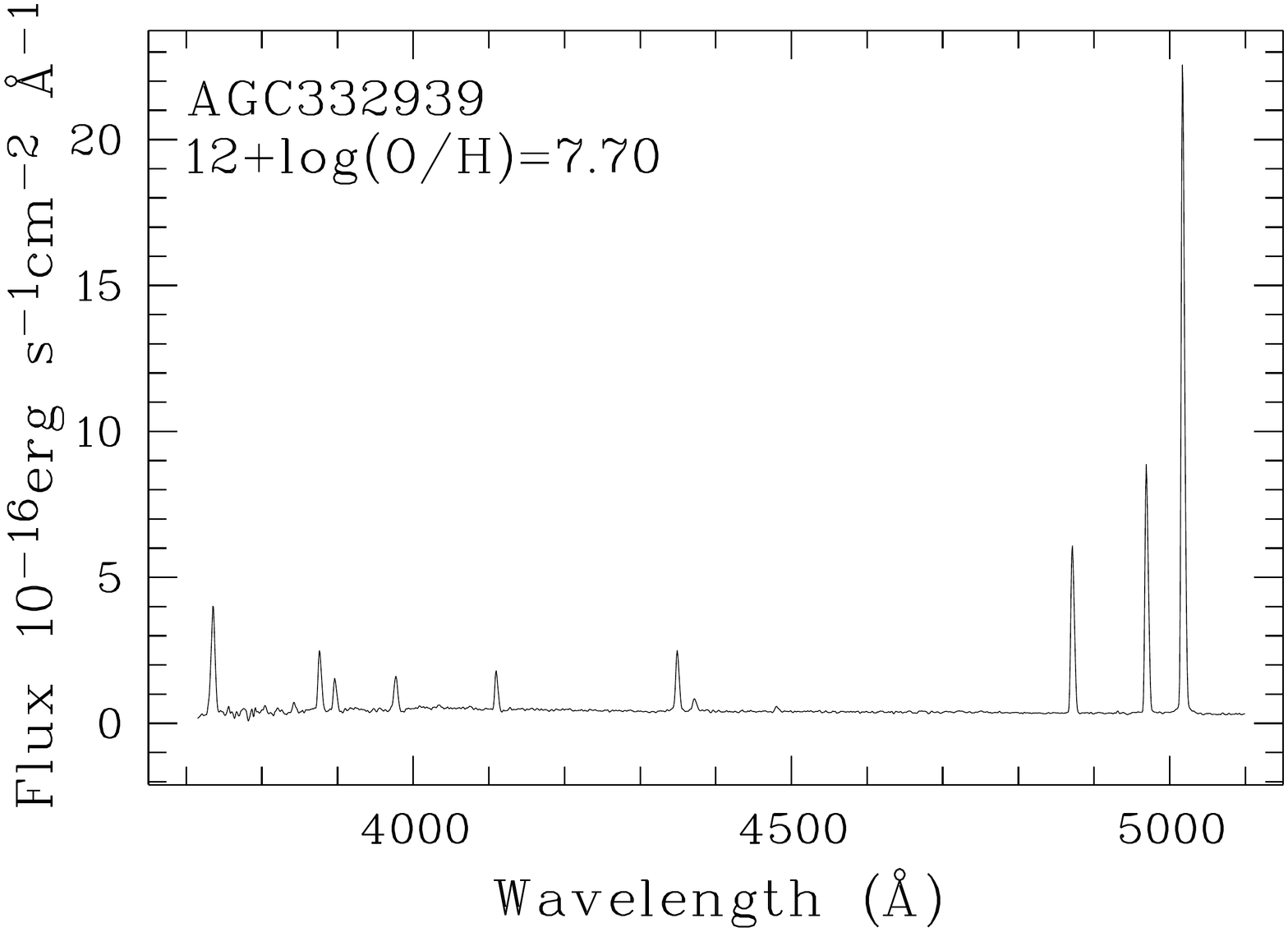}
\includegraphics[width=4.0cm,angle=-90,clip=]{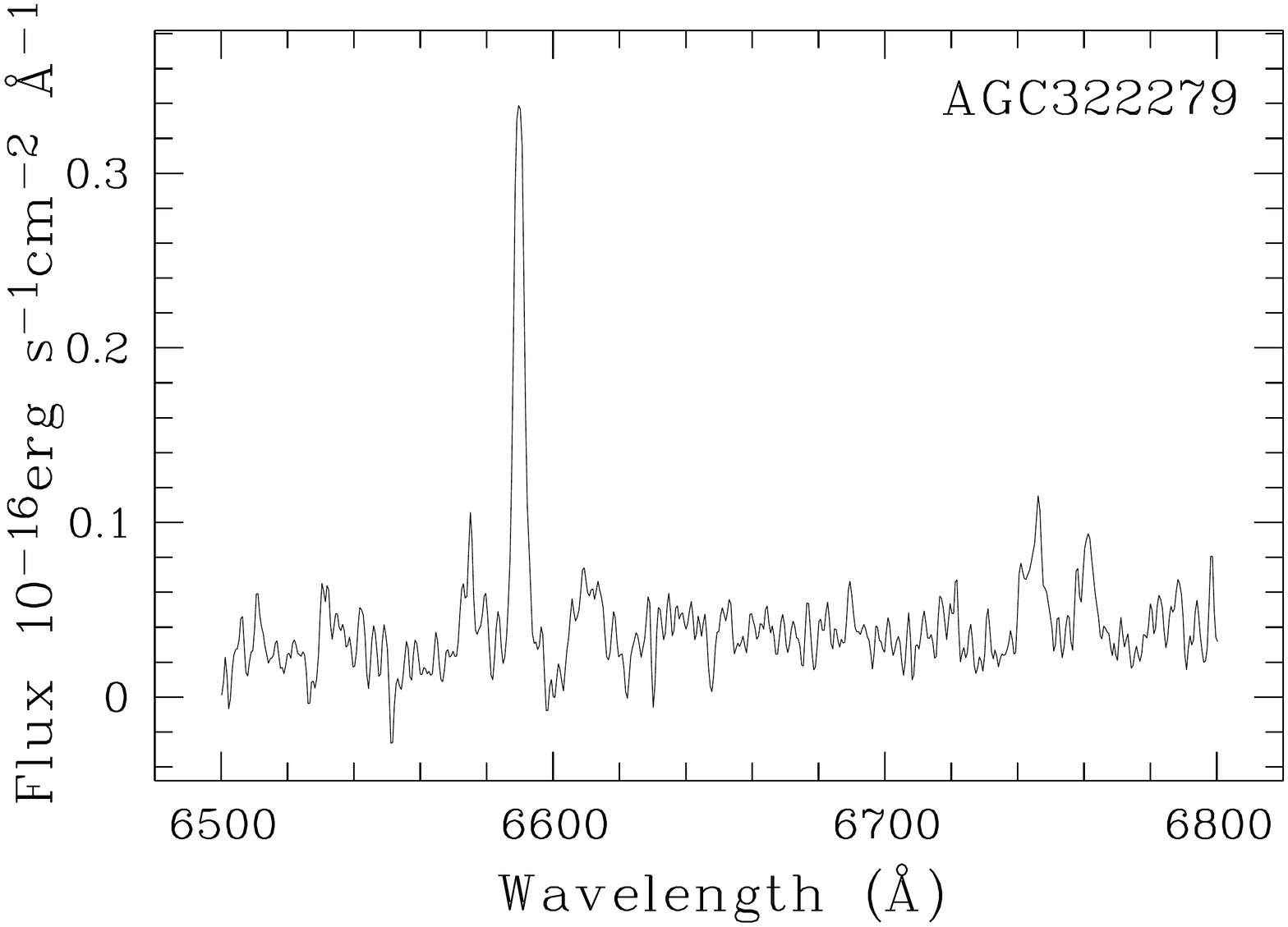}
\caption{1D spectra of void XMP candidates obtained with BTA.
The galaxy name and derived value of 12+$\log$(O/H) are shown at the top
of each box.
{\bf Top panel, from left to right:}
J1444+4242, O/H = 7.16 (average on 2 knots);
PGC2081790 = J1447+3630, O/H = 7.33; J1522+4201, O/H = 7.30;
{\bf Second  panel, left to right:}
J2103-0049, O/H = 7.21; AGC321307=J2214+2540, O/H = 7.89; AGC334513 =
J2348+2335, O/H = 7.24.
{\bf Bottom panel, from left to right:} {\bf Non-void galaxies:} red
spectrum of J0823+1748 with z = 0.092, AGC332939=J2308+3154, O/H = 7.69;
red spectrum of AGC322279=J2203+1747.
}
\label{fig:BTA_1Db}
\end{figure*}

\label{lastpage}

\end{document}